\documentclass[12pt,a4paper]{report}

\usepackage[top=0.70in, bottom=0.70in, left=0.8in,right=0.80in]{geometry} 
\usepackage[pdftex]{graphicx} 
\usepackage[final]{pdfpages} 
\usepackage{float} 
\usepackage{pslatex} 
\usepackage{array} 
\usepackage{setspace}
\usepackage{float}
\usepackage{enumerate}
\usepackage{longtable}
\newcommand{\eml}{\textsc{Embedded Matlab}}
\newcommand{\prover}{\textsc{Prover}}

\newcommand{\tdl}{\textsc{Tadl}}

\newcommand{\xtc}{\textsc{Xtc}}
\newcommand{\gt}[1]{\texttt{#1}}

\newcommand{\fp}{\textsc{$f_p$}}

\newcommand{\uppaal}{\textsc{Uppaal}}

\newcommand{\smc}{\textsc{Uppaal-smc}}

\newcommand{\simu}{\textsc{Simulink}}
\newcommand{\staf}{\textsc{Stateflow}}
\newcommand{\sldv}{\textsc{Simulink  Design Verifier}}
\newcommand{\sdv}{\textsc{SDV}}

\newcommand{\mt}{\textsc{Matlab}}
\newcommand{\ed}{\textsc{East-adl}}

\usepackage{makeidx}  
\usepackage{url}
\usepackage{graphicx}
\usepackage{subfigure}
\usepackage{makecell}
\usepackage{cite}
\usepackage{listings}
\usepackage{paralist}
\let\subparagraph\paragraph
\usepackage{cite}
\usepackage{amsfonts, amsmath, amssymb}
\usepackage{mathptmx}
\usepackage{algorithm}
\usepackage{algorithmic}
\usepackage{epstopdf}
\usepackage{float}
\usepackage{url}
\usepackage{enumerate}
\usepackage{color}
\usepackage{multirow}
\usepackage{verbatim}
\usepackage{bm}
\usepackage{longtable}
\usepackage[figuresright]{rotating}
\usepackage{geometry}
\usepackage{authblk}

\usepackage[font=small,labelfont=bf]{caption}

\usepackage{listings}
\usepackage{color}

\definecolor{dkgreen}{rgb}{0,0.6,0}
\definecolor{gray}{rgb}{0.5,0.5,0.5}
\definecolor{mauve}{rgb}{0.58,0,0.82}

\usepackage{fancyhdr}
\fancypagestyle{plain}{%
\fancyfoot[L]{\emph{Technical Report}} 
\fancyfoot[R]{\thepage}

}

\usepackage{pgf}
\usepackage{pgfpages}

\pgfpagesdeclarelayout{boxed}
{
  \edef\pgfpageoptionborder{0pt}
}
{
  \pgfpagesphysicalpageoptions
  {%
    logical pages=1,%
  }
  \pgfpageslogicalpageoptions{1}
  {
    border code=\pgfsetlinewidth{2pt}\pgfstroke,%
    border shrink=\pgfpageoptionborder,%
    resized width=.95\pgfphysicalwidth,%
    resized height=.95\pgfphysicalheight,%
    center=\pgfpoint{.5\pgfphysicalwidth}{.5\pgfphysicalheight}%
  }%
}
\begin{document}
\renewcommand\bibname{References}
\lhead{ }



\newpage
\begin{center}
\thispagestyle{empty}
\setlength{\voffset}{2in}
\vspace{2cm}
\LARGE{\textbf{Formal Analysis of Non-functional Properties for a Cooperative Automotive System}}\\
\vspace{2cm}
{\Large{\textbf{Eun-Young Kang$^{12}$, Li Huang$^{2}$ and Dongrui Mu$^{2}$}}}\\
\vspace{0.3cm}
\large{\textbf{$^1$PReCISE Research Centre,
University of Namur, Belgium}}\\
\vspace{0.3cm}
\large{\textbf{$^2$School of Data and Computer Science, \\Sun Yat-sen University, Guangzhou, China}}\\
\vspace{0.3cm}
\Large{{eykang@fundp.ac.be}}\\
\Large{{\{huangl223, mudr\}@mail2.sysu.edu.cn}}\\
\vspace{3cm}
\newpage

\end{center}

\newpage



\begin{center}
\thispagestyle{empty}
\vspace{2cm}
\LARGE{\textbf{ABSTRACT}}\\[1.0cm]
\end{center}
\thispagestyle{empty}
\large{\paragraph{}
Modeling and analysis of nonfunctional requirements is crucial in
automotive systems. \ed\ is an architectural language dedicated to
safety-critical automotive system design. We have previously modified
\ed\ to include energy constraints and transformed energy-aware timed
(ET) behaviors modeled in \simu/ \staf\ into \uppaal~models amenable to  formal verification. Previous work is extended in this paper by
including support for \sldv\ (SDV), i.e., the ET constraints are
translated into \emph{proof objective models} that can be verified using \sdv.
Furthermore, probabilistic extension of \ed\ constraints is defined and the semantics of the extended constraints is translated into verifiable  \uppaal\ models with stochastic semantics for formal verification. A  set of mapping rules are proposed to facilitate the guarantee of  translation. Verification \& Validation are performed on the extended  timing and energy constraints using \sdv\ and \smc. Our approach is demonstrated on a cooperative automotive system case study.}


\newpage

\pagenumbering{roman} 

\pagestyle{empty}
\addtocontents{toc}{\protect\thispagestyle{empty}}
\tableofcontents 

\addtocontents{lof}{\protect\thispagestyle{empty}}
\listoffigures 
\cleardoublepage

\pagestyle{fancy}

\newpage
\pagenumbering{arabic} 

\chapter{Introduction}

Model-driven development is rigorously applied in automotive systems in  which the software controllers interact with physical environments. The  continuous time behaviors (evolved with various energy rates) of those systems often rely on complex dynamics as well as on stochastic behaviors. Formal verification and validation (V\&V) technologies are indispensable and highly recommended for development of safe and  reliable automotive systems \cite{iso26262,iec61508}. Conventional V\&V, i.e., testing and model checking have limitations in terms of assessing the reliability of hybrid systems due to both the stochastic and  non-linear dynamical features. To ensure the reliability of safety critical hybrid dynamic systems, \emph{statistical model checking (SMC)}
techniques have been proposed
\cite{smc-challenge,smc-david-12,david2015uppaal}. These techniques for  fully stochastic models validate probabilistic performance properties of  given deterministic (or stochastic) controllers in given stochastic
environments.

\ed\ \cite{EAST-ADL,Blom2013EAST} is a domain specific language that
provides support for architectural modeling of automotive embedded
systems. A system in \ed\ is described by {\gt{Functional
Architectures (FA)}} at different abstraction levels. The {\small
\gt{FA}} are composed of a number of interconnected
\emph{functionprototypes} (\fp s), and the \fp s have ports and connectors
for communication. \ed\ relies on external tools for the analysis of
specifications related to requirements. For example, behavioral
description in \ed\ is captured in external tools, i.e., \simu/\staf
\cite{slsf}. The latest release of \ed\ has adopted the time model
proposed in the Timing Augmented Description Language (\tdl2)
\cite{TADL2}. \tdl2 allows for the expression and composition of timing
constraints, e.g., periodic, end-to-end delay, and synchronization
constraints on top of \ed\ models.

In this paper, we propose a model-driven approach to support
  formal analysis of energy and timed (ET) requirements for automotive
systems at the design level:
\begin{inparaenum} \item The ET constraints in \ed\ are interpreted in
\simu/\staf\ (S/S) and provide corresponding modeling extensions in
S/S;  \item  The ET requirements, specified in temporal logics, are
translated into \emph{proof objective models} to perform formal
verification using \sdv; \item Probabilistic extension of
\ed/\tdl\ constraints is defined and the semantics of the extended
constraints (\xtc) is translated into verifiable \smc\ \cite{smc} models with stochastic semantics for formal verification; \item A set of mapping  rules are proposed to facilitate the guarantee of translation; \item  Simulation and V\&V are performed on the \xtc\ and energy constraints  using \sdv\ and \smc. \end{inparaenum} Our approach is demonstrated on  the cooperative automotive system (CAS) case study.

The paper is organized as follows: Chap. \ref{sec:preliminary}
presents an overview of \simu, \sdv, and \smc. The CAS is introduced as  a  running example in Chap. \ref{sec:case-study}. Chap. \ref{sec:mdl-ss}, \ref{sec:sdv-trans}, \ref{sec:mdl-smc} and \ref{sec:smc-trans} describe a set of  translation patterns and how our approaches provide support for formal  analysis at the design level. The applicability of our method is  demonstrated by performing verification on the CAS case study in Chap.  \ref{sec:experiment}. Chap. \ref{sec:r-work} and  Chap. \ref{sec:conclusion} present related work and conclusion.

\chapter{Preliminary}
\label{sec:preliminary}
In our framework, Simulink and \eml\ (EML) are utilized for modeling purposes. Verification and experiment are performed by SDV and \smc.
\vspace{0.05in}

\begin{figure}[htbp]
\centerline{{\includegraphics[width=4.6in]{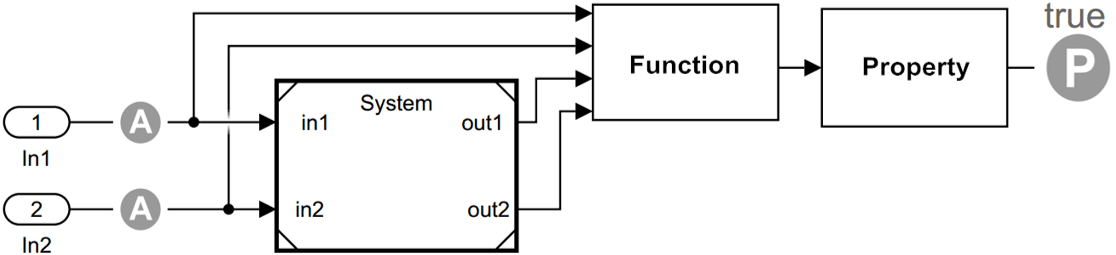}}}
\caption{General verification models in SDV}
\label{fig_general_verification_model}
\hfil
\end{figure}

\noindent \textbf{Simulink and SDV}: Simulink \cite{slsf} is a synchronous data flow language, which provides different types of \emph{blocks}, i.e., primitive-, control flow-, and temporal-blocks with predefined libraries for modeling and simulation of dynamic systems and code generation. Simulink supports the definition of custom blocks via Stateflow diagrams or \emph{user-defined function} blocks written in EML, C, and C++.
SDV is a plug-in to \prover, which is a formal verification tool that performs reachability analysis on Simulink/Stateflow model. The satisfiability of each reachable state is determined by a SAT solver. A proof objective model is specified in Simulink/SDV and illustrated in Fig.\ref{fig_general_verification_model}. A set of data (predicates) on the input flows of \emph{System} is constrained via {\scriptsize $\ll$}Proof Assumption{\scriptsize $\gg$} blocks during proof construction. A set of proof obligations is constructed via a function $F$ block and the output of $F$ is specified as input to a property $P$ block.
$P$ passes its output signal to an {\scriptsize $\ll$}Proof Objective{\scriptsize $\gg$} block and returns \emph{true} when the predicates set on the input data flows of the outline model are satisfied. Whenever {\scriptsize $\ll$}Proof Objective{\scriptsize $\gg$} (P block) is utilized, SDV verifies whether the specified input data flow is always \emph{true}.
The underlying \prover\ engine allows the formal verification of properties for a given model.
Any failed proof attempt ends in the generation of a counterexample representing an execution path to an invalid state.
A harness model is generated to analyze the counterexample and refine the model.

\begin{figure}[htbp]
\centering
  \subfigure[Implies]{
  \includegraphics[width=1.2in]{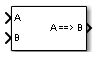}}
  \subfigure[Within Implies]{
  \includegraphics[width=1.2in]{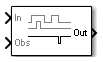}}
  \subfigure[From and Goto]{
  \includegraphics[width=2.5in]{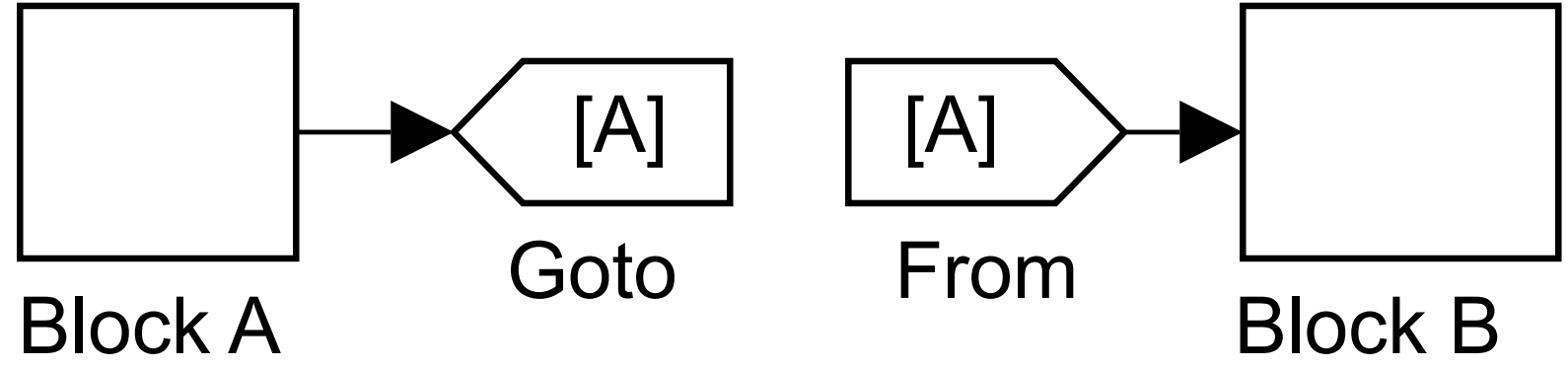}}
  \caption{Simulink Blocks used for property modeling}
\label{fig:sdvblock}
\end{figure}

Fig.\ref{fig:sdvblock} shows blocks in SDV that can be employed in properties specification.
{\scriptsize $\ll$}Implies{\scriptsize $\gg$} block (Fig.\ref{fig:sdvblock}.(a)) allows to specify a condition to produce a certain response: if A is true and B is false, the output is false; for all other pairs of A and B, the outputs are true.
{\scriptsize $\ll$}Within Implies{\scriptsize $\gg$} block (Fig.\ref{fig:sdvblock}.(b)) verifies the property that the response occurs within a given duration. This block captures the within implication by observing whether the \emph{Obs} input is true for at least one step within each true duration of \emph{In}. If \emph{Obs} is always false within a particular input true duration of \emph{In}, the output becomes false for one time step following the input true duration.
 As shown in Fig.\ref{fig:sdvblock}.(c), the two blocks always appear in pairs with the same tags. The output signal from Block A goes to the {\scriptsize $\ll$}Goto{\scriptsize $\gg$} block tagged with \emph{A}. {\scriptsize $\ll$}Goto{\scriptsize $\gg$} block then passes that signal to the {\scriptsize $\ll$}From{\scriptsize $\gg$} block tagged with \emph{A}. Block B receives the signal from the {\scriptsize $\ll$}From{\scriptsize $\gg$} block.

\vspace{0.05in}
\noindent \textbf{\smc} provides probabilistic analysis of properties based on statistical estimation of system models. By monitoring simulations of an hybrid system and using the statistic result to determine whether the properties are satisfied on top of the system model, an exhaustive exploration of the state-space of the system can be avoided.  \smc\ represents systems via networks of Timed Automata (TA) whose behaviours depend on both stochastic and non-linear dynamical features.
 Its clocks evolve with various rates, such rates can be specified with ordinary differential equations. \smc\ provides a number of queries related to the stochastic interpretation of TA (STA) \cite{david2015uppaal}.
 The following four queries are sufficient to perform V\&V on timing and energy constraints in \ed, where $N$ and $bound$ indicate the number of simulations and the simulation time for each simulation respectively.
\begin{enumerate}
\item \emph{Probability Estimation} estimates the probability that a requirement property $\phi$ is satisfied for a given STA model within the time bound: $Pr[bound]$ $\phi$;
\item \emph{Hypothesis Testing} checks if the probability of $\phi$ is satisfied over a given probability $P_0$: $Pr[bound]$ $\phi$ $\ge$ $P_0$;
\item \emph{Simulations}: \smc\ runs multiple simulations on the system model and the $k$ (state-based) properties/expressions $\phi_1,..., \phi_k$ are monitored and visualized along the simulations: $simulate$ $N$ $[\leq$ $bound]\{\phi_1,..., \phi_k\}$;
\item \emph{Expected Value} evaluates the minimal or maximal value of a clock or an integer value while \smc\ checks the STA model: $E[bound; N](min:\phi)$ or $E[bound; N](max:\phi)$.

\end{enumerate}

\chapter{Case Study}
\label{sec:case-study}
\begin{figure}[htbp]
  \centering
  \includegraphics[width=4.5in]{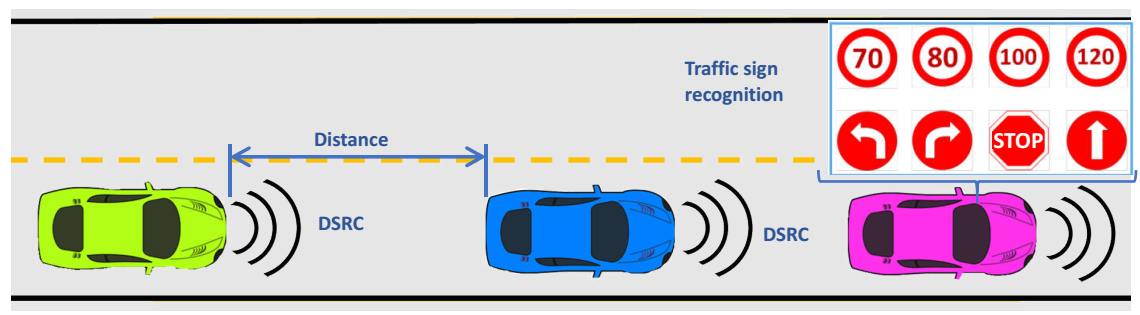}
  \caption{Cooperative Automotive System}
  \label{cass}
\end{figure}

A Cooperative Automotive System (CAS) is adopted as our running example where sensing and actuation are distributed over a number of vehicles, which is illustrated in Fig.\ref{cass}. A vehicle can be either a lead vehicle (\emph{leadVe}) or a following vehicle (\emph{followVe}) depends on its absolute position. We name the first vehicle as v1, the vehicle in the middle as v2 and the last vehicle v3. Communication between vehicles are realized by the Dedicated Short-Range Communication (DSRC), which is a wireless communication technology specifically dedicated for message transmission among automotive systems.
The lead vehicle can either run automatically (\emph{auto} mode) by detecting traffic signs on the road or be driven by the driver (\emph{userCtrl} mode). Any two adjacent vehicles should be distant enough to avoid rear-end collision and close enough to guarantee the communication quality.

The requested vehicle motion is based on driver's input, detected environment (e.g., traffic sign) and running situation.
The vehicle movement can be either horizontal or vertical and the position of each vehicle is represented by two-dimension coordinates (x, y) in Cartesian coordinate system, where x and y are distances measured from vehicle to two fixed perpendicular directed lines respectively.
Ideally, the following vehicle should maintain the same movement as the vehicle ahead and behaves similarly, e.g., when a lead vehicle accelerates, the following vehicles should be able to accelerate to a similar extent.

\begin{figure*}
\centering
  \subfigure[Architecture of cooperative automotive system model]{
  \includegraphics[width=5.8in]{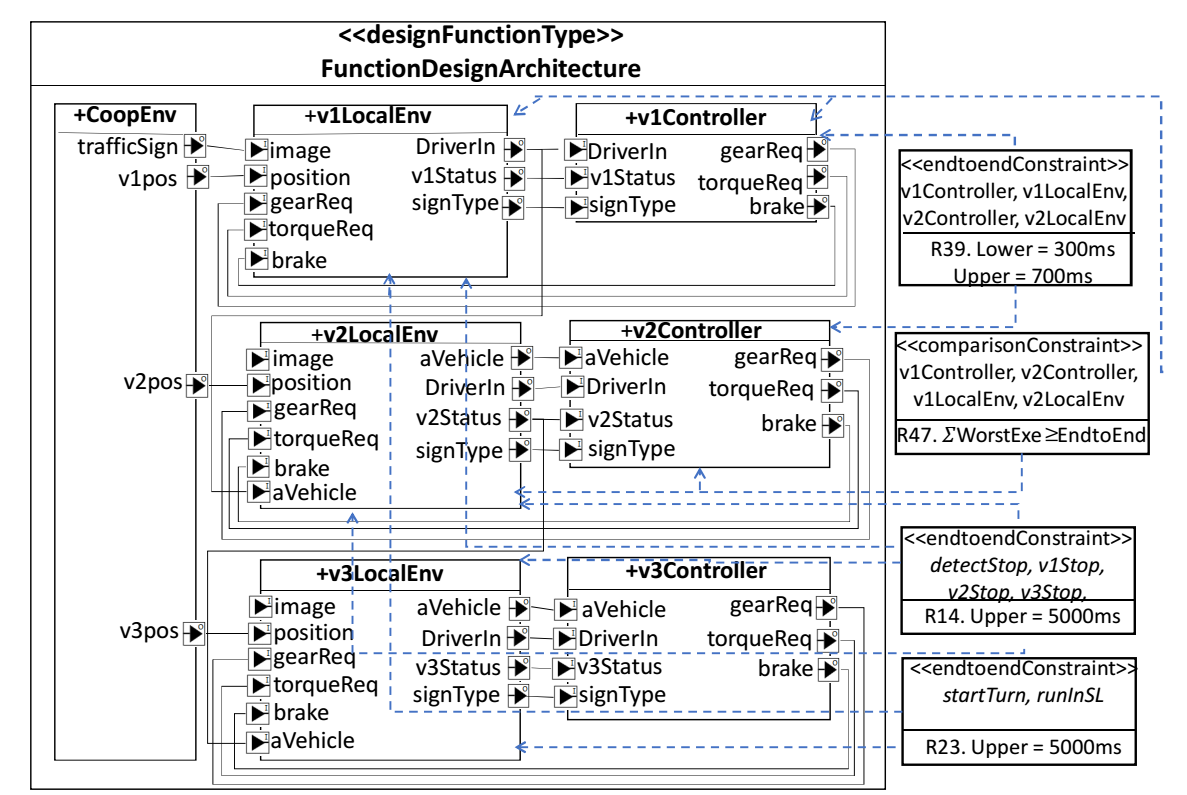}
  \label{fig:topview}}
  \subfigure[Function architecture of a vehicle ]{
  \includegraphics[width=4.8in]{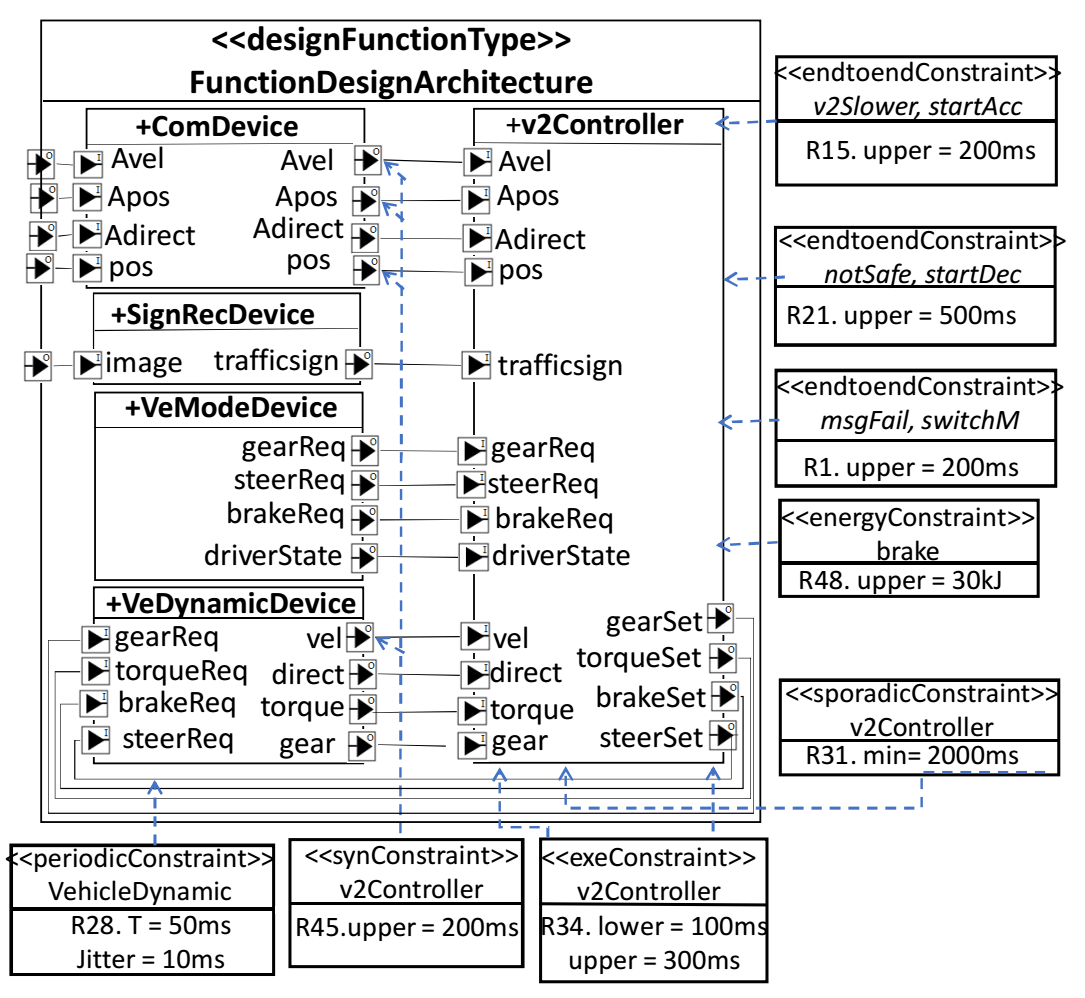}
  \label{fig:individualvehicle}}
  \caption{\ed\ Architecture of CAS}
\label{fig:east}
\end{figure*}

Fig.\ref{fig:topview} shows the \ed\ model of the CAS, which consists of cooperative environment ({\gt{CoopEnv}}) and local environments, left and middle, and the software controllers of three vehicles, right. The three {\gt{LocalEnv}} \fp s receive information of positions and the traffic sign from the {\gt{CoopEnv}}. Each lead vehicle sends its own position and velocity to its adjacent following vehicle.
The functionality of an individual vehicle, augmented with timing/energy constraints and viewed as {\gt{FunctionDesignArchitecture} (FDA)} in \ref{fig:individualvehicle}, consists of the following \fp s (components): {\gt{ComDevice}} represents the DSRC device for vehicle to send and receive messages. {\gt{SignRecDevice}} analyzes the detected images and computes desired images (sign types). {\gt{VeModeDevice}} reads the driver's requests by means of pedals (\emph{brakeReq}), steering wheel (\emph{steerReq}), gear (\emph{gearReq}) and a switch (\emph{driverState}) which indicates whether the driver intends to drive. {\gt{v1Controller}} determines how the torque and gear of the vehicle are adjusted based on the received information. {\gt{VeDynamicDevice}} specifies the kinematics behaviors of the vehicle based on engine torque, brake, steer and gear. The value of {\gt{signType}} ranges from 0 to 5, which represents ``straight'', ``min/max speed limit'', ``left/right turn ahead'' and ``stop'' signs respectively.

We consider the following functional and non-functional properties, which includes {\gt{Execution}}, {\gt{End-to-End}}, {\gt{Synchronization}}, {\gt{Periodic}}, {\gt{Sporadic}} and \\{\gt{Comparison}} timing and {\gt{Energy}} constraints, on top of the CAS EAST-ADL model, which are sufficient to capture the constraints described in Fig.\ref{fig:east}.
{\gt{Energy}} constraint refers to the battery consumption of an individual vehicle in CAS.
\begin{itemize}[]
 \item  R1. When v1 runs automatically and if there is no message received by the vehicle for a certain time, the running mode of the vehicle should be changed to \emph{manual} mode within a certain time.
  \item R2. When v2 runs automatically and if there is no message received by the vehicle for a certain time, the running mode of the vehicle should be changed to \emph{manual} mode within a certain time.
 \item R3. When v3 runs automatically and if there is no message received by the vehicle for a certain time, the running mode of the vehicle should be changed to \emph{manual} mode within a certain time.
 \item R4. When v1 runs automatically at a constant speed, if it detects a stop sign, it should start to brake in 500ms.
 \item R5. When v1 runs automatically at a constant speed, if it detects left turn sign, it should start to turn left in 200ms.
 \item R6. When v1 runs automatically at a constant speed, if it detects right turn sign, it should start to turn right in 200ms.
\item R7. If v1 is driven by the driver and it is running with a constant speed, the driver steers to the left, the vehicle will turn left within 200ms.
\item R8. If v1 is driven by the driver and it is running at a constant speed, the driver steers to the right, the vehicle will turn right within 200ms.
\item R9. If v1 is driven by the driver and it is running at a constant speed, the driver brake the car, the vehicle will slow down.
\item R10. When v1 is driven by the driver and it is running at a constant speed, if the driver gears up, the vehicle will accelerate.
\item R11. When v1 is driven by the driver and it is running at a constant speed, if the driver gears down, the vehicle will decelerate.
\item R12. v2 should not run ahead of v1.
\item R13. v3 should not run ahead of v2.
\item R14. When the lead vehicle detects a stop sign, the three vehicles should stop in 5s.
\item R15. When the vehicles are running straight with a constant speed, if v2 runs slower than v1, it should start to accelerate within a certain time.
\item R16. When the vehicles are running straight with a constant speed, if v3 runs slower than v2, it should start to accelerate within a certain time.
\item R17. When the vehicles are running straight with a constant speed, if v2 runs faster than v1, it should start to decelerate within a certain time.
\item R18. When the vehicles are running straight with a constant speed, if v3 runs faster than v2, it should start to decelerate within a certain time.
\item R19. If the distance between v1 and v2 is larger than  500m, v2 should accelerate within a certain time, e.g. 200ms.
\item R20. If the distance between v2 and v3 is larger than  500m, v3 should accelerate within a certain time, e.g. 200ms.
\item R21. If the distance between v1 and v2 is smaller than the safety distance, v2 should decelerate within a certain time, e.g. 500ms.
\item R22. If the distance between v2 and v3 is smaller than the safety distance, v3 should decelerate within a certain time, e.g. 500ms.
\item R23. When v1 starts to turn left, v1 and v2 should finish  turning and run in the same lane within a given time.
\item R24. When v2 starts to turn left, v2 and v3 should finish  turning and run in the same lane within a given time.
\item R25. When v1 starts to turn right, v1 and v2 should finish  turning and run in the same lane within a given time.
\item R26.  When v2 starts to turn right, v2 and v3 should finish  turning and run in the same lane within a given time.
\item R27. A {\gt{Periodic}} acquisition of {\gt{VehicleDynamic}} of v1 must be carried out for every 50 ms with a jitter 10 ms.
\item R28. A {\gt{Periodic}} acquisition of {\gt{VehicleDynamic}} of v2 must be carried out for every 50 ms with a jitter 10 ms.
\item R29. A {\gt{Periodic}} acquisition of {\gt{VehicleDynamic}} of v3 must be carried out for every 50 ms with a jitter 10 ms.
\item R30. {\gt{Sporadic}} constraint: After the running mode of v1 is changed to manual mode because of the failed message transmission, the driver should not change it into \emph{auto} mode within 20 seconds.
\item R31. {\gt{Sporadic}} constraint: After the running mode of v2 is changed to manual mode because of the failed message transmission, the driver should not change it into \emph{auto} mode within 20 seconds.
\item R32. {\gt{Sporadic}} constraint: After the running mode of v3 is changed to manual mode because of the failed message transmission, the driver should not change it into \emph{auto} mode within 20 seconds.
\item R33. An {\gt{Execution}} constraint applied on {\gt{v1Controller}}, which measured from the input port \emph{Avel} to output ports \emph{gearReq}, \emph{torqueReq} and \emph{brake}, should be between 100 ms and 300 ms;
\item R34. An {\gt{Execution}} constraint applied on {\gt{v2Controller}}, which measured from the input port \emph{Avel} to output ports \emph{gearReq}, \emph{torqueReq} and \emph{brake}, should be between 100 ms and 300 ms;
\item R35. An {\gt{Execution}} constraint applied on {\gt{v3Controller}}, which measured from the input port \emph{Avel} to output ports \emph{gearReq}, \emph{torqueReq} and \emph{brake}, should be between 100 ms and 300 ms;
\item R36.  An {\gt{Execution}} constraint applied on {\gt{ComDevice}} of v1, which measured from the input ports \emph{Avel}, \emph{Apos}, \emph{Adirect} and \emph{pos} to output ports \emph{Avel}, \emph{Apos}, \emph{Adirect} and \emph{pos} should be between 50 ms and 100 ms;
\item R37. An {\gt{Execution}} constraint applied on {\gt{ComDevice}} of v2, which measured from the input ports \emph{Avel}, \emph{Apos}, \emph{Adirect} and \emph{pos} to output ports \emph{Avel}, \emph{Apos}, \emph{Adirect} and \emph{pos} should be between 50 ms and 100 ms;
\item R38. An {\gt{Execution}} constraint applied on {\gt{ComDevice}} of v3, which measured from the input ports \emph{Avel}, \emph{Apos}, \emph{Adirect} and \emph{pos} to output ports \emph{Avel}, \emph{Apos}, \emph{Adirect} and \emph{pos} should be between 50 ms and 100 ms;
\item R39. An {\gt{End-to-End}} is measured from {\gt{v1Controller}} to {\gt{v2Controller}}. The time interval is bounded with a minimum value of 300 ms and a maximum value of 700 ms;
\item R40. An {\gt{End-to-End}} is measured from {\gt{v2Controller}} to {\gt{v3Controller}}. The time interval is bounded with a minimum value of 300 ms and a maximum value of 700 ms;
\item R41.  An {\gt{End-to-End}} constraint measured from input port \emph{position} of {\gt{v1LocalEnv}} to output ports \emph{gearReq}, \emph{torqueReq} and \emph{brake} of {\gt{v1Controller}} should be limited within [200, 500]ms;
\item R42.  An {\gt{End-to-End}} constraint measured from input port \emph{position} of {\gt{v2LocalEnv}} to output ports \emph{gearReq}, \emph{torqueReq} and \emph{brake} of {\gt{v2Controller}} should be limited within [200, 500]ms;
\item R43.  An {\gt{End-to-End}} constraint measured from input port \emph{position} of {\gt{v3LocalEnv}} to output ports \emph{gearReq}, \emph{torqueReq} and \emph{brake} of {\gt{v3Controller}} should be limited within [200, 500]ms;
\item R44. {\gt{Synchronization}} constraint: The position and velocity of the vehicle and its lead vehicle (\emph{pos, vel, Apos and Avel} ports on {\gt{v1Controller}}) must be detected by {\gt{v1Controller}} within a given time window, i.e., the tolerated maximum constraint = 200 ms;
\item R45. {\gt{Synchronization}} constraint: The position and velocity of the vehicle and its lead vehicle (\emph{pos, vel, Apos and Avel} ports on {\gt{v2Controller}}) must be detected by {\gt{v2Controller}} within a given time window, i.e., the tolerated maximum constraint = 200 ms;
\item R46. {\gt{Synchronization}} constraint: The position and velocity of the vehicle and its lead vehicle (\emph{pos, vel, Apos and Avel} ports on {\gt{v3Controller}}) must be detected by {\gt{v3Controller}} within a given time window, i.e., the tolerated maximum constraint = 200 ms;
\item R47. A {\gt{Comparison}} constraint: The execution time interval measured from \\{\gt{v1Controller}} to {\gt{v2Controller}} needs to be larger than or equal to the sum of the worst-case execution times of the \fp s.
\item R48. {\gt{Energy}} constraint: The maximum battery energy consumption for the braking of a vehicle should be less than 30kJ;
\item R49. {\gt{Energy}} constraint: The maximum battery energy consumption for the controller to decide its movement for once
should be less than 30J;
\item R50. {\gt{Energy}} constraint: The maximum battery energy consumption for {\gt{ComDevice}} to get/send signals for once should be less than 5J;
\end{itemize}

According to the \ed\ meta-model, the timing constraint describes a design constraint, but has the role of a property, requirement or validation result, based on its {\gt{Context}} \cite{EAST-ADL}. The \tdl\ meta-model is integrated with the \ed\ meta-model and is supplemented with structural concepts from \ed. The \ed/\tdl\ constraints contain the identifiable state changes as \emph{Events}. The causality related events are contained as a pair by \emph{EventChains}. Based on \emph{Event} and \emph{EventChains}, data dependencies, control flows, and critical execution paths are represented as additional constraints for the \ed\ functional architectural model, and apply timing constraints on these paths.

{\gt{Sporadic}} constraint specifies a minimum delay \emph{min} between any two consecutive occurrences of an event.
{\gt{Execution}} constraint restricts the time interval between inputs and outputs of a \fp, which can be specified using \emph{lower, upper} values given as {\gt{Execution}} constraint. {\gt{End-to-End}} constraint gives duration bounds (minimum and maximum) between two events \emph{source} and \emph{target}. The time intervals between \emph{source} event and \emph{target} event should satisfy \emph{lower, upper} values given as {\gt{End-to-End}} constraint.
{\gt{Synchronization}} constraint describes how tightly the occurrences of a group of events follow each other. All events must occur within a sliding window, specified by the \emph{tolerance} attribute, i.e., the maximum time interval allowed between events.
{\gt{Comparison}} constraint specifies the comparison relation (including $<$, $\leq$, $=$, $\geq$ and $>$) among timing expressions.
{\gt{Periodic}} constraint states that the period of successive occurrences of a single event must have a length of at least a \emph{lower} and at most an \emph{upper} time interval. The interval is given as {\gt{Periodic}} constraint. There are two types of {\gt{Periodic}} timing constraints:  cumulative ({\gt{Cumu\_Periodic}}) and non-cumulative timing constraints ({\gt{Noncumu\_Periodic}}).  Those (non)-functional properties are formally specified and verified in the  following sections.

\chapter{Modeling Non-functional Behaviors in \simu~\& \staf}
\label{sec:mdl-ss}
\section{Behavioral Modeling of \fp s in \simu~\& \staf~}
\begin{figure}[htbp]
  \centering
  \includegraphics[width=4.3in]{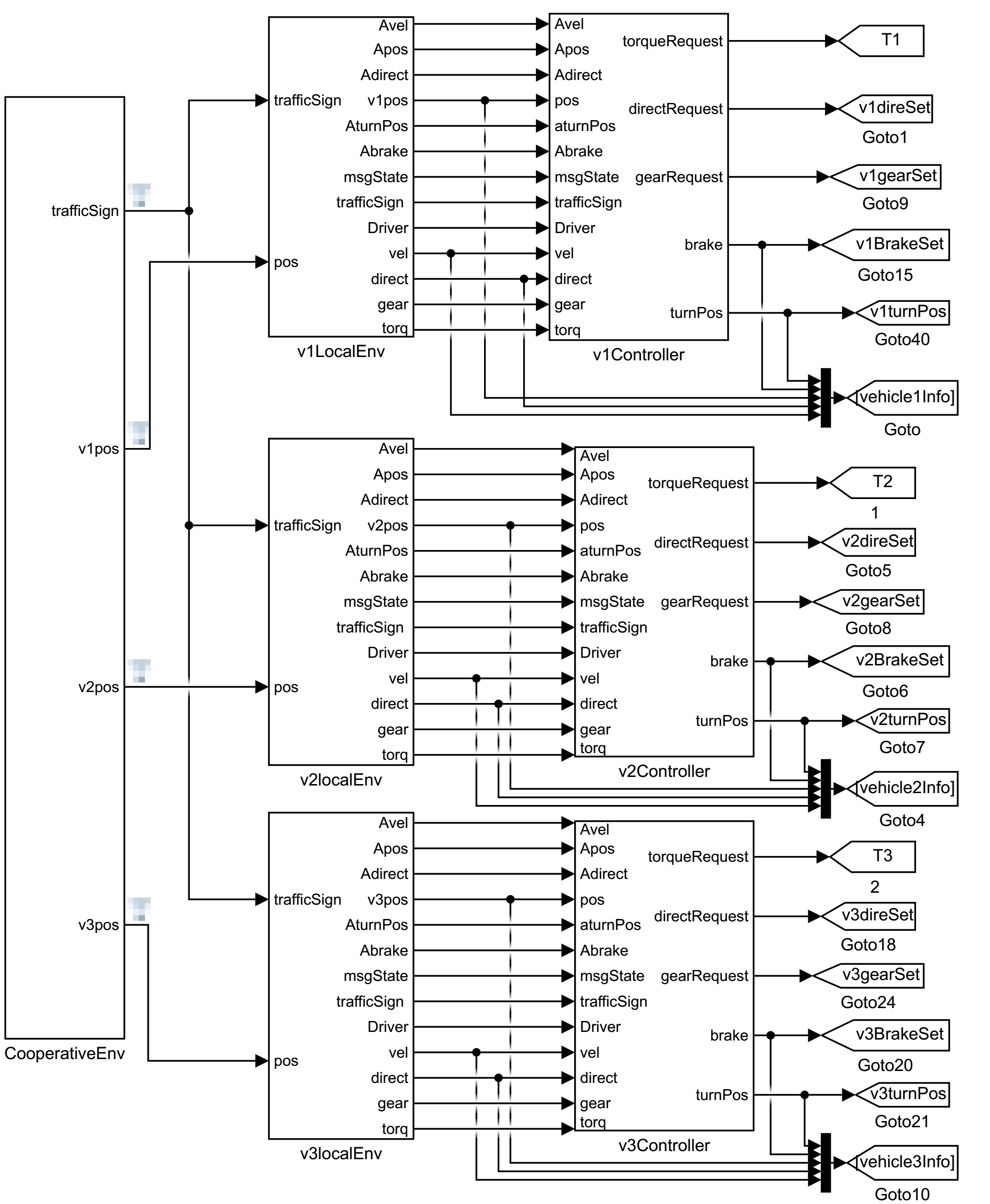}
\caption{Top-level diagram of CAS \simu~\& \staf\ model}
  \label{fig:cs_ss}
\end{figure}

Based on the \ed\ architectural model of CAS, the functional and non-functional behaviours of \fp s is constructed with Stateflow chart and Simulink blocks. The architecture of Simulink/Stateflow model can be seen in Fig.\ref{fig:cs_ss}. {\gt{v1Controller}} \fp~is modeled as a Stateflow chart while {\gt{localEnv}}, {\gt{CoopEnv}}  are modeled as Simulink subsystems with a number of mathematical and descriptive blocks provided in Simulink block library.

\begin{figure}[htbp]
  \centering
  \includegraphics[width=5in]{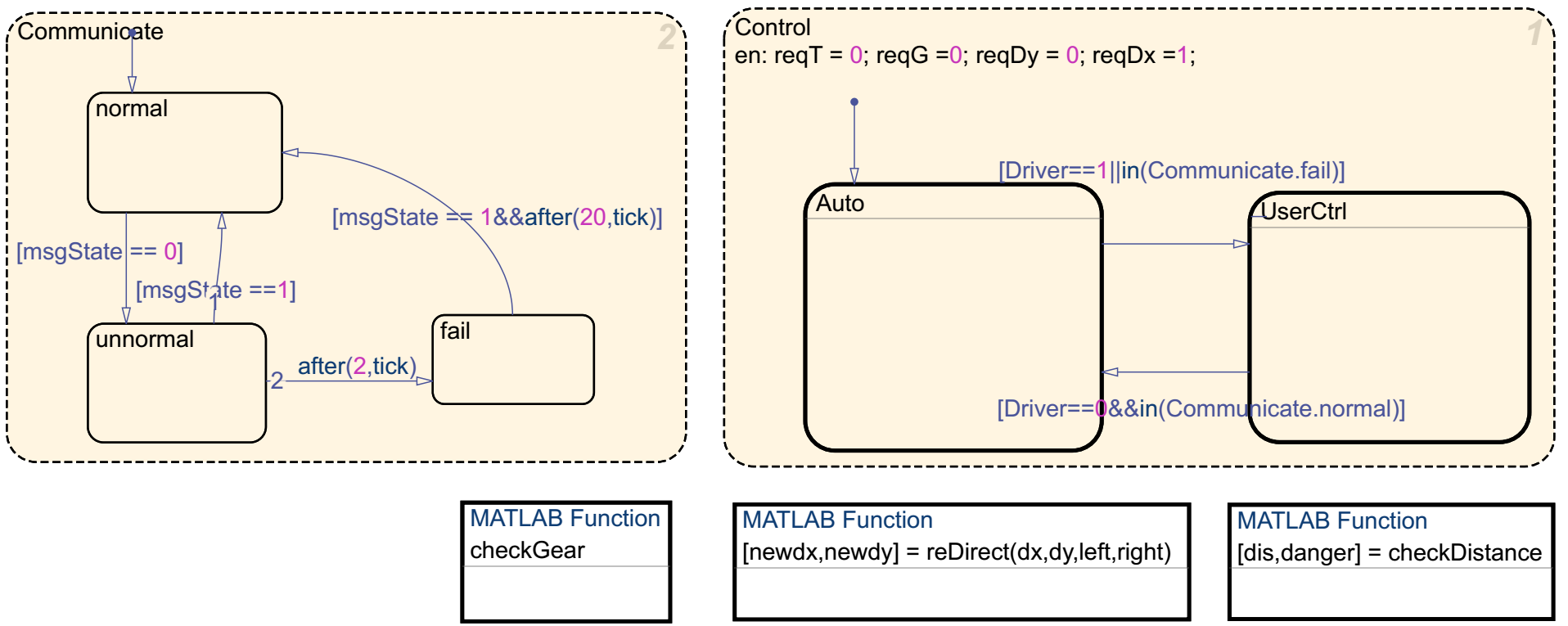}
\caption{ \staf\ chart of {\gt{v1Controller}}}
  \label{fig:topctrl_ss}
\end{figure}

The top-level diagram of  {\gt{v1Controller}} Stateflow chart is illustrated in Fig.\ref{fig:topctrl_ss}, where three Matlab functions blocks are embedded: \emph{checkGear()} is used to ensure that \emph{gear} is in the reasonable value range; \emph{reDirect()} adjusts vehicle's direction when vehicle turns; \emph{checkDis} calculates the distance between the vehicle and the vehicle in front of it. A single Stateflow chart, consisting of two parallel states, implements the control logic of the system in its entirety. In \emph{Control} state, \emph{Auto} and \emph{userCtrl} refer two different running modes of a vehicle depend on whether the driver controls the vehicle.

\begin{figure}[htbp]
  \centering
  \includegraphics[width=6.3in]{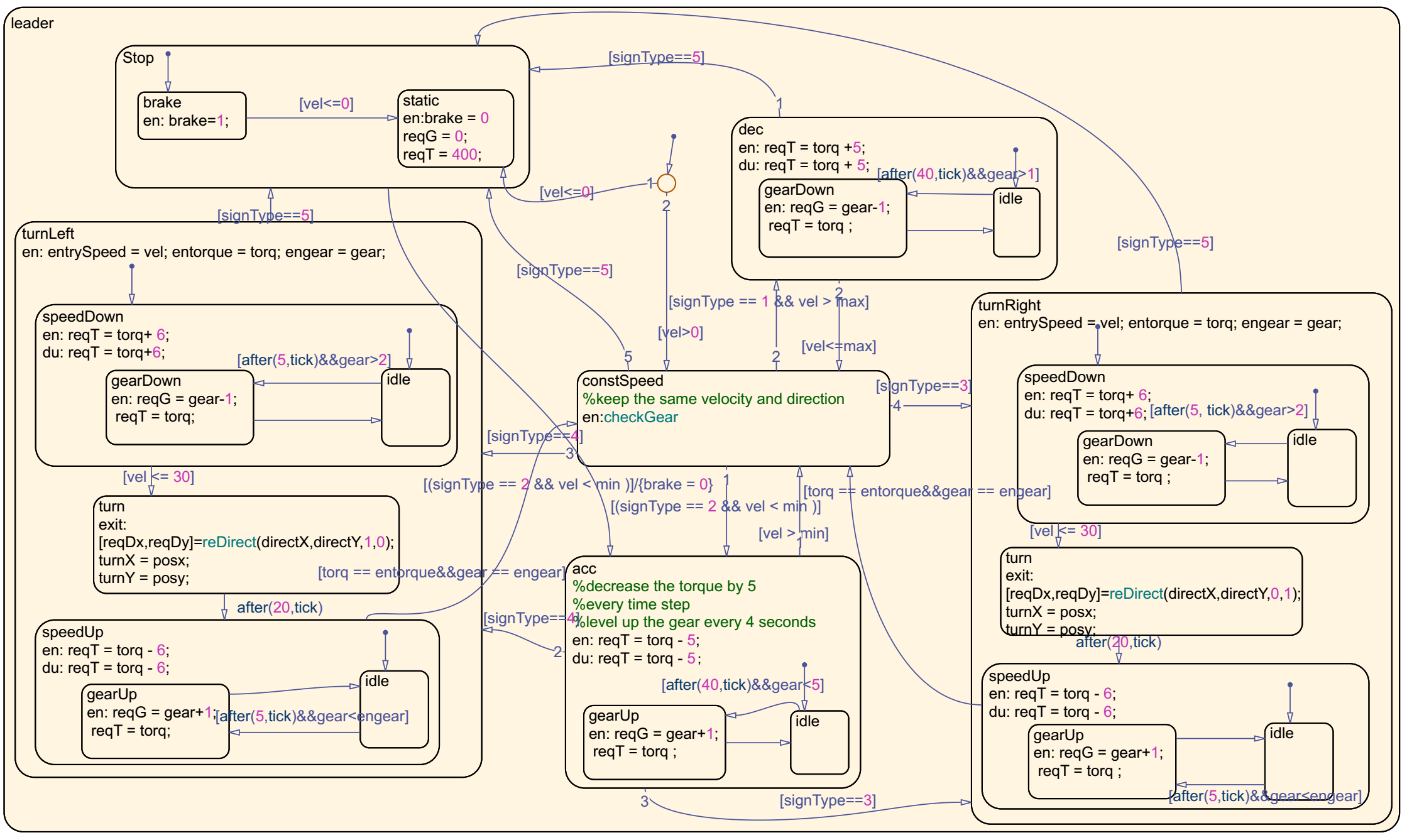}
\caption{\emph{leader} state in {\gt{v1Controller}}}
  \label{fig:leader_ss}
\end{figure}

\begin{figure}[htbp]
  \centering
  \includegraphics[width=5in]{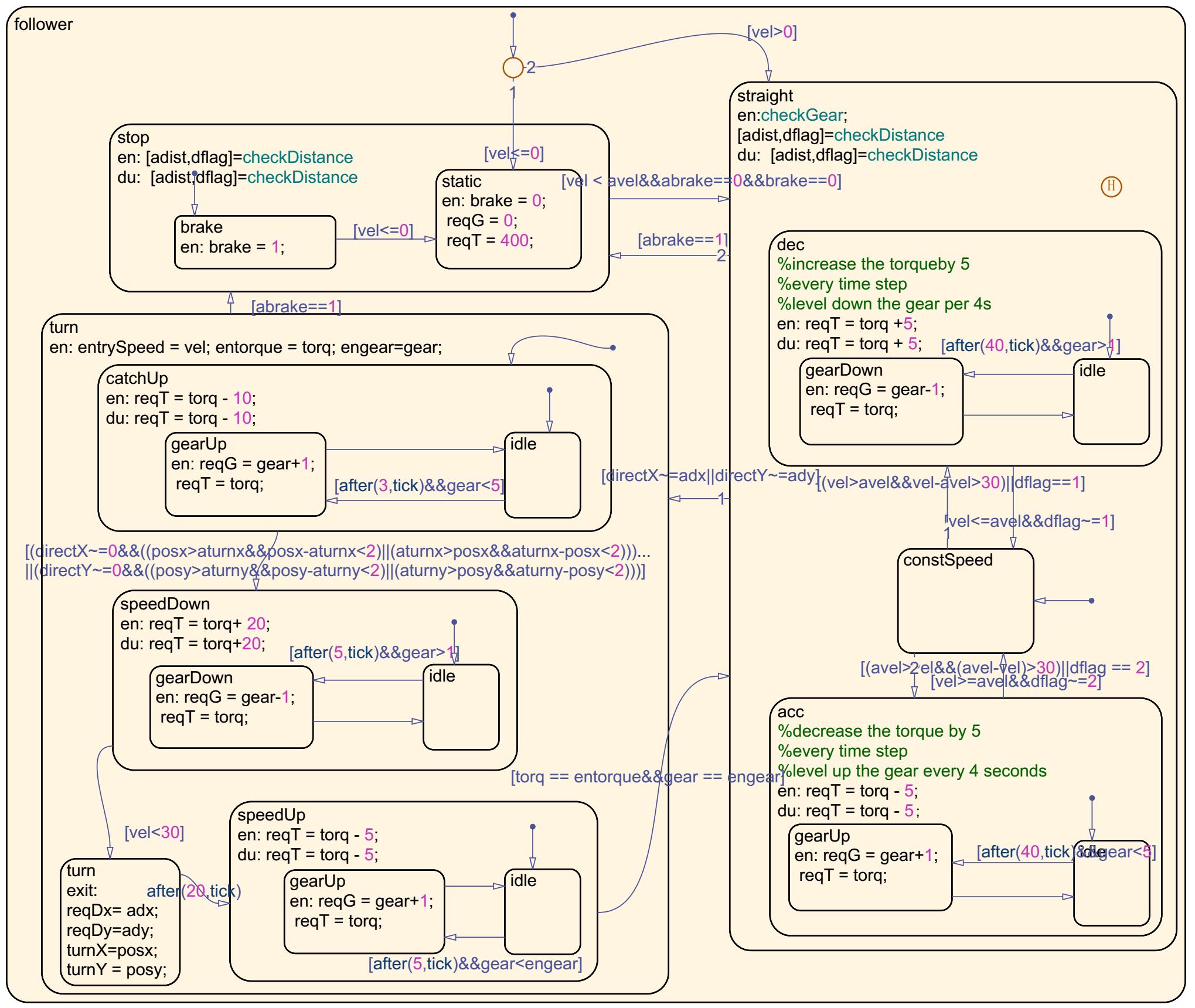}
\caption{\emph{follower} state in {\gt{v1Controller}}}
  \label{fig:follower_ss}
\end{figure}

If the vehicle runs automatically, either \emph{leader} (shown in Fig.\ref{fig:leader_ss}) or \emph{follower} (shown in Fig.\ref{fig:follower_ss}) will be activated according to the absolute positions of the three vehicles.  Inside the \emph{leader} state, there are six superstates \emph{turn\_right}, \emph{turn\_left}, \emph{constSpeed}, \emph{acc}, \emph{dec} and \emph{stop} together with their child states. The edges represent transitions between the states with the conditions on the edges. Initially, The chart transits to either \emph{straight} or \emph{stop} based on the initial speeds of the wheels. The transitions will be taken according to the current speed of the wheels and the value of {\gt{signType}}, whose various evaluation represents the different traffic sign. For example, when the vehicle is in \emph{constSpeed} state and the detected {\gt{signType}} is 5 (stop sign is encountered), \emph{braking} and \emph{stop} will be activated and \emph{brake} will become 1 indicating that the velocity of vehicle should be decreased steeply. Whenever the vehicle detects a stop sign, the vehicle will start to brake. If the straight sign is detected ({\gt{signType}} is 0), the vehicle will maintain the current state and speed. If the vehicle recognizes a left turn sign, it will first decrease its speed to a low level (less than 30 km/h) and start to turn. After turning left, the vehicle will finally go straight and maintain the speed. \emph{turnX} and \emph{turnY} specify the location at which the vehicle turns.

Since a following vehicle should follow the traces and movements of the vehicle in front of it, in \emph{follower} state, the transitions will be taken once the safety condition is violated, e.g., the distance between the two adjacent vehicles is either larger than 500m or less than the safety distance. As illustrated in \emph{turn} state on the left of Fig.\ref{fig:follower_ss}, when the following vehicle is turning left/right, it will first accelerate to reach to the turning location at which the lead vehicle turns.

\begin{figure}[htbp]
  \centering
  \includegraphics[width=5in]{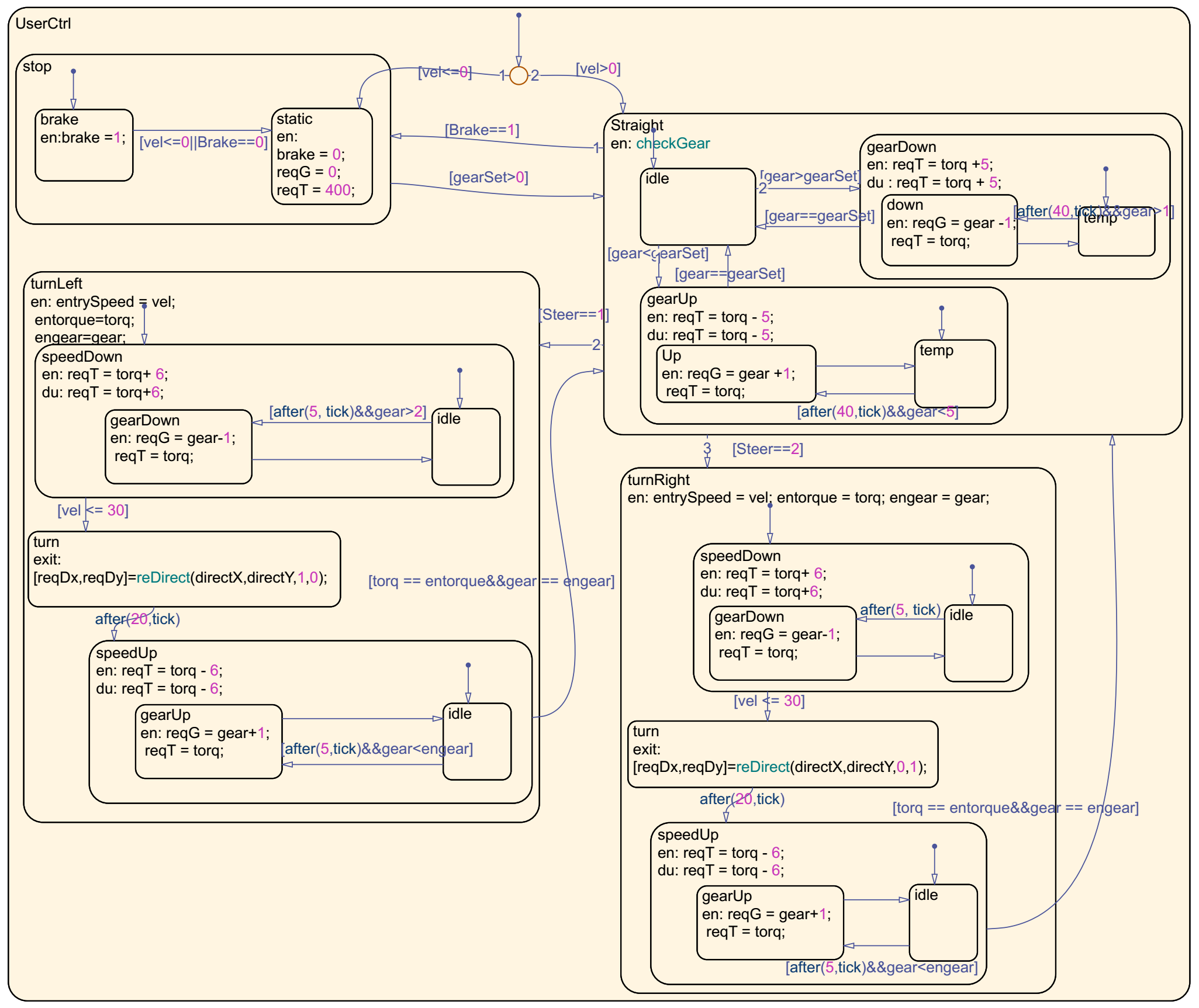}
\caption{\emph{userCtrl} state in {\gt{v1Controller}}}
  \label{fig:user_ss}
\end{figure}

If the vehicle is driven by its driver, \emph{userCtrl} state (as illustrated in Fig.\ref{fig:user_ss}) will be active. The running mode (\emph{stop}, \emph{turnLeft}, \emph{turnRight} and \emph{straight}) of the vehicle will be changed based on the driver's operations. \emph{steer} represents the steering operation is conducted, i.e., steering to left (\emph{steer} = 1), right (\emph{steer} = 2) and straight (\emph{steer} = 3).

\begin{figure}[htbp]
  \centering
  \includegraphics[width=4in]{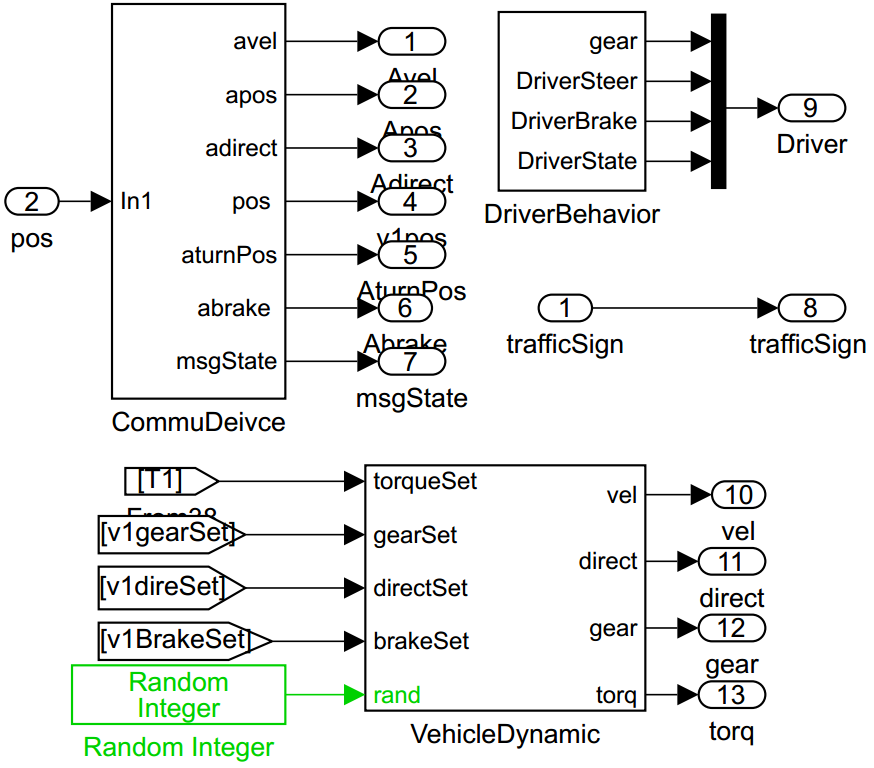}
\caption{ \simu~\& \staf\ model of {\gt{localEnv}}}
  \label{fig:localenv_ss}
\end{figure}

Fig.\ref{fig:localenv_ss} illustrates the Simulink and Stateflow model of {\gt{localEnv}}, whose functionality is captured in three subsystems:
{\gt{ComDevice}} is responsible for signal propagation and  message receiving; {\gt{DriverBehavior}} records the value of signals (\emph{gear, steer, brake and DriverState}) that represent various operations conducted by the driver; {\gt{VehicleDynamic}} specifies the how the velocity and running direction of the vehicle are changed according to the input \emph{torque} and \emph{gear} value. In this figure,  blocks/entities in green format are employed to model timing constraints, whose modeling process will be explained detailly in the following section.

\begin{figure}[htbp]
  \centering
  \includegraphics[width=6in]{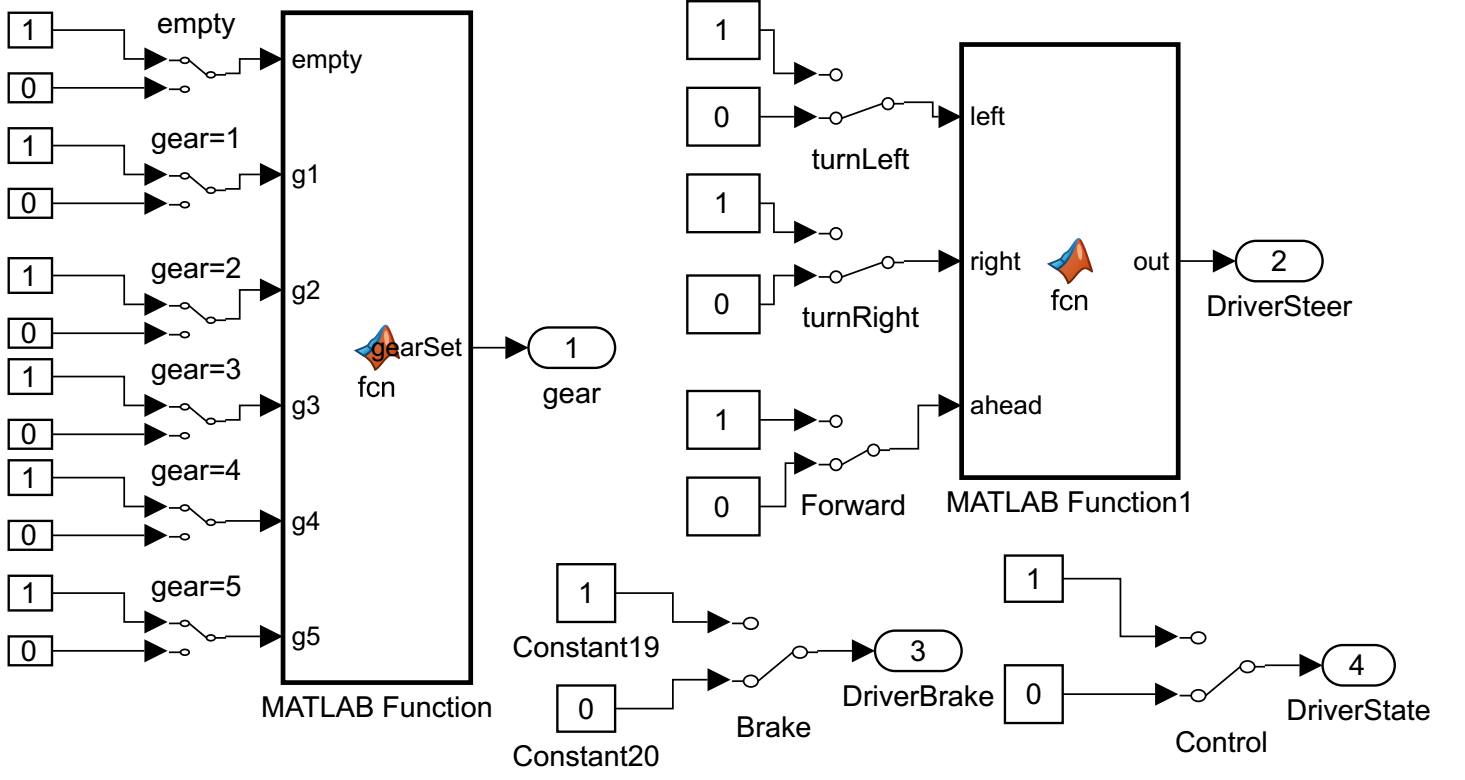}
\caption{ \simu~\& \staf\ model of {\gt{DriverBehavior}}}
  \label{fig:driver_ss}
\end{figure}

Fig.\ref{fig:driver_ss} shows the Simulink and Stateflow model of {\gt{DriverBehavior}} \fp. This subsystem obtains inputs from the driver by {\scriptsize $\ll$}Manual Switch{\scriptsize $\gg$} block, which represents a toggle switch where  one of its two inputs (1 or 0) will be selected to pass through and become the value of the output.

\begin{figure}[htbp]
  \centering
  \includegraphics[width=5in]{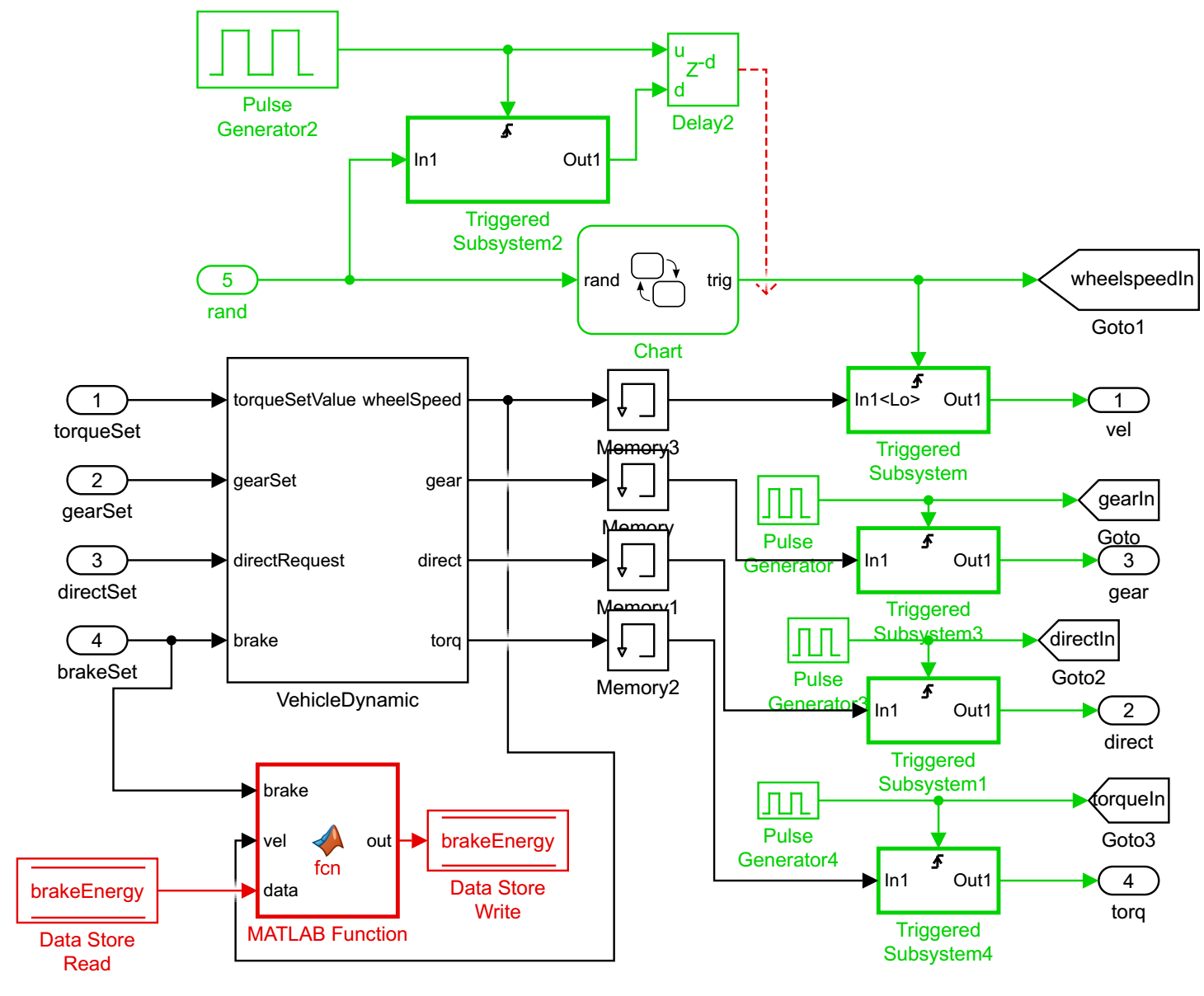}
\caption{ \simu~\& \staf\ model of {\gt{VehicleDynamic}}}
  \label{fig:vd_ss}
\end{figure}

Fig.\ref{fig:vd_ss} shows the Simulink and Stateflow model of {\gt{VehicleDynamic}} \fp. In {\gt{VehicleDynamic}} subsystem, a {\scriptsize $\ll$}Look Up Table{\scriptsize $\gg$} block maps inputs (\emph{gear} and \emph{torqueSet}) to an output value (\emph{wheelspeed}) by looking up or interpolating the predefined values in a table. {\scriptsize $\ll$}DataStore{\scriptsize $\gg$} block records the energy consumption in the duration of braking.

\begin{figure}[htbp]
  \centering
  \includegraphics[width=5.5in]{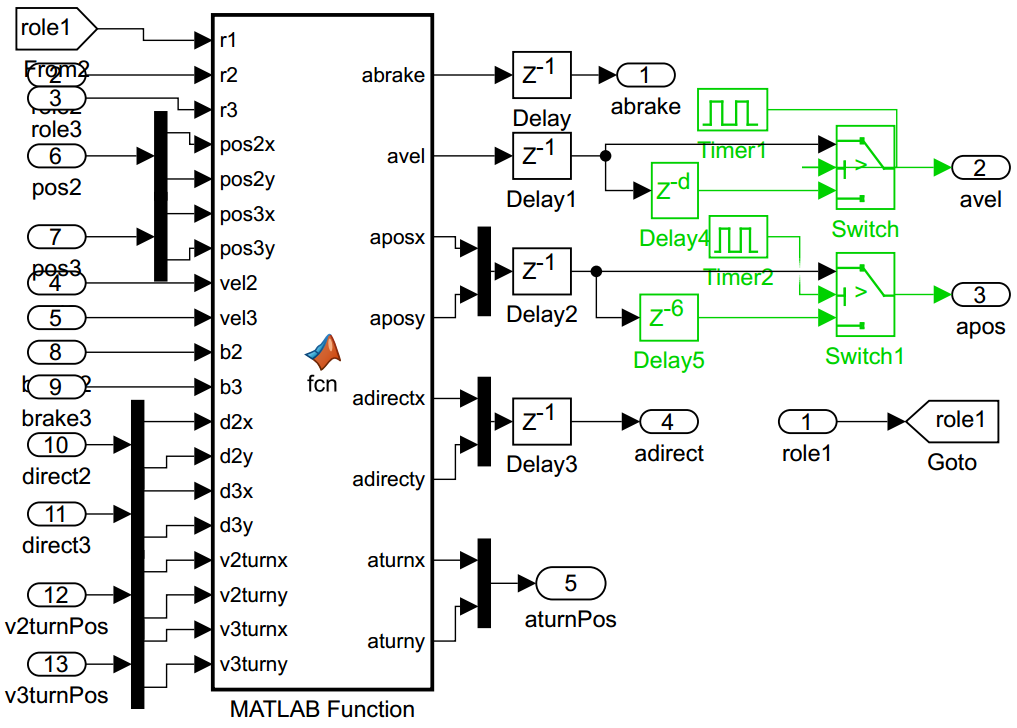}
  \caption{ \simu~\& \staf\ model of {\gt{ComDevice}}}
  \label{fig:commu_ss}
\end{figure}

\section{Modeling Timing and Energy Constraints in Simulink/Stateflow}
In this section, we investigate how to model timing and energy constraints in S/S. In our previous work, we have shown how those constraints are interpreted in S/S and provided corresponding modeling extensions in S/S. However, the previous modeling extensions contain blocks that are unsupported in SDV, which leads to incompatibility problem. To perform verification of timing and energy constraints in SDV, the modeling extensions are modified to ensure the compatibility.

\begin{figure}[htbp]
\centering
  \subfigure[Execution]{
  \includegraphics[width=2.4in]{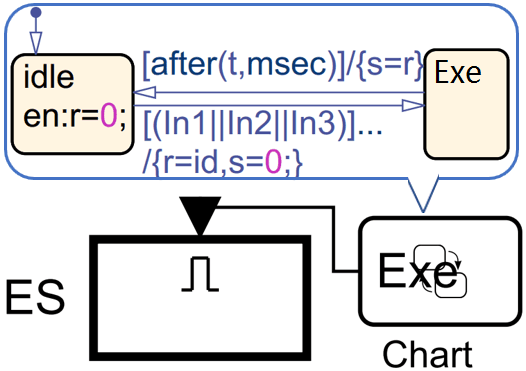}}
  \subfigure[End-to-End]{
  \includegraphics[width=2.3in]{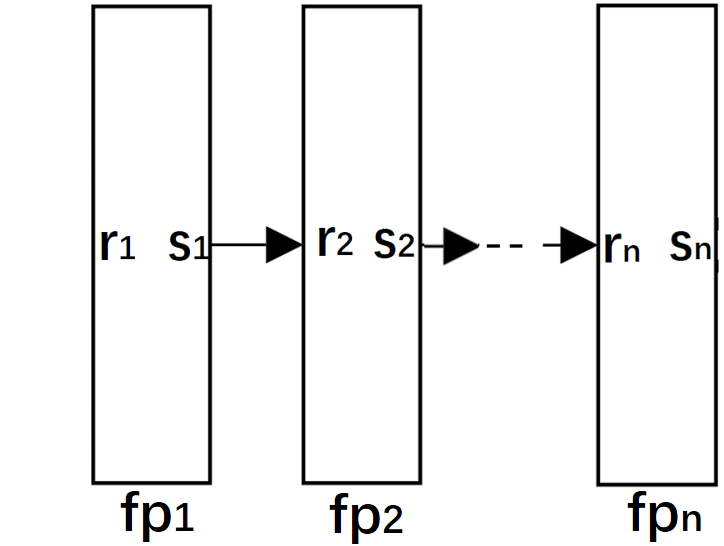}}
  \caption{Timing constraints modeling}
\label{fig:ssee}
\end{figure}

\vspace{0.05in}
{\gt{Sporadic}} timing constraint is modeled using ``after(min, msec)'' expression, which returns true after \emph{min} milliseconds (msec) have elapsed since activation of the associated state. {\gt{Synchronization}} timing constraint
restricts the time duration among a set of events
(i.e., maximum allowed time between the arrival of the event occurrences). We use {\emph{In1}}, {\emph{In2}} and {\emph{In3}} (in Fig.\ref{fig:ssee}(a)) to indicate the arrivals of the three inputs of \fp. To verify {\gt{Synchronization}} constraint, we check whether the interval between occurrences of earliest and latest input event is within {\emph{tolerance}}, which is described in Chap. \ref{sec:sdv-trans}.

As shown in Fig.\ref{fig:ssee}(a), to model {\gt{Execution}} timing constraint of an \fp, a Stateflow chart is used to create a true duration to execute the \fp. The {\scriptsize $\ll$}Enable Subsystem{\scriptsize $\gg$} {\gt{ES}} (representative of \fp) executes when {\emph{Exe}} state is active. {\emph{t}} is the {\gt{Execution}} constraint applied on the \fp. {\emph{id}} is a monotonically increased value. To ensure the one-to-one correspondence of {\gt{input}} and {\gt{output}} event, we use two variables, {\emph{r}} and {\emph{s}}, to represent the value of {\gt{input}} and {\gt{output}} event respectively. When {\gt{input}} event occurs, \fp\ will start to execute and  {\emph{r}} will be assigned by the value of current {\emph{id}}. After {\emph{t}} milliseconds, \fp\ finishes execution and {\emph{s}} becomes identical with {\emph{r}}, which indicates the occurrence of {\gt{output}} event.

Fig.\ref{fig:ssee}(b) shows an \emph{EventChain} of {\emph{n}} \fp s, we model each \fp\ based on the pattern of Fig.\ref{fig:ssee}(a). For each \fp, $r_i$ and $s_i$ represent the values of {\gt{input}} and {\gt{output}} event.
Hence in this \emph{EventChain}, {\emph{r$_1$}} is the value of {\gt{source}} event and s$_n$ is the value of {\gt{target}} event of {\gt{End-to-End}} constraint. To identify the one-to-one correspondence of {\gt{source}} and {\gt{target}} event, {\emph{r$_1$}} and {\emph{s$_n$}} should have the same assigned value.

\begin{figure}[htbp]
  \centering
  \includegraphics[width=3.5in]{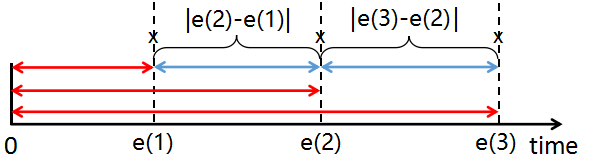}
  \caption{{\gt{Cumulative}} and {\gt{Noncumulative}} constraint}
  \label{diff}
\end{figure}

We consider two types of {\gt{Periodic}} timing constraints: cumulative constraint and noncumulative constraint. {\emph{T}} and {\emph{j}} indicate the period and jitter of {\gt{Periodic}} constraint. In Fig.\ref{diff}, {\gt{e(i)}} denotes the time point of the $i^{th}$ occurrence of a single event {\gt{e}}. e$_{\gt{i}}$  represents the $i^{th}$ occurrence of event {\gt{e}}. {\gt{Cumulative Periodic}} constraint limits that the time intervals between any two consecutive occurrences of {\gt{e}} (e.g, {\gt{$\mid${\gt{e(2)}}-{\gt{e(1)}}$\mid$}} and {\gt{$\mid${\gt{e(3)}-{\gt{e(2)}}}$\mid$}}) should be within [{\emph{T-j}}, {\emph{T+j}}]. {\gt{Noncumulative Periodic}} constraint limits that {\gt{e(i)}} should be within [{\emph{i*T-j}}, {\emph{i*T+j}}].

\begin{figure}[htbp]
\centering
  \subfigure[Cumulative]{
  \includegraphics[width=1.8in]{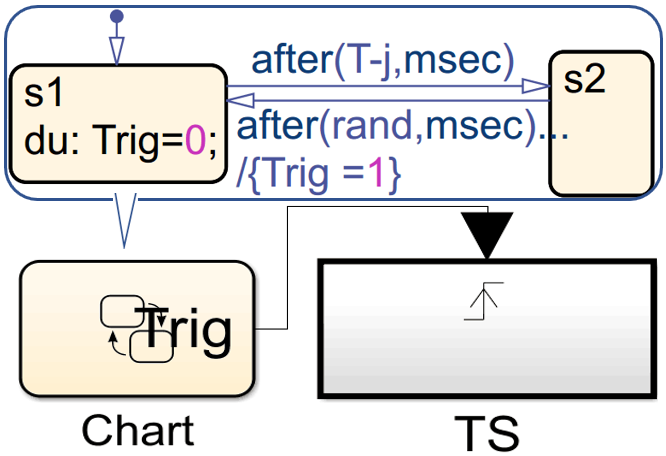}}
  \subfigure[Noncumulative]{
  \includegraphics[width=2.2in]{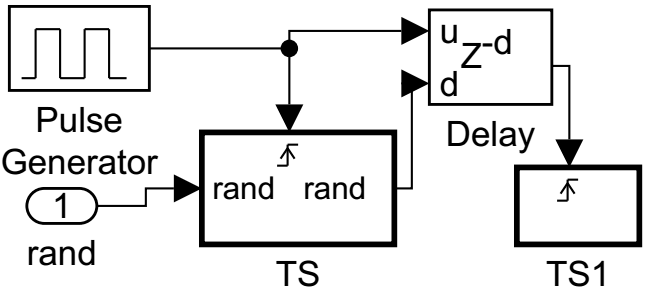}}
  \caption{{\gt{Periodic}} constraint modeling}
\label{periodicss}
\end{figure}

Fig.\ref{periodicss}(a) shows the model of {\gt{Cumulative Periodic}} timing constraint of \fp. The behaviour of \fp\ is modeled in {\gt{TS}} (Trigger subsystem)
The Stateflow chart generates a signal {\emph{Trig}} to trigger {\gt{TS}}. {\emph{rand}} is a random number within [0, 2*j]. Once \fp\ is triggered, it will be triggered again after {\emph{T-j+rand}} ms such that the time interval between any two consecutive triggerings of \fp~is in [{\emph{T-j}}, {\emph{T+j}}].
Fig.\ref{periodicss}(b) shows the model of
{\gt{Noncumulative Periodic}} timing constraint of \fp\ (modeled in {\gt{TS1}}). A {\scriptsize $\ll$}Pulse Generator{\scriptsize $\gg$} (PG) generates square waves based on {\emph{Period}} and {\emph{ Phase delay}} parameters. {\emph{Phase delay}} specifies the delay before the pulse is generated.
PG generates a square wave signal whose {\emph{Period}} is {\emph {T}} and {\emph{Phase delay}} is {\emph{T-j}} such that the $i^{th}$ rising edge of the signal arrives at the time point of {\emph{i*T-j}}. In the mean time, {\gt{TS}} will be triggered to pass {\emph{rand}} and {\scriptsize $\ll$}Delay{\scriptsize $\gg$} block delays the rising edges by {\emph{rand}} ms s.t. the time point of the $i^{th}$ rising edge will be in {\emph{[i*T-j, i*T+j]}}.

{\gt{Energy}} constraints are modeled with {\scriptsize $\ll$}Data Store{\scriptsize $\gg$} and integrated \mt\ blocks, which calculate and record energy consumption in different modes of a vehicle. According to the various modes of vehicles in CAS, the amount of battery consumed for the mechanical motion of wheels is calculated: $energy$ $=$ $\int_0^t$ $a * v$ $dx$ where, $t$, $a$, and $v$ denote running time, coefficient reflects an energy rate associated to the current mode of an individual vehicle, and velocity respectively.

\chapter{Timing \& Energy Requirements Translation in SDV}
To enable the verification of the timing and energy requirements illustrated in Chap. \ref{sec:mdl-ss}, we specify timing and energy requirements as LTL formulas and provide the approach to model linear temporal logic (LTL) \cite{LTL} formulas into proof objective models.

\label{sec:sdv-trans}
\section{LTL Properties}

An LTL formula usually consists of predicates and propositional connectives, together with temporal operators (e.g. { \gt{Always}}, {\gt{Eventually}} and {\gt{Until}}).

\noindent\textbf{{{Always}} p} (\textbf{G} {\gt{p}}) states that property {\gt{p}} always holds along the execution, which is modeled as the proof objective model described in Chap. \ref{sec:preliminary}.

\begin{figure}[htbp]
  \centering
  \includegraphics[width=2.7in]{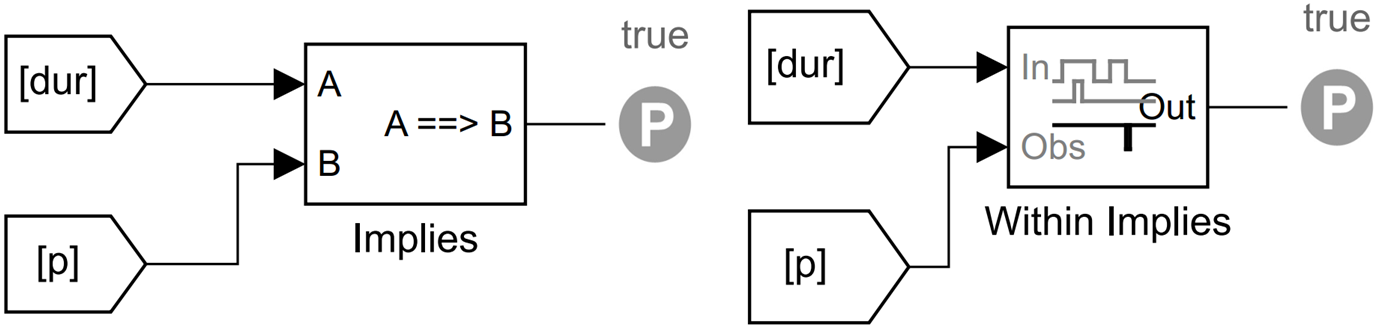}
  \caption{Bounded {\gt{Always}} and {\gt{Eventually}} property}
  \label{sdvae}
\end{figure}

\noindent\textbf{{{Always}} and {{Eventually}} within t time steps}: {\gt{Always p}} within {\emph{t}} time steps, denoted by \textbf{G}$_{[0,\ t]}$ {\gt{p}}, is valid if property {\gt{p}} holds over the first {\emph{t}} time steps during execution. {\gt{Eventually p}} within {\emph{t}} time steps, denoted as \textbf{F}$_{[0,\ t]}$ {\gt{p}}, holds if {\gt{p}} occurs within the first {\emph{t}} time steps during the execution.
The model of {\gt{Always}} property and {\gt{Eventually}} are illustrated on the left and right of Fig.\ref{sdvae}: {\emph{dur}} is a signal which is true within the first {\emph{t}} time steps. To checks whether {\gt{p}} always holds true within the true duration, {\gt{p}} is connected to {\scriptsize $\ll$}Implies{\scriptsize $\gg$} block.
{\scriptsize $\ll$}Within Implies{\scriptsize $\gg$} block is employed to verify whether {\gt{p}} holds for at least one time step within the true duration {\emph{dur}}. P block is applied to prove whether the property is satisfied.

\begin{figure}[htbp]
  \centering
  \includegraphics[width=2.7in]{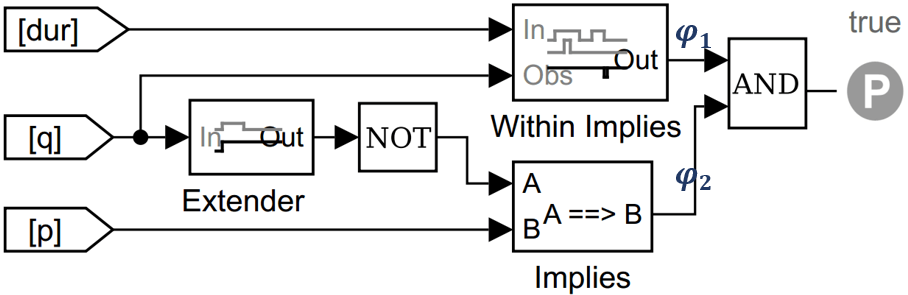}
  \caption{ p {\gt{Until}} q: \textbf{G} (\textbf{F}$_{[0,\ t]}$ {\gt{q}}  $\wedge$  ($\neg${\gt{q}} $\implies$ {\gt{p}}))}
  \label{sdvuntil}
\end{figure}

\vspace{0.05in}
\noindent\textbf{{{p}} {{Until}} {{q}} within t time steps}.
Denoted as {\gt{p}} \textbf{U}$_{[0,t]}$ {\gt{q}}, this property states that {\gt{p}} should always hold before {\gt{q}} becomes true within the first {\emph{t}} time steps during the execution. The satisfaction of this property requires
$\varphi_1$: {\gt{q}} must hold at least one time step within the first {\emph{t}} time steps;
$\varphi_2$: before {\gt{q}} holds, {\gt{p}} must always be true.
This property can be interpreted as: \textbf{F}$_{[0,\ t]}$ {\gt{q}} $\wedge$ \textbf{G}$_{[0,\ t]}$ ($\neg${\gt{q}} $\implies$ {\gt{p}}).
The implementation of this property is shown in Fig.\ref{sdvuntil}:
{\emph{dur}} has a true duration with length {\emph{t}}.
First, a {\scriptsize $\ll$}Within Implies{\scriptsize $\gg$} block check whether $\varphi_1$: F$_{[0,\ t]}$ {\gt{q}} holds.
An {\scriptsize $\ll$}Extender{\scriptsize $\gg$} block is applied to extend the true duration of {\gt{q}} to {\emph{t}} time steps after {\gt{q}} occurs. {\scriptsize $\ll$}Not{\scriptsize $\gg$} block is then employed to capture the true duration of $\neg${\gt{q}}.
{\scriptsize $\ll$}Implies{\scriptsize $\gg$} block checks whether ``not {\gt{q}} implies {\gt{p}}'' holds.
Afterwards, {\scriptsize $\ll$}And{\scriptsize $\gg$} and {\scriptsize $\ll$}Proof Objective{\scriptsize $\gg$} validate whether ``$\varphi_1$ $\wedge$ $\varphi_2$'' is always true.

\begin{figure}[htbp]
  \centering
  \includegraphics[width=3.5in]{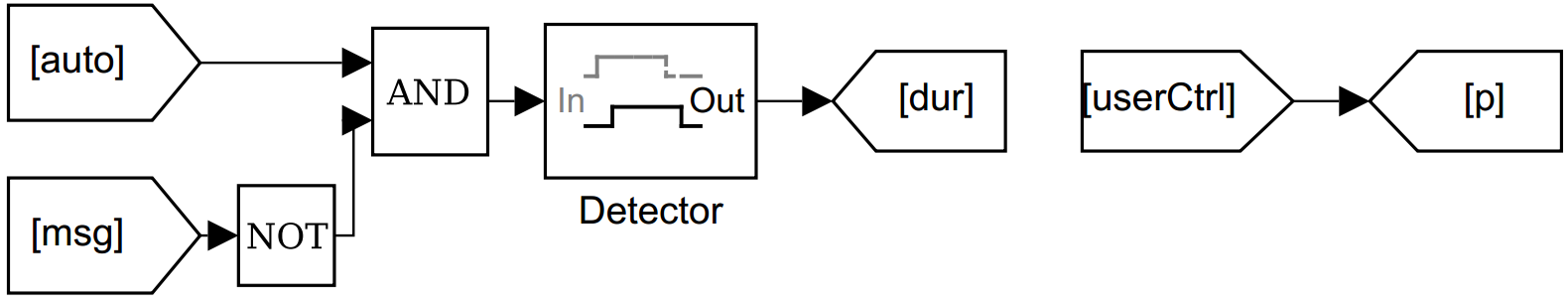}
  \caption{Functional Property: When \emph{auto} is true and \emph{msg} is false for 200ms, \emph{userCtrl} should become true within 200ms. }
  \label{evenexample}
\end{figure}

Fig.\ref{evenexample} illustrates the functional property (R1) which can be specified by using \emph{Always} and \emph{Eventually} operators: \textbf{G}$_{[0,200]}$(\emph{auto} == true $\wedge$ \emph{msgNormal} == false) $\implies$ \textbf{F}$_{[0,200]}$ \emph{userCtrl} == true.
\emph{auto} (\emph{userCtrl}) represents whether the vehicle is running automatically (driven manually). \emph{msgNormal} indicates whether the message is transmitted normally.
In Fig.\ref{evenexample}, {\scriptsize $\ll$}Detector{\scriptsize $\gg$} constructs a signal with a 200ms true duration after ``\emph{auto} == true and \emph{msg} == false'' becomes true for 200ms. Within this true duration, ``\emph{userCtrl} == true'' should be satisfied at least one time step.
To check this, the output signal from {\scriptsize $\ll$}Detector{\scriptsize $\gg$} goes to the {\scriptsize $\ll$}Goto{\scriptsize $\gg$} block tagged with \emph{dur}. Then it is passed to the corresponding \emph{dur} tag on the right of Fig.\ref{sdvae}. Similarly, the signal that represents ``\emph{userCtrl} == true'' is passed to \emph{p} tag on the right of  Fig.\ref{sdvae} (the input of {\scriptsize $\ll$}Within Imply{\scriptsize $\gg$}) block.

\begin{figure}[htbp]
  \centering
  \includegraphics[width=3.3in]{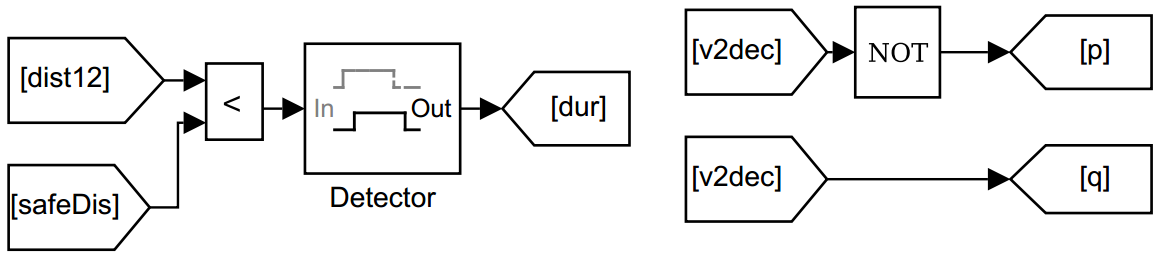}
  \caption{Property: If the \emph{dis12} is smaller than \emph{safeDis}, \emph{v2dec} should be true within 500ms.}
  \label{fig:R4}
\end{figure}

Fig.\ref{fig:R4} illustrates the model of the functional property (R21), which can be specified with \emph{Util} and \emph{Always} temporal operators: \textbf{G}  (\emph{dist12} $<$ \emph{safeDis}  $\implies$ \emph{v2.dec} == false \textbf{U}$_{[0,500]}$ \emph{v2.dec} == true). \emph{dist12} is the distance between the lead vehicle (\emph{v1}) and the following vehicle (\emph{v2}). \emph{safeDis} is the required safety distance between them. \emph{v2.dec} represents whether \emph{v2} is decelerating.
{\scriptsize $\ll$}Detector{\scriptsize $\gg$} creates a signal with a 500ms true duration after ``\emph{dist12} $<$ \emph{safeDis}'' becomes true. This signal then goes to the {\scriptsize $\ll$}From{\scriptsize $\gg$} block tagged with \emph{dur} in Fig.\ref{sdvuntil}. In this example, \emph{p} represents the property ``\emph{v2.dec} == false'' and \emph{q} represents ``\emph{v2.dec} == true'' as illustrated on the right of Fig.\ref{fig:R4}.

\vspace{0.05in}
\vspace{0.05in}

\section{Timing Constraints in SDV}
Timing constraints in \ed\ are modeled by means of constraints specified on \emph{Events} and \emph{EventChains}. A constraint is expressed as a proof objective model in SDV. To proof the correctness of the timing behaviours of Simulink/Stateflow model,
we show how to construct the proof objective models for the design model based on the timing constraints.
We focus on {\gt{Synchronization, Execution, End-to-End, Periodic, Sporadic}} constraints (R27-R46) that are associated to \fp s in \ed.

\begin{figure}[htbp]
  \centering
  \includegraphics[width=3.1in]{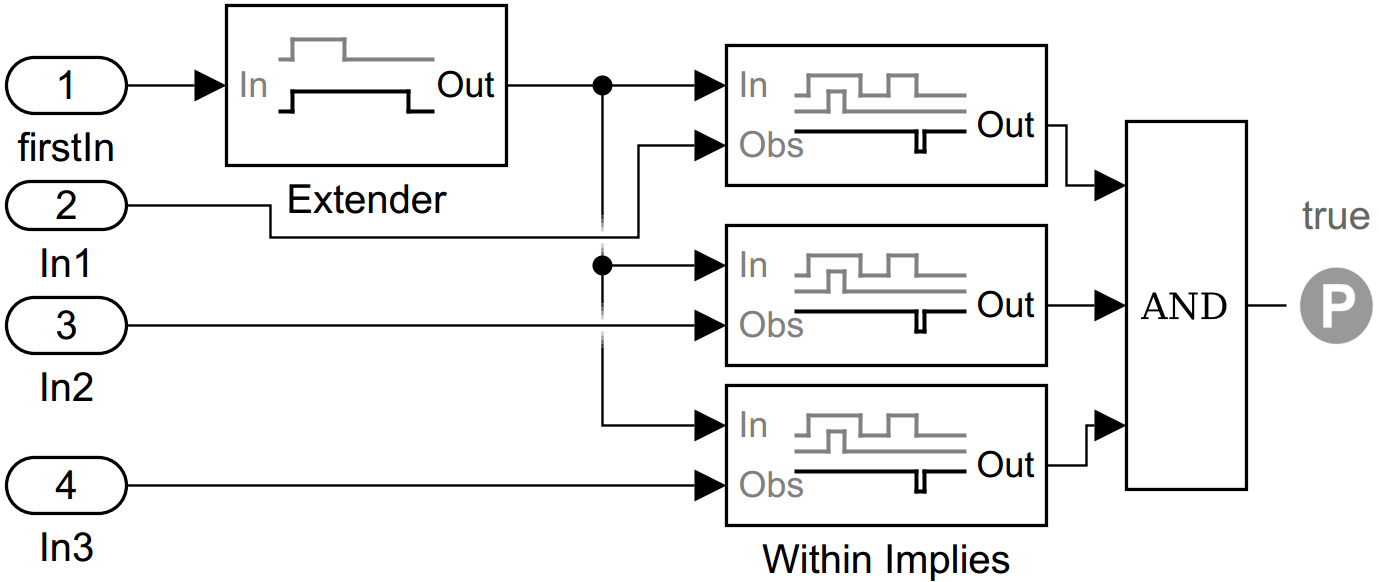}
  \caption{Synchronization timing constraint}
  \label{sdvsyn}
\end{figure}

\vspace{0.05in}
\noindent\textbf{Synchronization timing constraint} can be interpreted as: all the input signals should be received within \emph{tolerance} after the first (earliest) input signal arrives.
The model of the timing constraint is shown in Fig.\ref{sdvsyn}:
Assume that there are three input signals of an \fp. \emph{firstIn} represents whether the first (earliest) input signal is received. \emph{In1}, \emph{In2} and \emph{In3} indicate the arrival of three input signals of \fp~respectively.
{\scriptsize $\ll$}Extender{\scriptsize $\gg$} is employed to construct a true duration with length \emph{tolerance} from the time step in which \emph{firstIn} becomes true.
Three {\scriptsize $\ll$}Within Implies{\scriptsize $\gg$} block are applied to capture whether \emph{In1}, \emph{In2} and \emph{In3} becomes true within the true duration.
{\scriptsize $\ll$}And{\scriptsize $\gg$} block then validates whether \emph{In1}, \emph{In2} and \emph{In3} all become true within the true duration and pass the output signal to a {\scriptsize $\ll$}Proof Objective{\scriptsize $\gg$} block, which proves whether the constraint is valid.

\begin{figure}[htbp]
\centering
  \subfigure[Execution]{
  \includegraphics[width=4in]{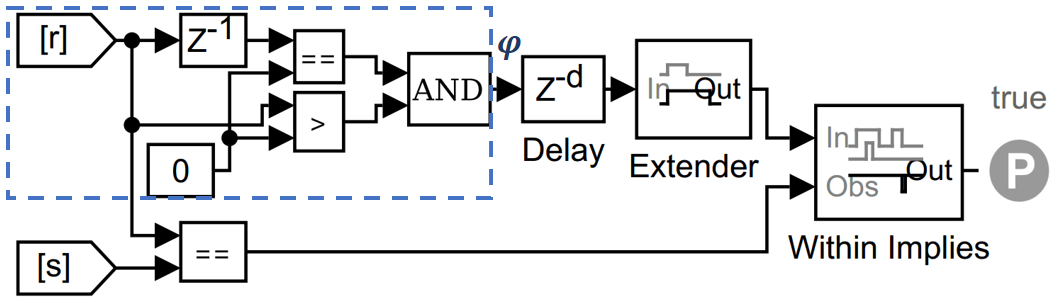}}
  \subfigure[End-to-end]{
  \includegraphics[width=4in]{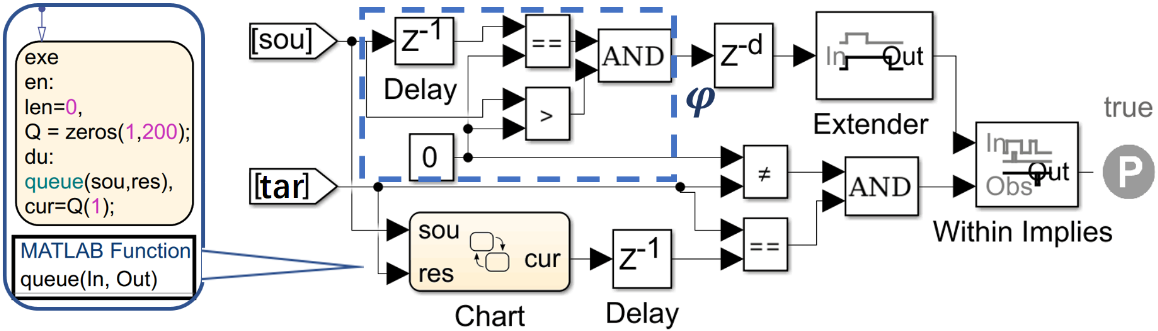}}
  \caption{Timing constraints translation}
\label{fig:sdvee}
\end{figure}

\vspace{0.05in}
\noindent\textbf{Execution timing constraint} limits that after the input signals are received, the output data should be sent out within \emph{upper} time steps. The construct of this timing constraint is shown on the left of Fig.\ref{fig:sdvee} (a): \emph{dataIn} (\emph{dataOut})
indicates the input (output) signal received (sent out). {\scriptsize $\ll$}Extender{\scriptsize $\gg$} is applied to construct a true duration with length \emph{upper} after \emph{dataIn} becomes true. Afterwards {\scriptsize $\ll$}Within Implies{\scriptsize $\gg$} and {\scriptsize $\ll$}Proof Objective{\scriptsize $\gg$} blocks are used to verify whether the \emph{dataOut} becomes true within the true duration.

\vspace{0.05in}
\noindent\textbf{End-to-End timing constraint} restricts that after \emph{source} event occurs, \emph{target} event should happen within \emph{tolerance}. The model of this timing constraint is illustrated on  the right of Fig.\ref{fig:sdvee} (b): \emph{source} and \emph{target} indicate the \gt{source} and \gt{target} event occurs (true) or not (false). {\scriptsize $\ll$}Extender{\scriptsize $\gg$} constructs a true duration with length \emph{tolerance} after \emph{source} becomes true. Afterwards {\scriptsize $\ll$}Within Implies{\scriptsize $\gg$} and {\scriptsize $\ll$}Proof Objective{\scriptsize $\gg$} blocks are used to verify whether the \emph{target} becomes true within the true duration.

\begin{figure}[htbp]
  \centering
  \includegraphics[width=2.7in]{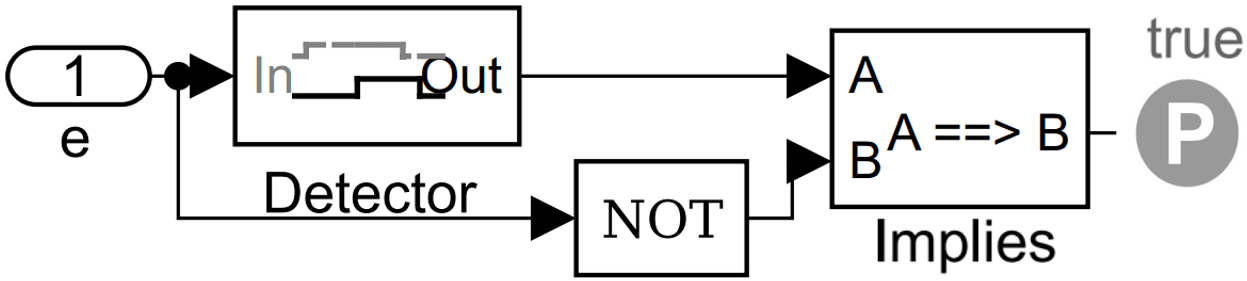}
  \caption{Sporadic timing constraint}
  \label{sdvsporadic}
\end{figure}

\begin{figure}[htbp]
  \centering
  \includegraphics[width=3.2in]{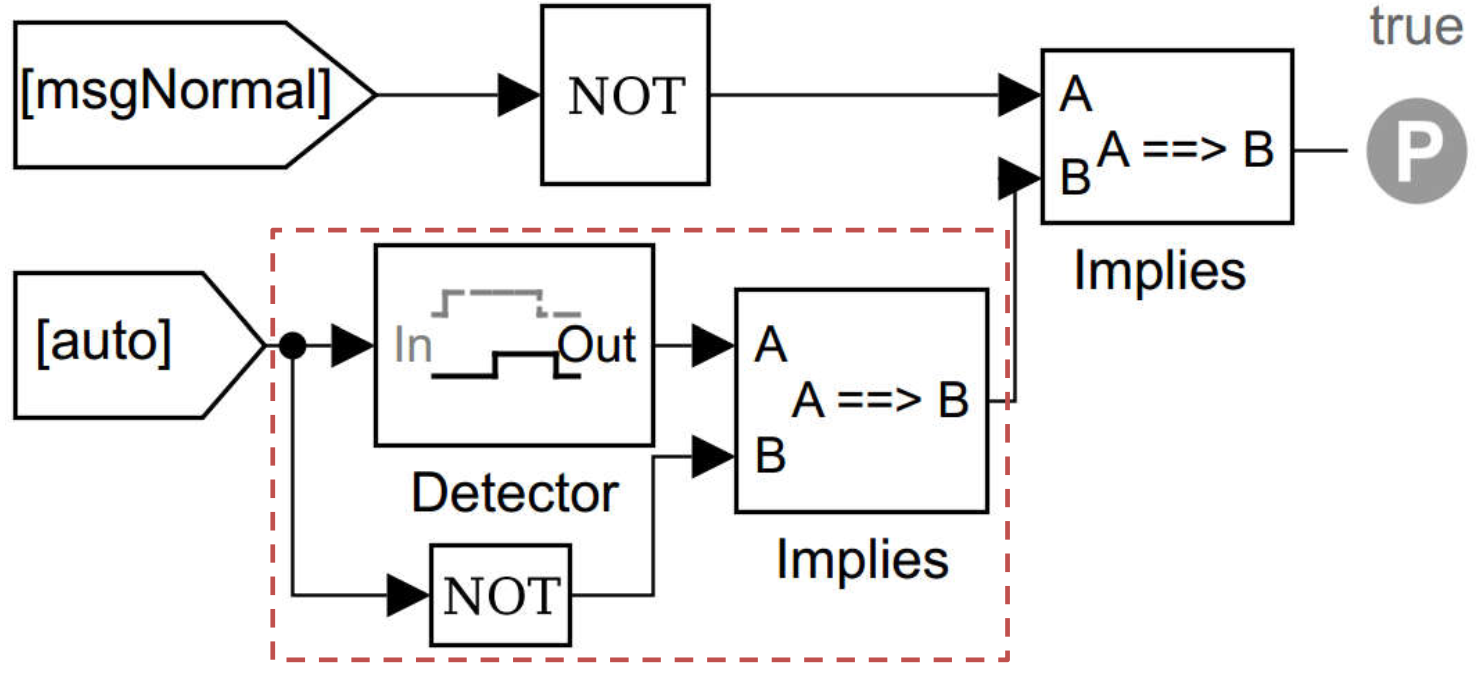}
  \caption{R30: if the message transmission is failed (\emph{msgNormal == false}) when the vehicle runs in auto-mode, for the next 20 seconds, the vehicle should not run automatically, i.e., \emph{auto == false}}
  \label{sporadicexp}
\end{figure}

\noindent\textbf{Sporadic timing constraint} specifies that after the event occurs, the next occurrence should not happen within \emph{min} time steps. This timing constraint is constructed as the model in Fig.\ref{sdvsporadic}. \emph{event} indicates the occurrence of an event.
{\scriptsize $\ll$}Detector{\scriptsize $\gg$} constructs a
true duration with length \emph{min} after \emph{event} becomes true. Afterwards, {\scriptsize $\ll$}Negation{\scriptsize $\gg$} (Not) and {\scriptsize $\ll$}Implies{\scriptsize $\gg$} blocks are employed to validate whether \emph{p} is false over the true duration. At last, {\scriptsize $\ll$}Proof Objective{\scriptsize $\gg$} block checks whether the property is satisfied. Fig.\ref{sporadicexp} illustrates an example of sporadic timing constraint (R30), where the constraint is modeled in the dash box. In this example, \emph{event} is ``auto == true''.

\begin{figure}[htbp]
  \centering
  \includegraphics[width=2.6in]{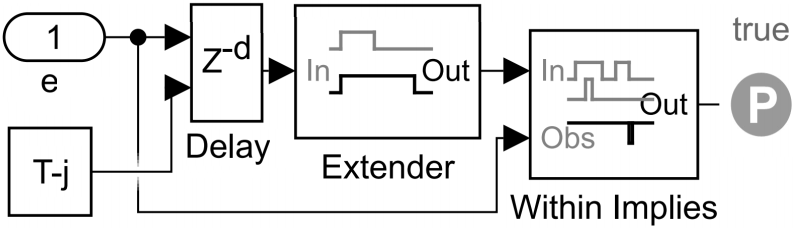}
  \caption{Cumulative Periodic Timing Constraint}
  \label{sdvcumu}
\end{figure}

\begin{figure}[htbp]
  \centering
  \includegraphics[width=2.4in]{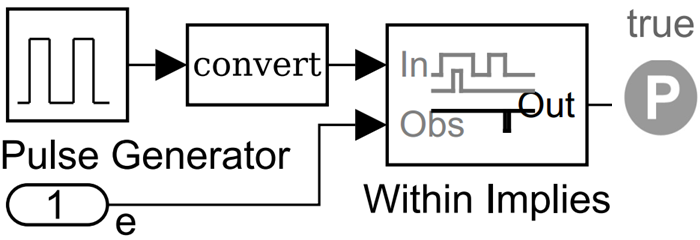}
  \caption{Non-cumulative Periodic Timing Constraint}
  \label{sdvperiodic}
\end{figure}

\vspace{0.05in}
\noindent\textbf{Periodic timing constraint} limits a single event \fp~should occur periodically with period \emph{T} and jitter \emph{j}.
 We consider two types of timing constraints in SDV:
 \begin{inparaenum}
\item {\gt{Cumulative Periodic}} timing constraint is modeled as the proof objective model illustrated in Fig.\ref{sdvcumu}. To verify this property, we check whether $t_i+1$ locates in [$t_i$+$T$-$j$, $t_i$+$T$+$j$]. As shown on the left of Fig.\ref{sdvcumu}, a {\scriptsize $\ll$}Delay{\scriptsize $\gg$} is to add a \emph{T-j} delay to the input signal \emph{event}. As illustrated on the right of Fig.\ref{sdvcumu}, an {\scriptsize $\ll$}Extender{\scriptsize $\gg$} then generates a true duration with length \emph{2*j} such that the range of true duration is [$t_i$+$T$-$j$, $t_i$+$T$+$j$].
Afterwards, {\scriptsize $\ll$}Within Implies{\scriptsize $\gg$} checks whether \emph{event} occurs within each true duration.
Take R27 as an example, if \emph{event} firstly occurs at 0.01s, a corresponding true duration [0.05, 0.07] is generated, where the existence of the second occurrence is checked.

\item {\gt{Noncumulative Periodic}} timing constraint is constructed with the blocks illustrated in Fig.\ref{sdvperiodic}: a {\scriptsize $\ll$}Pulse Generator{\scriptsize $\gg$} is applied to generate signals (square waves) based on \emph{Period, Pulse width, Phase delay} parameters (see Fig.\ref{sdvperiodic}). \emph{Pulse width} is a duty cycle specified as a percentage of \emph{Period}, i.e., \emph{Pulse width} = \emph{Width}/\emph{Period}. \emph{Phase delay} specifies the delay before the pulse is generated. {\scriptsize $\ll$}Convert{\scriptsize $\gg$} block is used to transform the signal type from \emph{double} to \emph{boolean}. The generated signal and \emph{event} signal from the input port, are then passed to {\scriptsize $\ll$}Within Implies{\scriptsize $\gg$} block. {\scriptsize $\ll$}Within Implies{\scriptsize $\gg$} checks whether \emph{event} becomes true within each true duration of the generated signal. Take R27 as an example,  the \emph{Phase delay} is specified as 0.04s, \emph{Period} is 0.05s and \emph{Pulse width} is 40\%, which is illustrated on the right of Fig.\ref{sdvperiodic}. To satisfy this timing constraint, {\gt{VehicleDynamic}} should be triggered within each true duration of the generated signal.
\end{inparaenum}

\vspace{0.05in}
\section{Energy Constraints in SDV}
According to the various modes of vehicles in CAS, the amount of battery consumed for the mechanical motion of wheels is calculated: $energy$ $=$ $\int_0^t$ $a * v$ $dx$ where, $t$, $a$, and $v$ denote running time, coefficient reflects an energy rate associated to the current mode of an individual vehicle, and wheel speed respectively.

\begin{figure}[htbp]
  \centering
  \includegraphics[width=3in]{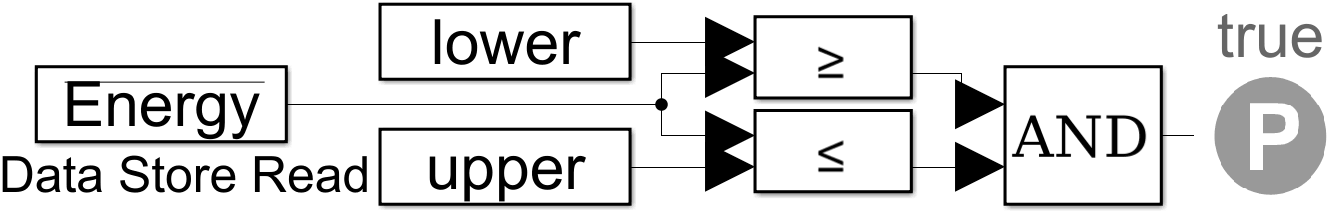}
  \caption{Energy constraint}
  \label{fig:sdvenergy}
\end{figure}

\textbf{Energy constraint} demarcates that the energy consumption on different modes of an individual vehicle is within $[lower, upper]$. It is specified as \textbf{G}$(lower$ $\leqslant$ $energy$ $\wedge$ $energy$ $\leqslant$ $upper)$ and its POM is depicted in Fig. \ref{fig:sdvenergy}, where the value of the energy consumption is obtained through {\scriptsize $\ll$}Data Store Read{\scriptsize $\gg$} block and compared with \emph{upper} and \emph{lower}.

\chapter{Modeling Functional Behaviours in \smc}
\label{sec:mdl-smc}
\begin{figure}[htbp]
  \centering
  \includegraphics[width=4.5in]{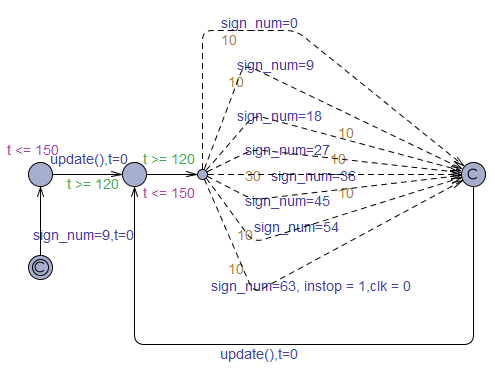}
  \caption{STA of \gt{SignRecognition}}
  \label{fig:signreg}
\end{figure}

We translate the \ed\ architectural model illustrated in Fig.\ref{fig:east} and Simulink/Stateflow model in Fig.\ref{fig:cs_ss} into a networks of STAs in \smc. In the model, there are three vehicles receiving the information of traffic signs from cooperative environment {\gt{CoopEnv}} \fp. The functionality traffic sign recognition is modeled as the STA shown in Fig.\ref{fig:signreg}.  The type of traffic signs include \emph{stop}, \emph{left/right turn}, \emph{straight}, \emph{minimum/maximum speed limit}. The dash line in Fig.\ref{fig:signreg} indicates the probabilistic distribution of different traffic signs, i.e., traffic sign occurs randomly with a certain probability. The lead vehicle (v1) will adjust its speed and running direction according to the detected sign types. The two follower vehicles (v2 and v3) will follow the lead vehicle.

\begin{figure}[htbp]
  \centering
  \includegraphics[width=4.5in]{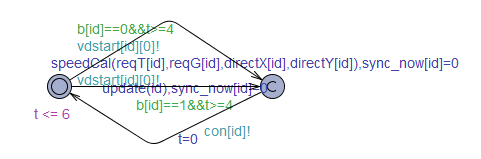}
  \caption{STA of {\gt{VehicleDynamic}}}
  \label{fig:VD}
\end{figure}

\begin{figure}
\centering
\subfigure[Sensor of sending signals]{
\label{fig_sending}
\includegraphics[width=4in]{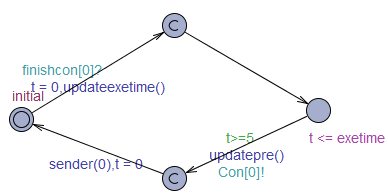}}
\subfigure[Sensor of receiving signals]{
\label{fig_receiving}
\includegraphics[width=5.2in]{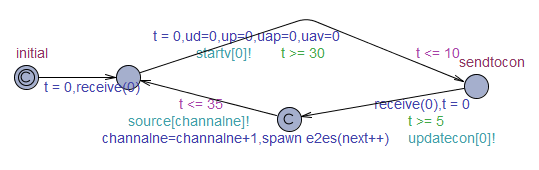}}

\caption{STA for modeling {\gt{ComDevice}}}
\label{fig:CD}
\end{figure}

The functionality of the three vehicles are similar. Each vehicle consists of a {\gt{localEnv}} \fp\ and a {\gt{Controller}}. Each {\gt{localEnv}} consists of three \fp s: {\gt{ComDevice}}, {\gt{VehicleDynamic}} and {\gt{DriverBehaviour}}.
{\gt{ComDevice}} is modeled in Fig.\ref{fig:CD}.
The inner behaviour of the vehicle calculating velocity according to the gears and torque is modeled in Fig.\ref{fig:VD}. This \fp\ will be triggered to execute periodically with period 50ms and jitter 10ms, i.e., STA of  {\gt{VehicleDynamic}} will be triggered to update the speed, gear and torque of the vehicle periodically. The computation is implemented as the function \emph{speedCal()}, which will be called to update the states and valuation of variables whenever the corresponding transition is taken.

Fig.\ref{fig:CD}(a) captures the behaviour of the sensor sending out signals of position, velocity and direction of vehicle. Fig.\ref{fig:CD}(b) models the behaviour of the sensor receiving signals from other vehicles. It receives information (the position, velocity and direction) from the lead vehicle, and detects whether the lead vehicle is braking periodically. The sensor will send out signals/information when the \gt{Controller} of the vehicle finishes execution.

\begin{figure}
\centering
\subfigure[Brake Pedal]{
\label{fig_bp}
\includegraphics[width=4.8in]{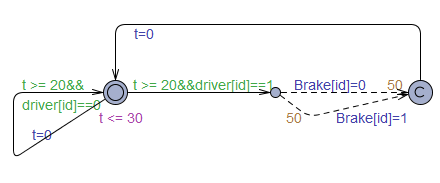}}
\subfigure[Control Status Button]{
\label{fig_c}
\includegraphics[width=2.4in]{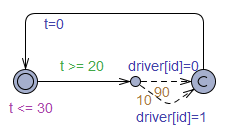}}
\subfigure[Gear]{
\label{fig_g}
\includegraphics[width=5.4in]{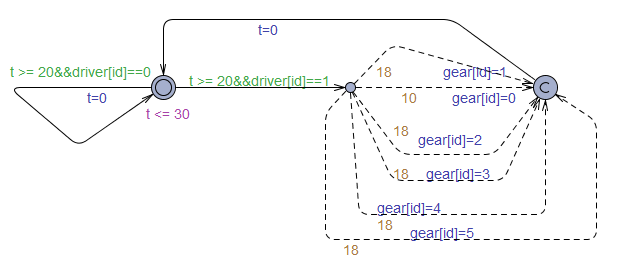}}
\subfigure[Steer]{
\label{fig_s}
\includegraphics[width=4.6in]{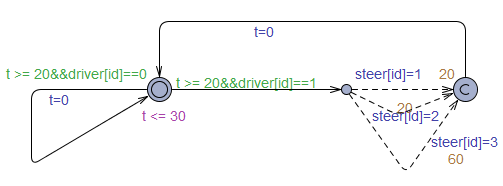}}
\caption{STA for modeling {\gt{DriverBehaviour}}}
\label{fig:DB}
\end{figure}

The behaviors of the driver is modeled as the STAs shown in Fig.\ref{fig:DB}. {\gt{Brake Pedal}} STA represents the behaviors of braking pedal (Fig.\ref{fig:DB}(a)). {\gt{Control status button}} STA records whether the driver intends to drive the vehicle manually (Fig.\ref{fig:DB}(b)). The behaviors of controlling gear and steer is modeled as the STA in Fig.\ref{fig:DB}(c) and Fig.\ref{fig:DB}(d) respectively. The driver can take charge of the running state of the vehicle by switching the running state to \emph{userCtrl} mode using Control Button. The driver can control the vehicle by steer (steer the running direction of vehicle), gear (control the speeds of the wheels) and the brake pedal (to stop the vehicle).

\begin{figure}[htbp]
  \centering
  \includegraphics[width=4.8in]{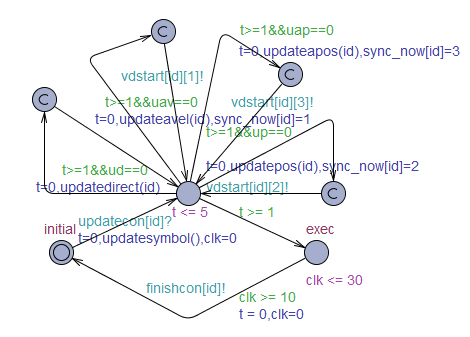}
  \caption{STA of {\gt{Controller}}}
  \label{fig:controltime}
\end{figure}

\begin{figure}[htbp]
  \centering
  \includegraphics[width=4in]{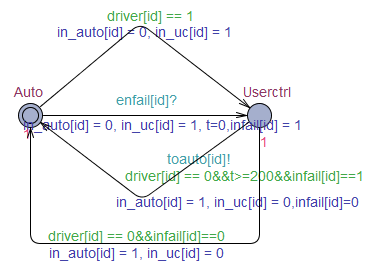}
  \caption{STA of {\emph{Control}} state}
  \label{fig:controller}
\end{figure}

\begin{figure}[htbp]
  \centering
  \includegraphics[width=4.3in]{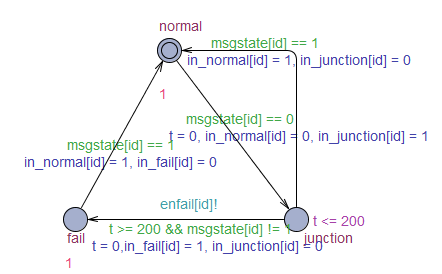}
  \caption{STA for modeling {\emph{Communication}} state}
  \label{fig:communication}
\end{figure}

The top view of the {\gt{Controller}} modeled in Fig.\ref{fig:topctrl_ss} is shown in Fig.\ref{fig:controltime}. The controller gets information from four different sensors. After receiving all input signals, the \gt{Controller} will decide which state should be activated. Fig.\ref{fig:controller} and Fig.\ref{fig:communication} describe two states that represent \emph{auto} mode and \emph{userCtrl} mode of vehicle. If the driver intends to control the vehicle, the Control Button will transit to \emph{userCtrl}, indicating that the vehicle is driven manually. When \emph{Auto} location is active, and the vehicle cannot get the information from the ahead vehicle, then \emph{userCtrl} will be activated automatically.

\begin{figure}[htbp]
  \centering
  \includegraphics[width=3.2in]{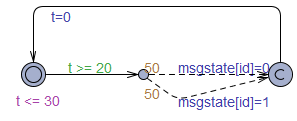}
  \caption{STA for deciding whether the message is missing}
  \label{fig:message}
\end{figure}

Fig.\ref{fig:communication} describes the behavior that it is of 50\% probability that the message transmission is failed. The message state is modeled by STA in Fig.\ref{fig:message}. When the message is missed for 2s,  \emph{userCtrl} mode will be activated.

\begin{figure}[htbp]
  \centering
  \includegraphics[width=6.5in]{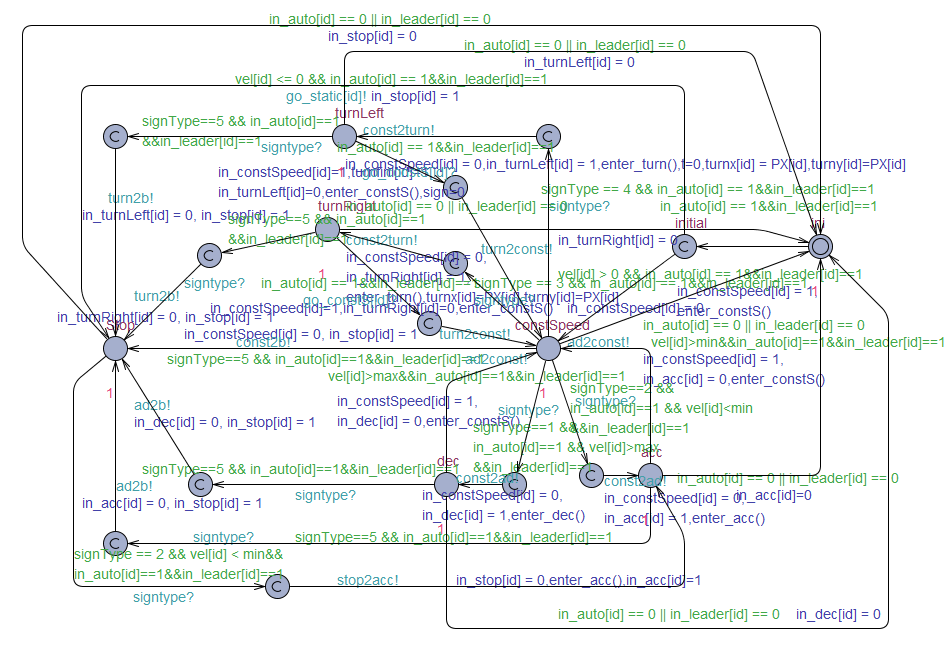}
  \caption{STA for modeling \emph{leader} state}
  \label{fig:leader_smc}
\end{figure}

\begin{figure}
\centering
\subfigure[Stop]{
\label{fig_Stop}
\includegraphics[width=4.5in]{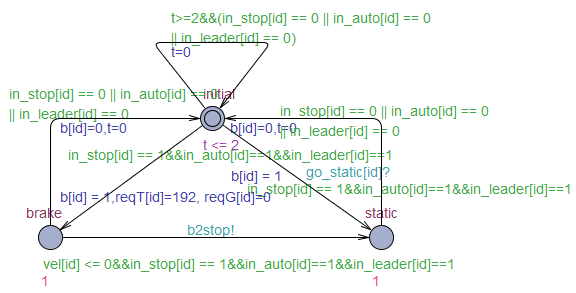}}
\subfigure[Turn Left]{
\label{fig_turnLeft}
\includegraphics[width=6.5in]{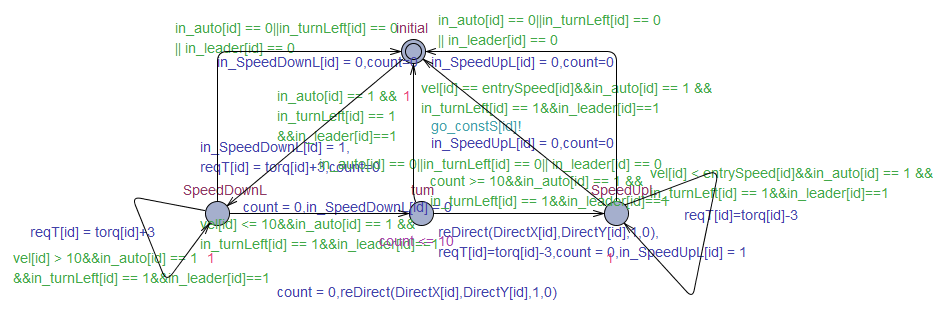}}
\subfigure[Turn Right]{
\label{fig_turnRight}
\includegraphics[width=6.5in]{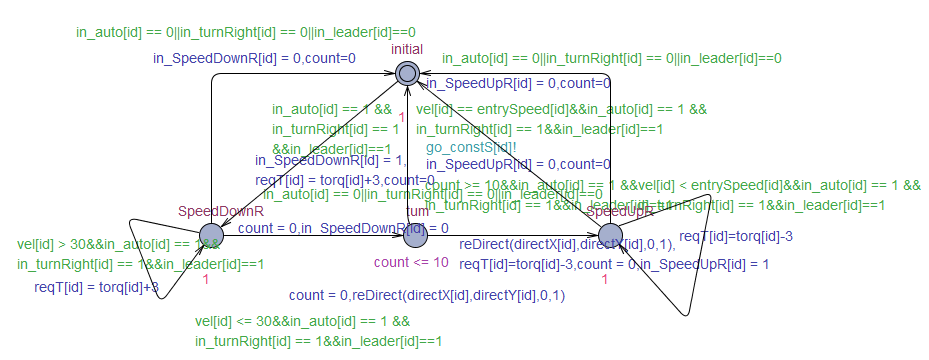}}
\caption{STA for modeling substates of \emph{leader} state}
\label{fig:subleader}
\end{figure}

The behaviour of \emph{leader} state in the Stateflow chart is captured by STA illustrated in Fig.\ref{fig:leader_smc}. The leader gets image information from {\gt{CoopEnv}} and changes its running mode according to {\gt{signType}}, the current running state and velocity. The substates of \emph{leader} state is modeled in Fig.\ref{fig:subleader}.

\begin{figure}
\centering
\subfigure[Accelerate]{
\label{fig_acc}
\includegraphics[width=3.8in]{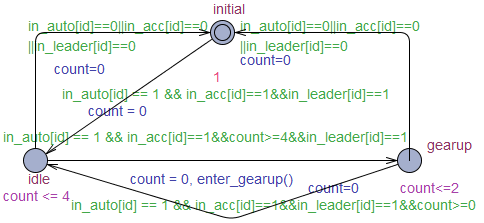}}
\subfigure[Decelerate]{
\label{fig_dec}
\includegraphics[width=3.8in]{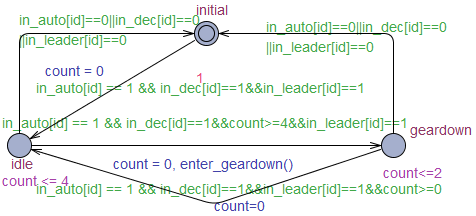}}
\caption{STA for modeling substates of \emph{leader} state}
\label{fig:accdec}
\end{figure}

The inner behaviours of \emph{Stop} can be captured in Fig.\ref{fig:subleader}(a). Similarly, Fig.\ref{fig:subleader} (b) and  Fig.\ref{fig:subleader} (c) describe the behaviour of \emph{turnLeft}, \emph{turnRight}. Fig.\ref{fig:accdec} (a) and Fig.\ref{fig:accdec} (b) illustrates the STA of \emph{acc} and \emph{dec} state respectively. When the leader vehicle detects a stop sign ({\gt{signType}} == 5) and if it is in \emph{Auto} mode, \emph{Stop} state in \emph{leader} STA will be activated and the torque will be set to maximum value (the velocity will be decreased rapidly). When the velocity of the vehicle is less than or equal to 0, it will stop and \emph{static} state (shown in Fig.\ref{fig:subleader}) will be active. When a minimum speed limit sign is detected or the driver takes control of the vehicle, the \emph{straight} state will be active.  When the lead vehicle detects a left turn sign and the \emph{straight} state is active, \emph{turnLeft} state will be active. The vehicle will first decelerate by decreasing the value of its gear until its velocity is less than or equal to 30km/h. Afterwards, the direction of the vehicle will be changed according to its current direction and the detected traffic sign. After turning, the vehicle will increase its velocity to the original velocity before turning.

\begin{figure}
\centering
\subfigure[speedDown]{
\label{fig_Stop}
\includegraphics[width=5.5in]{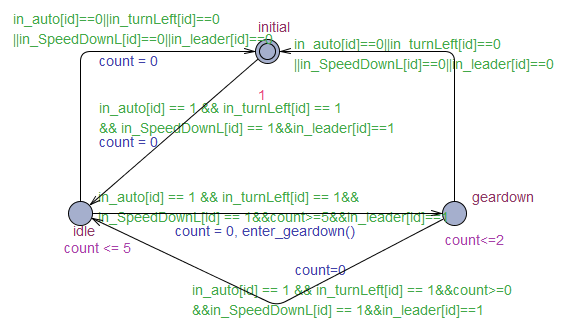}}
\subfigure[speedUp]{
\label{fig_turnLeft}
\includegraphics[width=5.5in]{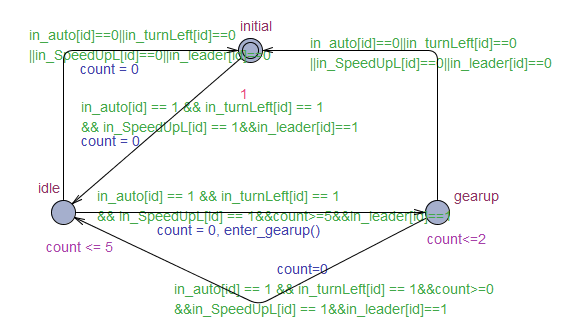}}
\caption{STA of \emph{turnLeft} state}
\label{fig:subturnLeft}
\end{figure}

The inner behaviour \emph{speedUp} and \emph{speedDown} is modeled in Fig.\ref{fig:subturnLeft}. When the vehicle is decelerating, the torque (\emph{torq}) of the vehicle will be increased and the gear (\emph{gear}) will be decreased. Since the velocity of one vehicle is positively proportional to \emph{gear}, the velocity will be increased when \emph{gear} is increased. When the vehicle finishes turning, the torque of the vehicle will be decreased and the gear will be increased to resume to the original velocity.

\begin{figure}
\centering
\subfigure[speedDown]{
\label{fig_Stop}
\includegraphics[width=5.5in]{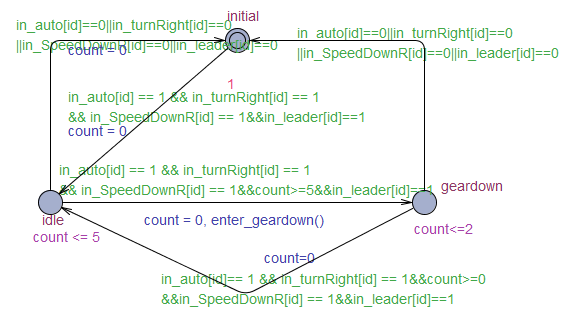}}
\subfigure[speedUp]{
\label{fig_turnLeft}
\includegraphics[width=5.5in]{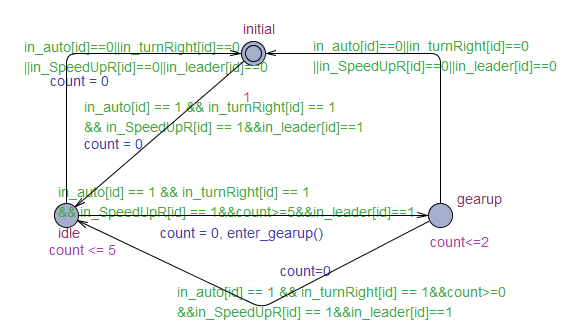}}
\caption{STA of \emph{turnRight} state}
\label{fig:subturnRight}
\end{figure}

The inner behaviour of \emph{turnLeft} state is captured in Fig.\ref{fig:subturnRight}. Similarly, when the vehicle detects a turn right sign, it will decelerate until its speed velocity is less than 30km/h. Afterwards, it will turn (its running directions will be changed) and increases its velocity. The \emph{acc} and \emph{dec} states (Fig.\ref{fig:subleader}(d), (e)) represent the states that the vehicle is accelerating or decelerating. When the lead vehicle (v1) detects a minimum speed limit, and the velocity of the vehicle is less than the limit, \emph{acc} state will be active. The \emph{torque} of the vehicle will be decreased to order to increase the velocity. In the mean time, the \emph{gear} of the vehicle will be increased every 40ms until it reaches the maximum value. When the lead vehicle detects a maximum speed limit and the velocity is larger than the limit, it will activate \emph{dec} state and its velocity will be decreased.

\begin{figure}[htbp]
  \centering
  \includegraphics[width=7in]{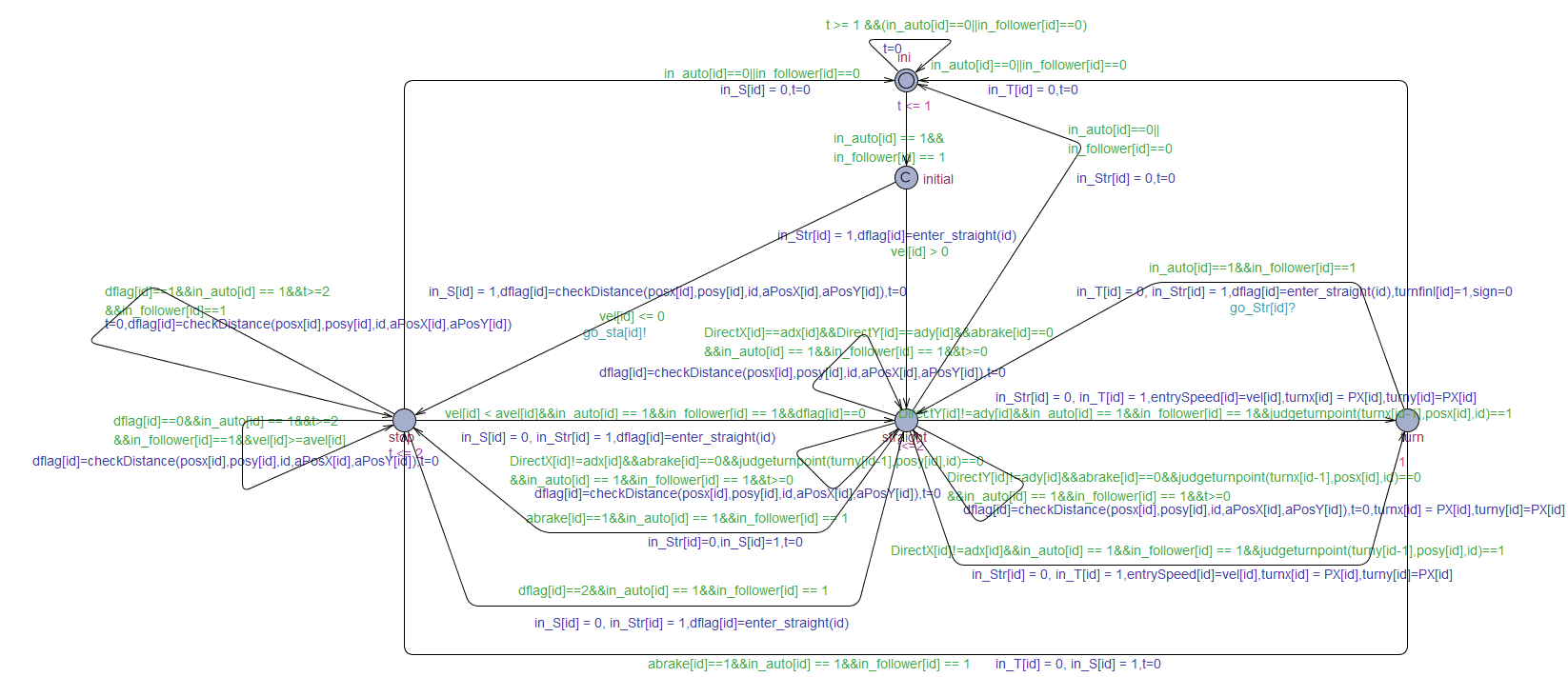}
  \caption{STA of \emph{follower} state}
  \label{fig:follower}
\end{figure}

\begin{figure}
\centering
\subfigure[stop]{
\label{fig_stop}
\includegraphics[width=4.7in]{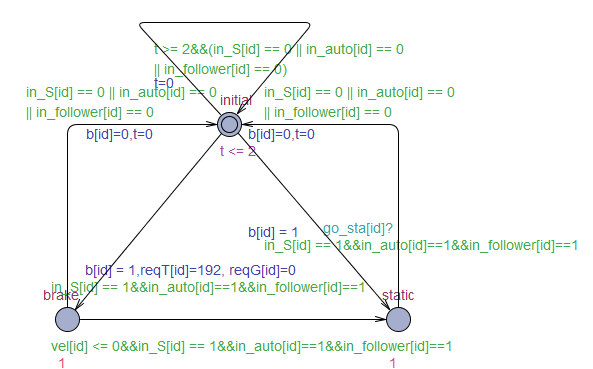}}
\subfigure[turn]{
\label{fig_turn}
\includegraphics[width=6.5in]{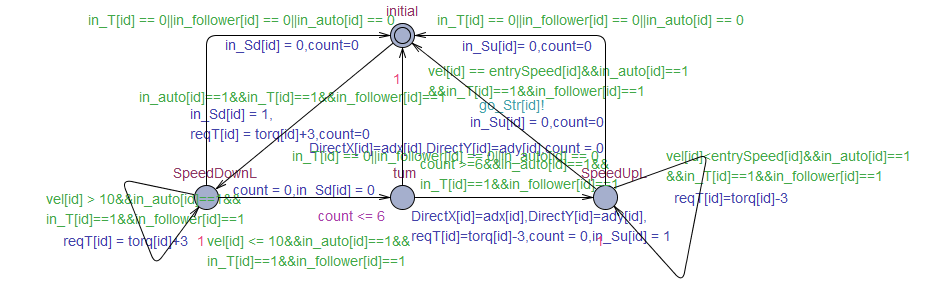}}
\subfigure[straight]{
\label{fig_straight}
\includegraphics[width=7.2in]{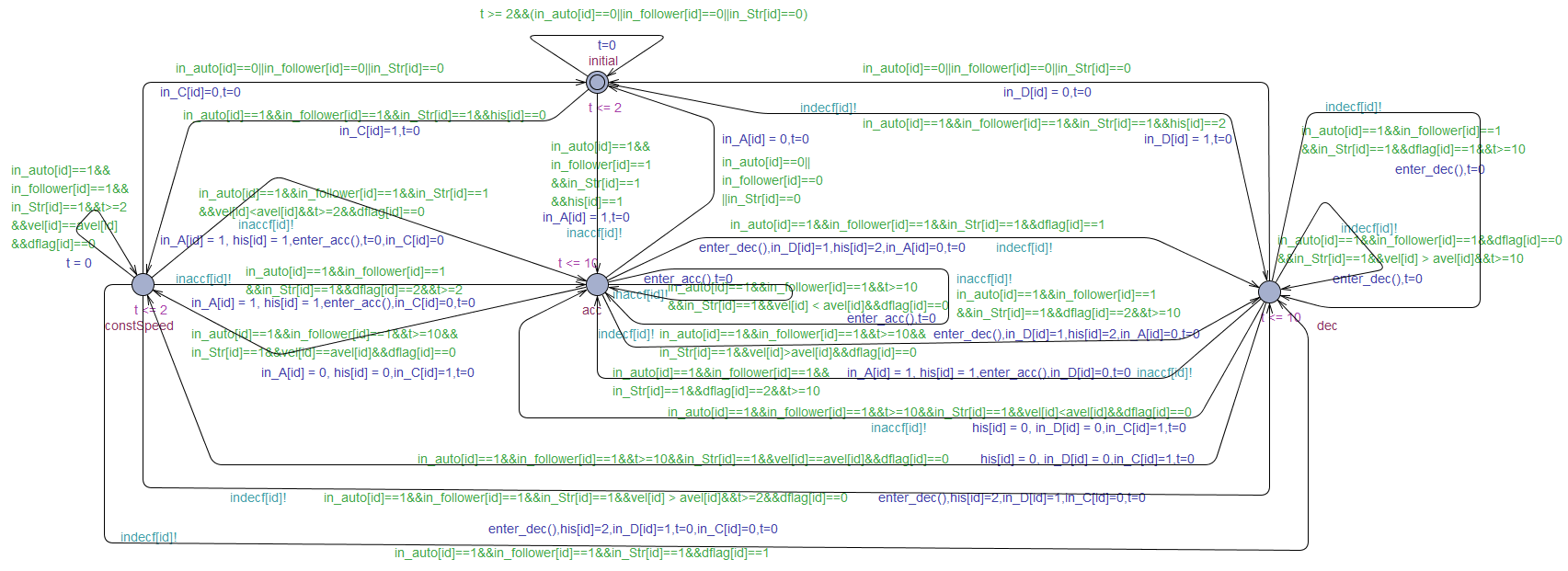}}

\caption{STA of \emph{follower} state}
\label{fig:subfollower}
\end{figure}

\begin{figure}
\centering
\subfigure[dec]{
\label{fig_decf}
\includegraphics[width=4.2in]{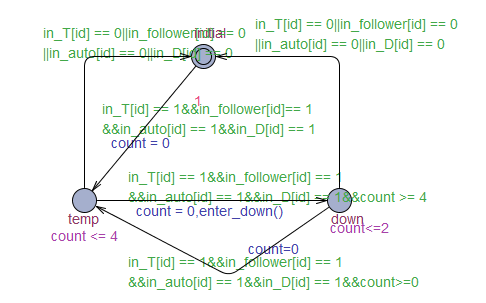}}
\subfigure[acc]{
\label{fig_accf}
\includegraphics[width=4.2in]{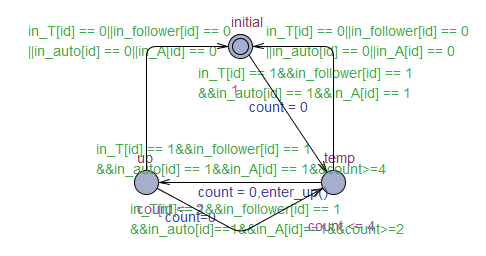}}
\caption{STA for modeling substates of \emph{straight} state}
\label{fig:substraight}
\end{figure}

The behaviour of the follower is captured in Fig.\ref{fig:follower}. The follow vehicle will decide its behaviour by three super-states: \emph{stop}, \emph{turn} and \emph{straight}, which will be active according to the velocity and running mode of the lead vehicle.
It is required to guarantee the distance between the following vehicle and the lead vehicle should be limited in order to ensure the safety. When the follower is in \emph{straight} state (Fig.\ref{fig_straight}), it will always check the distance departing from the leader vehicle by executing \emph{checkDistance()} function.
If the distance is less than the safety distance, then it will activate \emph{dec} state (Fig.\ref{fig:subfollower}(c)) inside \emph{straight} state.
If the following vehicle is in \emph{dec} state, which is shown in Fig.\ref{fig:substraight}(a), it will increase its torque or decrease its gear every 40ms until it reaches 0. If the distance between the following vehicle and the leader vehicle is greater than the safety distance, the vehicle will stop decelerating. When the distance between the following vehicle and the leader vehicle is greater than 500m, in this case,  the communication between the two vehicles may be lost. Hence the following vehicle should increase its speed in order to maintain the communication quality.

\begin{figure}
\centering
\subfigure[speedDown]{
\label{fig_speeddownf}
\includegraphics[width=4.5in]{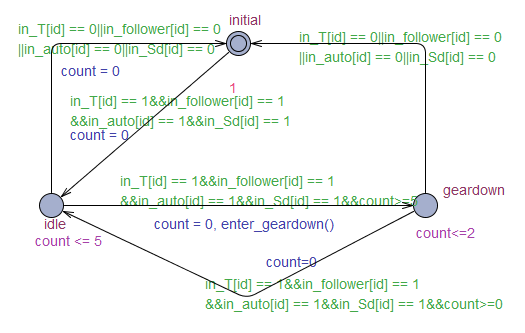}}
\subfigure[speedUp]{
\label{fig_speedupf}
\includegraphics[width=4.5in]{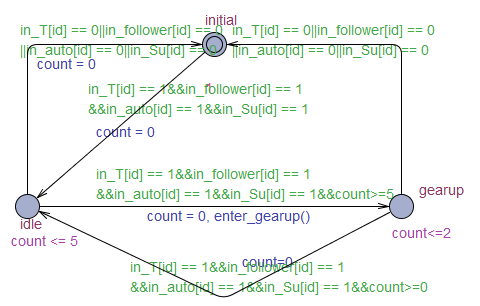}}
\caption{STA for modeling substates of \emph{turn} state}
\label{fig:subturn}
\end{figure}

As presented in Fig.\ref{fig:substraight}(b), when the vehicle accelerates, it will decrease its torque and increase its gear every 40ms. If the lead vehicle is braking, the follower will activate \emph{stop} state (shown in Fig.\ref{fig:subfollower}(a)).
When the follower stops, it will maintain the safety distance with the lead vehicle.
When the follow vehicle detects that its running direction and the running direction of the lead vehicle are different, it will turn left/right in order to keep the same running direction with the lead vehicle. In this case, the following vehicle will first reach the turning point of the lead vehicle (recorded in \emph{judgeturnpoint}), which is illustrated in Fig.\ref{fig:follower}. Similar to lead vehicle, the following vehicle will first decreases the velocity to turn and increase to finish turning. Detailed behaviors of \emph{turn} state is captured in Fig.\ref{fig:subturn}.

\begin{figure}[htbp]
  \centering
  \includegraphics[width=6.5in]{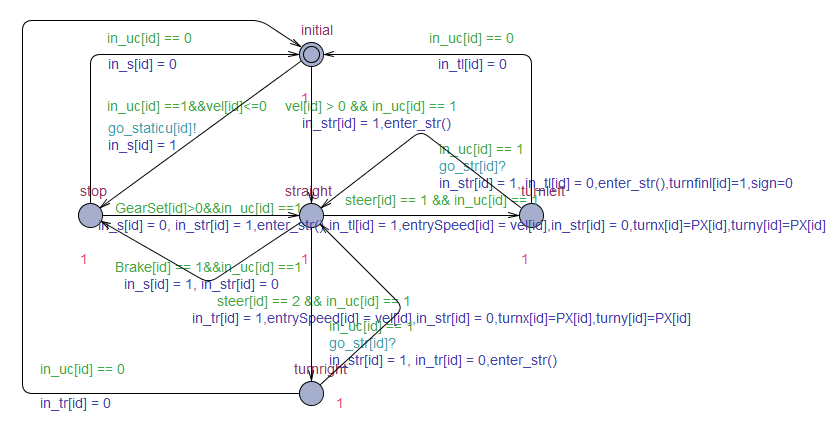}
  \caption{STA for modeling \emph{userctrl} state}
  \label{fig:userctrl}
\end{figure}

\begin{figure}
\centering
\subfigure[stop]{
\label{fig_stop}
\includegraphics[width=3.5in]{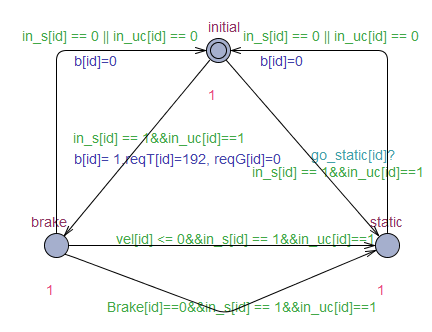}}
\subfigure[TurnLeft]{
\label{fig_turnleft}
\includegraphics[width=5in]{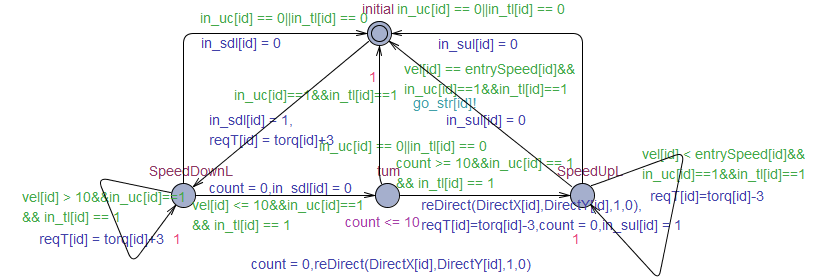}}
\subfigure[TurnRight]{
\label{fig_turnright}
\includegraphics[width=5.0in]{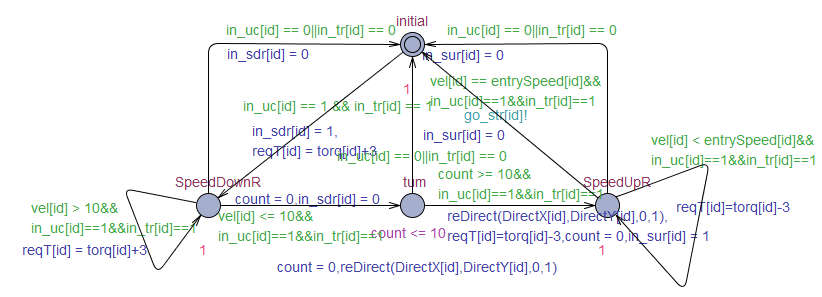}}
\subfigure[Straight]{
\label{fig_str}
\includegraphics[width=5.5in]{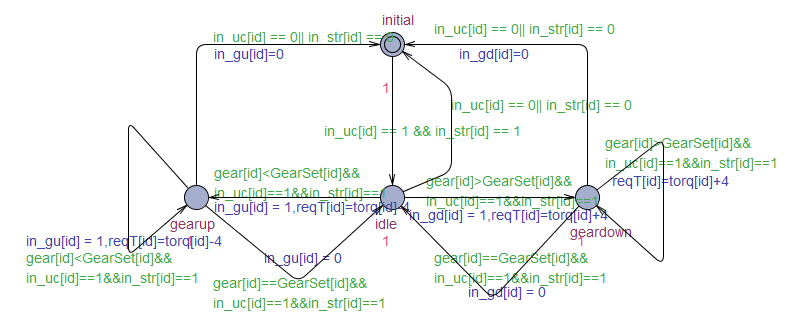}}
\caption{STA  of \emph{UserCtrl} state}
\label{fig:subuserctrl}
\end{figure}

If the driver takes control of the vehicle,  \emph{userCtrl} (Fig.\ref{fig:userctrl}) state will be active. The movement of the vehicle will  be controlled by the driver. Hence when the driver pressed the brake pedal, \emph{stop} state in  \emph{userCtrl}  will be active. Detailed behaviours of \emph{stop} state is captured in Fig.\ref{fig:subuserctrl}(a).

\begin{figure}
\centering
\subfigure[SpeedDownL]{
\label{fig_speeddownf}
\includegraphics[width=4.5in]{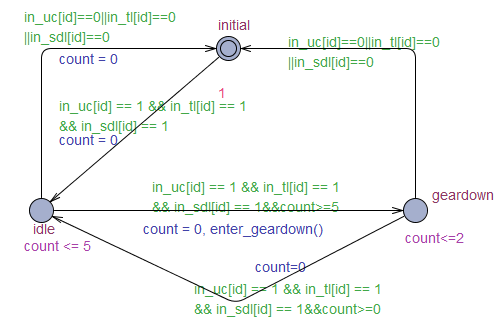}}
\subfigure[SpeedUpL]{
\label{fig_speedupf}
\includegraphics[width=4.5in]{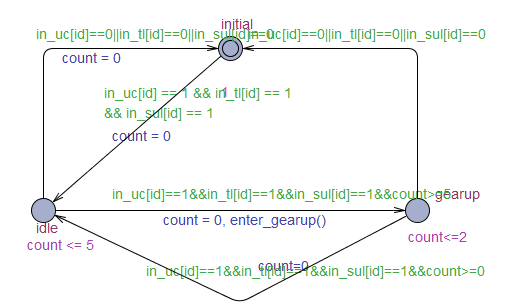}}
\caption{STA for modeling substates of \emph{turnleft} state}
\label{fig:subturnleft}
\end{figure}

\begin{figure}
\centering
\subfigure[SpeedDownR]{
\label{fig_speeddownf}
\includegraphics[width=4.5in]{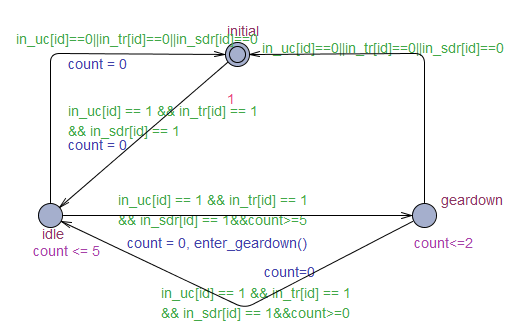}}
\subfigure[SpeedUpR]{
\label{fig_speedupf}
\includegraphics[width=4.8in]{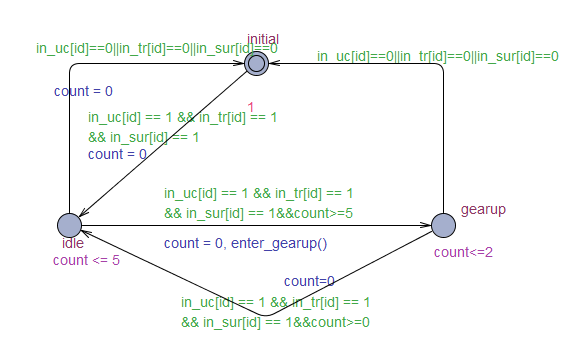}}
\caption{STA for modeling substates of \emph{turnright} state}
\label{fig:subturnright}
\end{figure}

However, when the driver increases the velocity of the vehicle, \emph{stop} state will transit to \emph{straight} state. In \emph{straight} mode,the speed of the vehicle can  be increased/decreased  by changing the gear (gear up or gear down). As shown in Fig.\ref{fig:substraight}, the vehicle will adjust its velocity according to the operation by the driver. If the driver steer to left or right, the vehicle will decrease its velocity to turn and increase the velocity to the original velocity after turning (as illustrated in Fig.\ref{fig:subturnleft} and Fig. \ref{fig:subturnright}).

\begin{figure}
\centering
\subfigure[gearUp]{
\label{fig_gearup}
\includegraphics[width=3.5in]{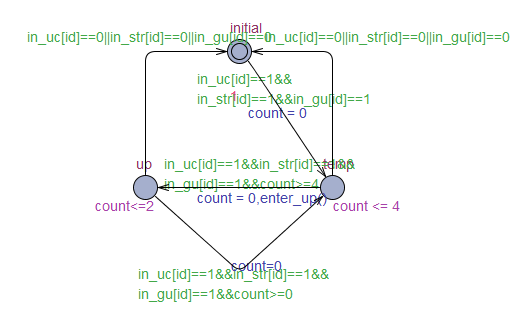}}
\subfigure[gearDown]{
\label{fig_geardown}
\includegraphics[width=3.0in]{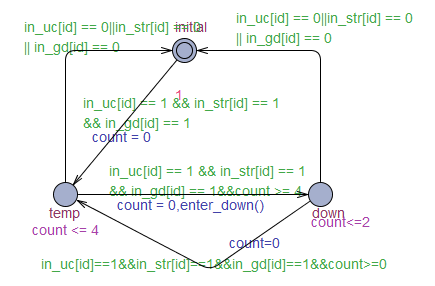}}
\caption{STA for modeling substates of \emph{straight} state}
\label{fig:substraight_}
\end{figure}

\chapter{(Non)-functional Property Translation in \smc}
\label{sec:smc-trans}
In this section, we discuss semantics of the extended {\gt{Periodic, Sporadic}}, and {\gt{Comparison}} timing constraints (\xtc) with probabilistic parameters according to \tdl \cite{TADL2}. Afterwards, we provide translation of \xtc\ in \smc. Finally, we present how the energy consumption of an individual vehicle in CAS can be modeled in \smc.

\section{Probabilistic Timing Constraints Translation}

\xtc\ that we verified follow the weakly-hard (\textbf{{\gt{WH}}}) approach \cite{TADL2}, which describes that a bounded number of occurrences among all are allowed to violate the constraints. The semantics of the \textbf{{\gt{WH(c,m,k)}}} is that a system behavior must satisfy a given timing constraint {\gt{c}} at least {\gt{m}} times out of {\gt{k}} consecutive occurrences. We use object-oriented notation to define the attributes of occurrences of an event except for {\gt{Comparison}} timing constraint, which does not related to any event. {\gt{\fp.e}} refers to an event {\gt{e}} of \fp. we apply {\gt{e$_{\gt{i}}$}} and {\gt{e(i)}} to represent the $i^{th}$ occurrence of {\gt{\fp.e}} and the time point of $i^{th}$ occurrence respectively.

\begin{figure}
\centering
\subfigure[Source]{
\label{fig_source}
\includegraphics[width=1.2in]{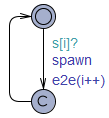}}
\subfigure[End-to-End (e2e)]{
\label{fig_e2e}
\includegraphics[width=3in]{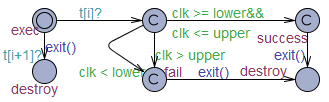}}
\subfigure[Noncumulative]{
\label{fig_noncuperiodic}
\includegraphics[width=2.8in]{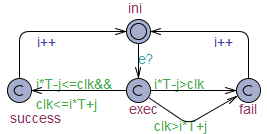}}
\subfigure[Sporadic]{
\label{fig_spo}
\includegraphics[width=2.2in]{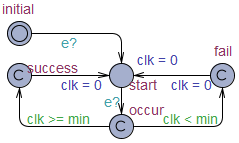}}

\caption{Probabilistic Extension of Timing Constraints in \smc}
\label{fig:smctemplate}
\end{figure}

\noindent \textbf{{\gt{WH(End-to-End,m,k)}}} limits the interval between the occurrences of {\gt{\fp.source}} and corresponding {\gt{\fp.target}}. For any {\gt{source$_{\gt{i}}$}} $\in$ {\gt{\fp.source}} ({\gt{i}} $\leq$ {\gt{$\mid$\fp.source$\mid$}}), out of {\gt{k}} consecutive occurrences of {\gt{\fp.source}}, i.e., {\gt{source$_{\gt{i}}$}} to {\gt{source$_{\gt{i+k}}$}}, at least {\gt{m}} satisfy {\emph{lower}} $\leq$ {\gt{$\mid$target(i)-source(i)$\mid$}} $\leq$ {\emph{upper}}, where {\gt{$\mid$target(i)-source(i)$\mid$}} denotes the corresponding delay of {\gt{\fp.source}}. We have previously proposed the translation pattern of probabilistic {\gt{End-to-End}} constraint, which is applicable for the system where {\gt{target$_{\gt{i}}$}} always happens prior to {\gt{source$_{\gt{i+1}}$}}.
To allow the case that {\gt{source$_{\gt{i+1}}$}} occurs prior to {\gt{target$_{\gt{i}}$}}, we model {\gt{End-to-End}} constraint as a spawnable STA \cite{david2015uppaal}. 
As shown in Fig.\ref{fig:smctemplate}(a), whenever {\gt{\fp.source}} occurs, {\gt{Source}} STA will activate a spawned STA ({\gt{STA$_{\gt{sp}}$(i)}}) from {\gt{End-to-End}} STA (in Fig.\ref{fig:smctemplate}(b)). 
When {\gt{STA$_{\gt{sp}}$(i)}} is activated, the clock $clk$ starts to count from 0 until {\gt{target$_{\gt{i}}$}} occurs. {\gt{STA$_{\gt{sp}}$(i)}} will judge whether the time interval from {\gt{source$_{\gt{i}}$}} to {\gt{target$_{\gt{i}}$}} is within the lower and upper bound. According to the judgment, either \emph{success} or \emph{fail} location will be activated, representing that the {\gt{End-to-End}} constraint is satisfied or not. Afterwards, {\gt{STA$_{\gt{sp}}$(i)}} will be inactive. To ensure the correspondence of {\gt{\fp.source}} and {\gt{\fp.target}}, if {\gt{target$_{\gt{i}}$}} doesn't occur, {\gt{STA$_{\gt{sp}}$(i)}} will be inactive  when {\gt{target$_{\gt{i+1}}$}} occurs. {{\gt{WH(End-to-End,m,k)}}} is specified as: $Pr[bound]$ ($[\ ]$ $\forall$ {\gt{source$_{\gt{i}}$}} $\in$ {\gt{\fp.source}} ($\neg$ {\gt{STA$_{\gt{sp}}$(i)}}.$fail$)) $\geqslant$ $P$.

\noindent \textbf{{\gt{WH(Periodic,m,k)}}} limits the period of the successive occurrences of an event including \emph{jitter}. As our previous work has provide probabilistic extension of  {\gt{Cumu\_\\Periodic}} and translation rules to STA, we focus on {\gt{Noncumu\_Periodic}}. {\gt{Noncumu\_\\Periodic}}: For any {\gt{e(i)}} $\in$ {\gt{\fp.start}}, at least {\gt{m}} times out of {\gt{k}} occurrences, {\gt{e(i)}} satisfies: ({\emph{i*T-jitter}} $\leq$ {\gt{e(i)}} $\leq$ {\emph{i*T+jitter}}).
In Fig.\ref{fig_noncuperiodic}, the STA enforces the $i^{th}$ event \emph{event} to occur within [{\emph{i*T-jitter}}, {\emph{i*T+jitter}}]. Initially, STA resets the local clock $clk$ when it counts to $jitter$, which indicates that the time moves backwards by $jitter$ time on the time line. When \emph{event} happen, STA will judge if $clk$ is within [{\emph{T-2*jitter}}, {\emph{T}}] and changes its current state to either (\emph{success} or \emph{fail}) based on the judgment. Afterwards, the STA will wait at \emph{reset} till period $T$ and reset $clk$. Finally the STA repeats the calculation. {\gt{WH(Noncumu\_Periodic,m,k)}} is specified as: $Pr[bound]$ ($[\ ]$ $\neg$ $STA_{Noncumu\_Periodic}.fail$) $\ge$ $P$.

\vspace{0.05in}
\noindent \textbf{{\gt{WH(Sporadic,m,k)}}} describes a single event \fp\ that occurs sporadically. For any {\gt{e(i)}} $\in$ {\gt{\fp.start}} s.t. its successive occurrence {\gt{e'(i)}} starts at {\gt{e(i)}}, out of {\gt{k}} consecutive occurrences {\gt{e(i)}}, i.e., {\gt{e(i)}} to {\gt{e(i+k)}}, at least {\gt{m}} sequences satisfy ({\gt{$\mid$e'(i)-e(i)$\mid$}} $\geq$ {\gt{min}}), where {\gt{min}} denotes the minimum time interval between consecutive occurrences.In Fig.\ref{fig_spo}, initially, STA resets the local clock $clk$ when it receives signal $event?$, which indicates the first occurrence of an event. When STA receives $event?$ again, it will judge if $clk$ is larger than $min$ and changes its current state to either (\emph{success} or \emph{fail}) based on the judgment. Finally STA reset $clk$ and repeat calculation. {\gt{WH(Sporadic,m,k)}} is specified as: $Pr[bound]$ ($[\ ]$ $\neg$ $STA_{Sporadic}.fail$) $\ge$ $P$.

\vspace{0.05in}
\noindent \textbf{{\gt{WH(Comparison,m,k)}}} limits the comparison relation (including $<$, $\leq$, $==$, $\geq$ and $>$) among timing expressions, where {\gt{WH(Comparison,m,k)}} can be expressed as arithmetic. {\gt{WH(Comparison,m,k)}} is defined as: out of {\gt{k}} runs, at least {\gt{m}} times the comparison relation between ${\gt{TimingExpr1}}$ and ${\gt{TimingExpr2}}$ is satisfied. \\ {\gt{WH(Comparison,m,k)}} is interpreted as: $Pr[bound]$($[]$ {\gt{TimingExpr1}} $\prec$ {\gt{TimingExpr2}}) $\ge$ $P$, where $\prec$ $\in$ \{$<$, $\leq$, $=$, $\geq$ or $>$\}.

\section{Energy Constraint Estimation}

\begin{figure}[htbp]
  \centering
  \includegraphics[width=5.5in]{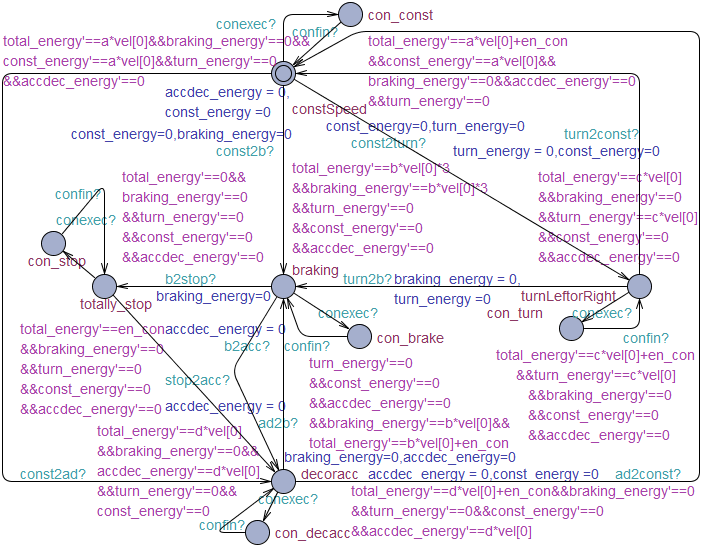}
  \caption{Energy estimation of {\gt{v1Controller}}}
  \label{energyofcon}
\end{figure}

Fig.\ref{energyofcon} shows the stochastic timed automata (STA) that estimates the energy consumption of the Controller of the lead vehicle with extended arithmetic on clocks: \emph{total\_energy} represents the total energy consumption of the Controller of the lead vehicle. Different modes of the vehicle are corresponded to different locations in the STA. As the energy consumption rate for turning right and turning left mode are the same, we combine the two modes into one location in the STA, i.e., \emph{turnLeftorRight}. Similarly, the acceleration and deceleration mode of the Controller are combined into \emph{decoracc}. Since the energy consumption rate varies with the running mode of the vehicle, e.g., the energy consumes faster when the vehicle is in braking mode than when the vehicle drives in a  constant speed, we define \emph{total\_energy'} (the rate of \emph{total\_energy}) with ordinary differential equation (ODE) and assign different values to \emph{total\_energy'} for different locations in the STA. Parameters \emph{a}, \emph{b}, \emph{c} and \emph{d} are user-defined coefficients to estimate energy consumption of the lead vehicle when it drives in a constant speed, brakes, turns or increases (decreases) its speed respectively. The values of the coefficients indicate the increasing rate of the energy consumption of the Controller and \emph{b $>$ d $>$ c $>$ a}. The estimation of energy consumption requirement of Controller (R48) will be evaluated in Chap.\ref{sec:experiment}.

\begin{figure}
\centering
\subfigure[Energy of {\gt{ComDevice}}]{
\label{fig_enc}
\includegraphics[width=3.5in]{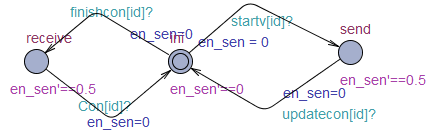}}
\subfigure[Energy of {\gt{VeModeDevice}}]{
\label{fig_env}
\includegraphics[width=3in]{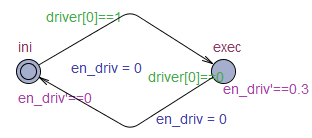}}
\subfigure[Energy of {\gt{SignRecDevice}}]{
\label{fig_ens}
\includegraphics[width=3in]{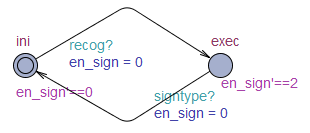}}

\caption{Energy constraints on sensors in \smc}
\label{fig:energysensor}
\end{figure}

Also, the four sensors in Fig.\ref{fig:east}(b) will consume energy. As {\gt{VeDynamicDevice}} is an imitation of how the velocity of an vehicle can be achieved, its energy consumption is more like the mechanical kinetic energy of the vehicle. While in our case, the velocity and the energy consumption of the vehicle is strongly related to the mode of the vehicle, thus we model the mechanical kinetic energy consumption in {\gt{Controller}} (Fig.\ref{energyofcon}). The energy of the other three \fp s, {\gt{ComDevice}}, {\gt{SignRecDevice}} and {\gt{VeModeDevice}}, are modeled in Fig.\ref{fig:energysensor}. In order to simplify calculation, we set constants as the increasing rate of energy consumption of each sensor. When the sensor is working, the energy consumption will increase uniformly. And when the sensors don't work anymore, the energy consumption will not increase.

\chapter{Experiment and Verification}
 \label{sec:experiment}

Formal verification of functional and non-functional properties are conducted by using SDV and \smc. Since SDV only supports a sub-set of blocks in Simulink block library, several blocks need to be replaced from the original model, e.g., \emph{look-up table} block and \emph{random value generator}. The \emph{proof objective models} that specify the functional and non-functional properties are constructed, which are illustrated in the following figures.
\begin{figure}[htbp]
  \centering
  \includegraphics[width=3.8in]{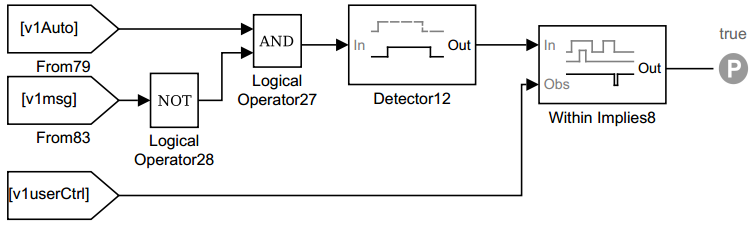}
  \caption{ When v1 runs automatically and if there is no message received by the vehicle for a certain time, the running mode of the vehicle should be changed to manual mode within a certain time: \textbf{G} (\textbf{G}$_{[0,200]}$ v1.Auto==true $\wedge$  v1.msg == false $\implies$ \textbf{F}$_{[0,2000]}$ v1.userCtrl == true)}
  \label{sdvr1}
\end{figure}

\begin{figure}[htbp]
  \centering
  \includegraphics[width=3.8in]{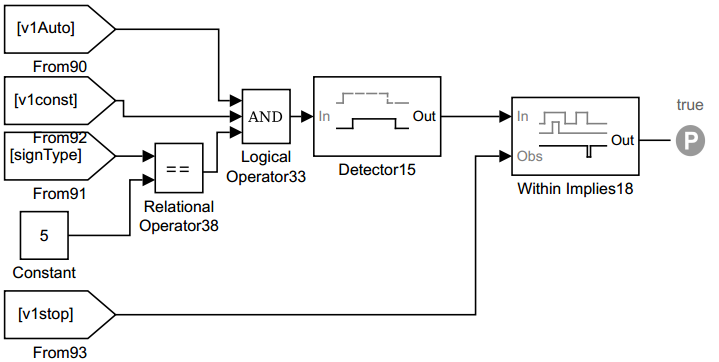}
  \caption{ When v1 runs automatically with a constant speed, if it detects a stop sign, it should start to brake in 500ms:
\textbf{G} (v1.Auto == true $\wedge$ v1.const == true $\wedge$ signType == 5 $\implies$ \textbf{F}$_{[0,500]}$ v1.stop == true)
}
  \label{sdvr4}
\end{figure}
\begin{figure}[htbp]
  \centering
  \includegraphics[width=3.8in]{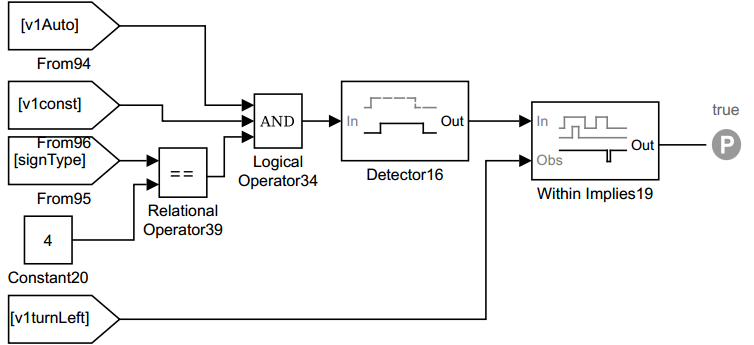}
  \caption{ When the lead vehicle (v1) runs automatically with a constant speed, if it detects left turn sign, it should start to turn left in 200ms:
\textbf{G} (v1.auto == true $\wedge$  v1.const == true $\wedge$ signType == 4 $\implies$ \textbf{F}$_{[0,200]}$ v1.turnLeft == true)}
  \label{sdvr5}
\end{figure}

\begin{figure}[htbp]
  \centering
  \includegraphics[width=3.8in]{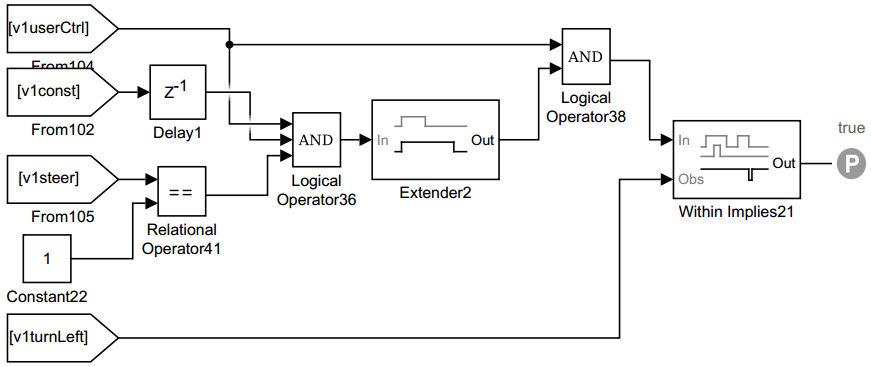}
  \caption{ If v1 is driven by the driver and it is running with a constant speed, the driver steers to the left, the vehicle will turn left within 200ms: \textbf{G} (v1.userCtrl == true $\wedge$  v1.const == true $\wedge$ v1.steerReq == 1 $\implies$ \textbf{F}$_{[0,200]}$ v1.turnLeft == true)
}
  \label{sdvr7}
\end{figure}

\begin{figure}[htbp]
  \centering
  \includegraphics[width=3.8in]{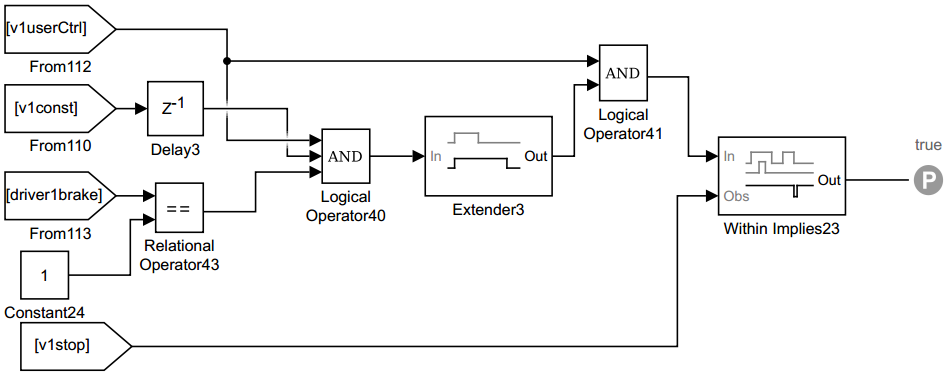}
  \caption{ If v1 is driven by the driver and it is running at a constant speed, the driver brakes the car, the vehicle will slow down. \textbf{G}: (v1.userCtrl == true $\wedge$  v1.const == true $\wedge$ v1.brakeReq == 2 $\implies$ \textbf{F}$_{[0,200]}$ v1.brake == true)}
  \label{sdvr9}
\end{figure}

\begin{figure}[htbp]
  \centering
  \includegraphics[width=3.8in]{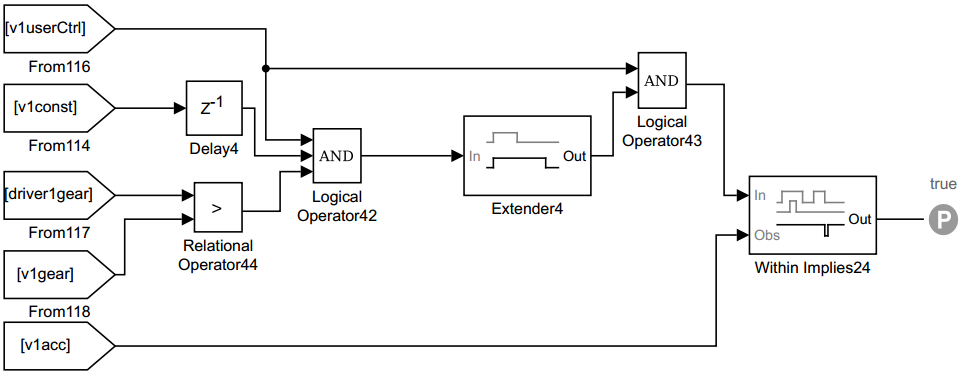}
  \caption{ When v1 is driven by the driver and it is running at a constant speed, if the driver gears up, the vehicle will accelerate.
\textbf{G} (v1.userCtrl == true $\wedge$  v1.const == true $\wedge$ v1.gearReq $>$ v1.gear $\implies$ \textbf{F}$_{[0,200]}$ v1.acc == true)}

  \label{sdvr10}
\end{figure}

\begin{figure}[htbp]
  \centering
  \includegraphics[width=5in]{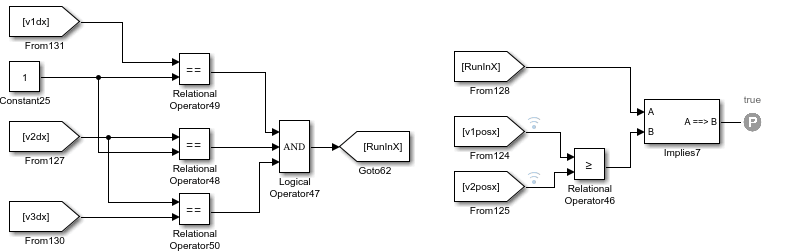}
  \caption{ When the vehicles are running horizontally, following vehicle should not run ahead of lead vehicle. \textbf{G} (v1.dx == 1 $\wedge$ v2.dx == 1 $\wedge$ v3.dx == 1 $\implies$ v1.x $>$ v2.x)}
  \label{sdvr12}
\end{figure}

\begin{figure}[htbp]
  \centering
  \includegraphics[width=3.8in]{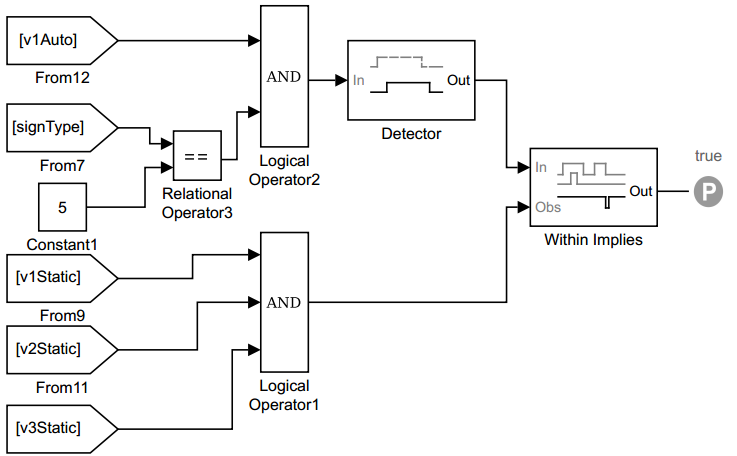}
  \caption{ When the lead vehicle detects a stop sign, the three vehicles should stop in 5s.
\textbf{G} (v1.auto == true $\wedge$ signType == 5 $\implies$ \textbf{F}$_{[0,5000]}$ v1.static == true $\wedge$ v2.static == true $\wedge$ v3.static==true)
}
  \label{sdvr13}
\end{figure}

\begin{figure}[htbp]
  \centering
  \includegraphics[width=3.8in]{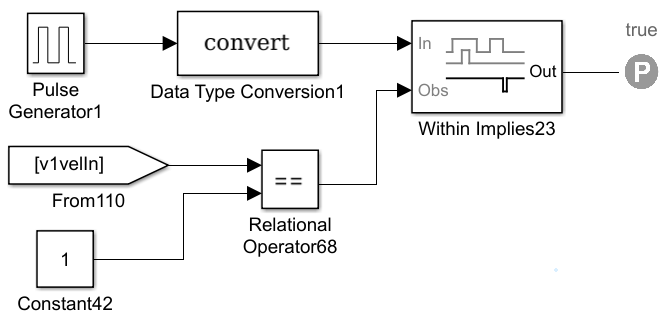}
  \caption{{\gt{Noncumulative Periodic}} timing constraint: A {\gt{Periodic}} acquisition of {\gt{VehicleDynamic}} of v1 must be carried out for every 50 ms with a jitter 10 ms.
\textbf{G} (v1.velIn == true $\implies$ v1.velIn == false \textbf{U}$_{[40,\ 60]}$ v1.velIn == true)
}
  \label{sdvr27}
\end{figure}
\begin{figure}[htbp]
  \centering
  \includegraphics[width=5in]{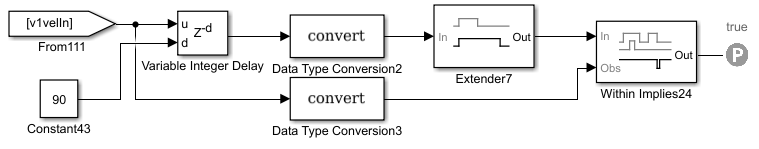}
  \caption{{\gt{Cumulative Periodic}} timing constraint of R27}
  \label{sdv27-1}
\end{figure}

\begin{figure}[htbp]
  \centering
  \includegraphics[width=3.8in]{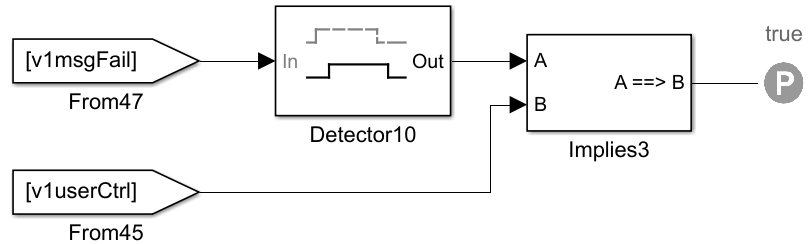}
  \caption{{\gt{Sporadic}} constraint: After the running mode of the lead vehicle is changed to manual mode because of the failed message transmission, the driver should not change it into \emph{auto} mode within 20 seconds.
\textbf{G} (v1.msgNormal == false $\implies$ v1.auto == false \textbf{U}$_{[2000,\ \infty]}$) v1.auto == true}
  \label{energyes}
\end{figure}
\begin{figure}[htbp]
  \centering
  \includegraphics[width=4.5in]{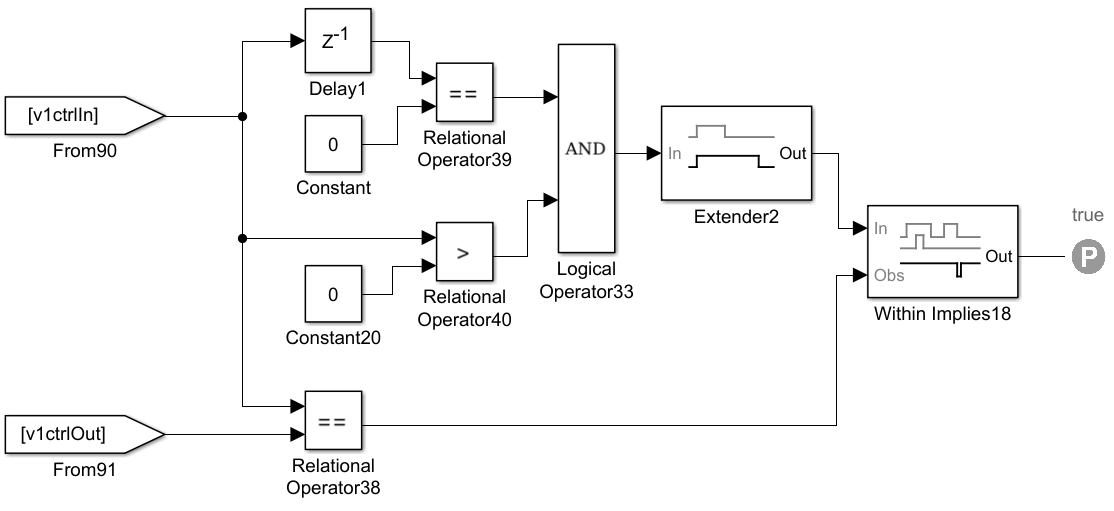}
  \caption{An {\gt{Execution}} constraint applied on {\gt{v3Controller}}, which measured from the inputs ports \emph{Avel} to output ports \emph{gearReq}, \emph{torqueReq}, and \emph{brake}, should be between 100 ms and 300 ms; \textbf{G} (v3.ctrlIn $\implies$ \textbf{F}$_{[100,300]}$ v3.ctrlOut)}
  \label{energyes}
\end{figure}

\begin{figure}[htbp]
  \centering
  \includegraphics[width=5.2in]{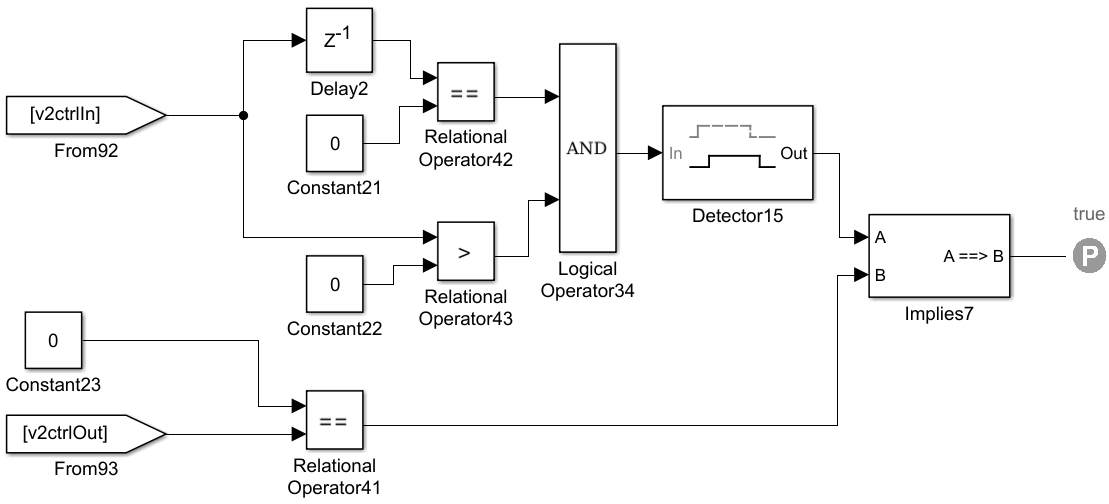}
  \caption{An {\gt{Execution}} constraint applied on {\gt{v3Controller}}, which measured from the inputs ports \emph{Avel} to output ports \emph{gearReq}, \emph{torqueReq}, and \emph{brake}, should be between 100 ms and 300 ms; \textbf{G} (v3.ctrlIn $\implies$ \textbf{F}$_{[100,300]}$ v3.ctrlOut)}
  \label{energyes}
\end{figure}

\begin{figure}[htbp]
\centering
\subfigure[Upper bound]{
\label{t8}
\includegraphics[width=4.5in]{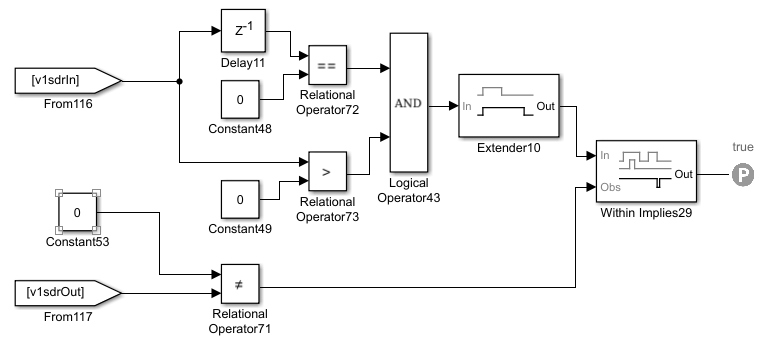}}
\subfigure[Lower bound]{
\label{t8-1}
\includegraphics[width=4.5in]{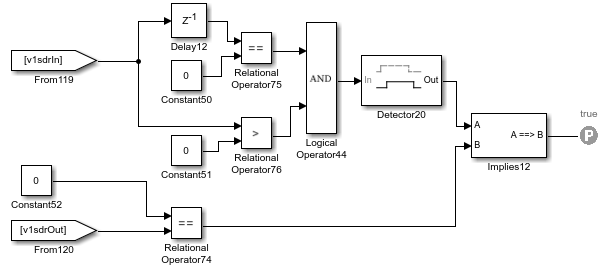}}
\caption{An {\gt{Execution}} constraint applied on {\gt{ComDevice}} of v1, which measured from the inputs ports \emph{torqueSet}, \emph{gearSet} and \emph{brake} to output ports \emph{wheelSpeed}, \emph{direct} should be between 100 ms and 150 ms;
\textbf{G} (v1.sdrIn $\implies$ \textbf{F}$_{[100,300]}$ v1.sdrOut}
\label{t88}
\end{figure}

\begin{figure}[htbp]
  \centering
\subfigure[R1]{
\label{fig_R1}
\includegraphics[width=3in]{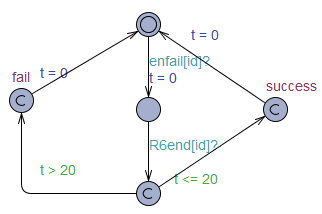}}
\subfigure[Trigger of R1]{
\label{fig_R1trigger}
\includegraphics[width=3in]{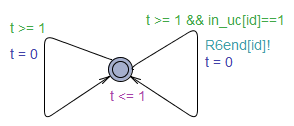}}
  \caption{Observer STA of R1 and its triggering}
  \label{fig:R1}
\end{figure}

Functional properties of CAS (R1-R26) can be measured using observer STAs. Observer STA describes the time delay between two transitions, i.e., two events. An event can be indicated by a synchronization channel, changes of states or data. \gt{Trigger} STA represents  the events that are represented by changes of states or data. Fig.\ref{fig:R1} shows how functional requirements can be modeled in observer STA. For R1, the observer STA will start to count the elapsed time when  communication failure occurs. It stops counting when the driver of this vehicle takes control of the vehicle. An STA is used to indicate the time instant at which the driver takes control of the car (the \emph{manual} mode is activated). Whenever the user takes control (in\_uc[0]==1), a signal will be sent to the STA in Fig.\ref{fig:R1}(a) and time counting will be ceased.

\begin{figure}
\centering
\subfigure[Cumulative Periodic]{
\label{fig_cumup}
\includegraphics[width=2.7in]{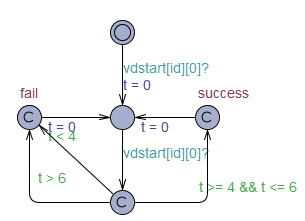}}
\subfigure[Noncumulative Periodic]{
\label{fig_nonp}
\includegraphics[width=2.7in]{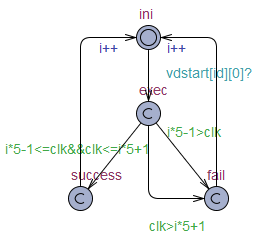}}
\subfigure[Sporadic]{
\label{fig_sp}
\includegraphics[width=2.5in]{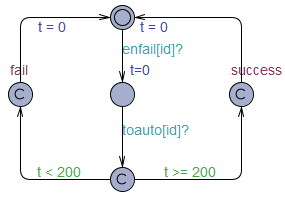}}
\subfigure[Synchronization]{
\label{fig_synchr}
\includegraphics[width=5in]{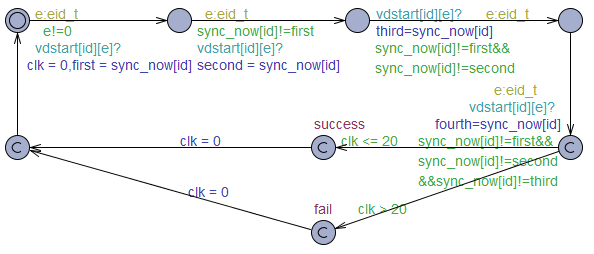}}
\caption{\smc\ STAs for timing constraints verification}
\label{fig:timing}
\end{figure}

Verification of timing properties (R27-R50) are done by adapting observer STAs in Fig.\ref{fig:timing}. R27-R29 are verified by using the STAs in Fig.\ref{fig:timing}(a) and (b), where the  {\gt{VeDynamicDevice}} is triggered periodly with jitter 10ms. R30-R32 are {\gt{Sporadic}} constraints: if communication is missing, the driver should take control of the vehicle for at least 2s (Fig.\ref{fig:timing}(c)). The {\gt{Execution}} timing constraint of {\gt{Controller}} and {\gt{ComDevice}} are considered in STAs shown in Fig.\ref{fig:timing}(d), (e). The {\gt{End-to-End}} constraints measured from the {\gt{Controller}} of the lead vehicles to the {\gt{Controller}} of the follow vehicles (R39 and R40) can be checked by employing the STAs in Fig.\ref{fig:timing}(f). R41-R43 measure the time duration from the {\gt{Controller}} of each vehicle receiving signal from the sensors to sendig information to other vehicles (Fig.\ref{fig:timing}(g)). The synchronization of the input ports of the {\gt{Controller}} can be modeled in STA in Fig.\ref{fig:timing}(h).

\begin{figure}
\centering
\subfigure[Execution on {\gt{Controller}}]{
\label{fig_execc}
\includegraphics[width=3.2in]{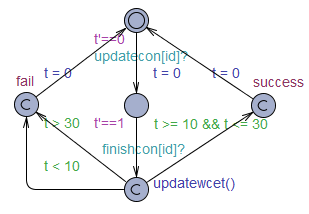}}
\subfigure[Execution on {\gt{Sensor}}]{
\label{fig_execs}
\includegraphics[width=3in]{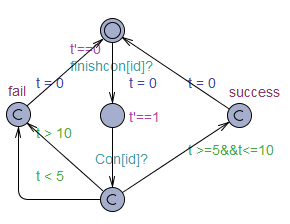}}
\caption{\smc\ STAs for \gt{End-to-End} timing constraints verification}
\label{fig:exe-timing}
\end{figure}

\begin{figure}
\centering
\subfigure[End-to-End between vehicles]{
\label{fig_e2ebv}
\includegraphics[width=3.2in]{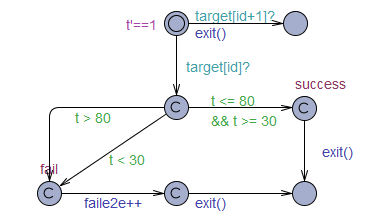}}
\subfigure[End-to-End for one vehicle]{
\label{fig_e2ev}
\includegraphics[width=3in]{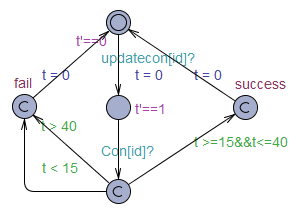}}
\caption{\smc\ STAs for \gt{End-to-End} timing constraints verification}
\label{fig:ee-timing}
\end{figure}

Table.\ref{table_verification_result} depicts the verification result of a list of functional (R1-R26) and non-functional (R27-R50) properties. The experiments are performed on a computer with Intel Core i3 CPU and 4GB of RAM. For {\gt{Periodic}} timing constraints (R27-R29), results are provided in two categories: cumulative (R27.1, R28.1, R29.1) and non-cumulative (R27.2, R28.2, R29.2).
The non-functional properties (R27-R50) are all satisfied in both \smc\ and SDV.
The functional properties R1-R22, R25-R26 are established as valid in SDV and are satisfied in \smc\ with confidence greater than 95\%, while R23-R24 is invalid in both SDV and \smc.

\begin{sidewaystable}[htbp]
\begin{longtable}[htbp]{|c|c|p{320pt}|c|c|c|c|}
  \caption{Logic expression and verification results of the requirements}\\ \hline
    Req & Type & Expression & Result & Time (min) & Memory (Mb) & CPU (\%))\\
    \hline
    \multirow{2}{*}{R1} & SDV & \textbf{G} (\textbf{G}$_{[0,2000]}$ v1.auto==true $\wedge$  v1.msgNormal==false $\implies$ \textbf{F}$_{[0,200]}$ v1.userCtrl==true) & valid & 14.95 & 1675.47 & 0.55 \\\cline{2-7}
    & \smc & Pr[$\leqslant$3000] ([ ] ($\neg$Commufail(0).fail)) & [0.902, 1] & 7.61 & 38.48 & 22.40\\
    \hline

    \multirow{2}{*}{R2} & SDV &  \textbf{G} (\textbf{G}$_{[0,200]}$ v2.auto==true $\wedge$  v2.msgNormal==false $\implies$ \textbf{F}$_{[0,200]}$ v2.userCtrl==true)
& valid & 18.45 & 1782.39 & 0.29 \\\cline{2-7}
    & \smc & Pr[$\leqslant$3000] ([ ] ($\neg$Commufail(1).fail)) & [0.902, 1] & 6.83 & 40.49 & 22.49\\
    \hline

    \multirow{2}{*}{R3} & SDV & \textbf{G} (\textbf{G}$_{[0,200]}$ v3.auto==true $\wedge$  v3.msgNormal==false $\implies$ \textbf{F}$_{[0,200]}$ v3.userCtrl==true) & valid &26.3 & 1932.35 & 0.69\\\cline{2-7}
    & \smc & Pr[$\leqslant$3000] ([ ] ($\neg$Commufail(2).fail)) & [0.902, 1] & 7.89 & 42.88 & 22.62\\
    \hline

    \multirow{2}{*}{R4} & SDV & \textbf{G} (v1.auto==true $\wedge$   v1.const==true $\wedge$ signType==5 $\implies$ \textbf{F}$_{[0,500]}$ v1.brake==true)  & valid & 46.8 & 1964.57 & 10.25\\\cline{2-7}
    & \smc & Pr[$\leqslant$3000] ([ ] ($\neg$R4TR.fail)) $\geqslant$ 0.95 & valid & 21.63 & 39.27 & 24.75  \\
    \hline

    \multirow{3}{*}{R5} & SDV & \textbf{G} (v1.auto==true $\wedge$  v1.const==true $\wedge$ signType==4 $\implies$ \textbf{F}$_{[0,200]}$ v1.turnLeft==true) & valid & 40.08 & 2072.23 & 11.21\\\cline{2-7}
    & \smc & Pr[$\leqslant$3000] ([ ] ($\neg$R5TR.fail)) $\geqslant$ 0.95 & valid & 25.33 & 39.20 & 24.52 \\
    \hline

    \multirow{3}{*}{R6} & SDV & \textbf{G} (v1.auto==true $\wedge$  v1.const==true $\wedge$ signType==3 $\implies$ \textbf{F}$_{[0,200]}$ v1.turnRight==true) & valid & 40.28 & 2026.57 & 6.84\\\cline{2-7}
    & \smc & Pr[$\leqslant$3000] ([ ] ($\neg$R6TR.fail)) $\geqslant$ 0.95 & valid & 25.67 & 39.20 & 24.78 \\
    \hline

    \multirow{3}{*}{R7} & SDV & \textbf{G} (v1.userCtrl==true $\wedge$  v1.const==true $\wedge$ v1.steerReq==1 $\implies$ \textbf{F}$_{[0,200]}$ v1.turnLeft==true)
& valid & 99.95 & 2144.16 & 8.64\\\cline{2-7}
    & \smc & Pr[$\leqslant$3000] ([ ] ($\neg$R7TR.fail)) $\geqslant$ 0.95 & valid & 27.6 & 39.10 & 24.67 \\
    \hline

    \multirow{3}{*}{R8} & SDV & \textbf{G} (v1.userCtrl==true $\wedge$  v1.const==true $\wedge$ v1.steerReq==2 $\implies$ \textbf{F}$_{[0,200]}$ v1.turnRight==true) & valid & 157.68 & 1852.91 & 1.78\\\cline{2-7}
    & \smc & Pr[$\leqslant$3000] ([ ] ($\neg$R8TR.fail)) $\geqslant$ 0.95 & valid & 24.48 & 39.09 & 24.65 \\
     \hline

    \multirow{3}{*}{R9} & SDV & \textbf{G} (v1.userCtrl==true $\wedge$  v1.const==true $\wedge$ v1.brakeReq==2 $\implies$ \textbf{F}$_{[0,200]}$ v1.brake==true)
 & valid & 80.46 & 1667 &1.46\\\cline{2-7}
    & \smc & Pr[$\leqslant$3000] ([ ] ($\neg$R9TR.fail)) $\geqslant$ 0.95 & valid & 28.2 & 39.26 & 24.63 \\
    \hline

    \multirow{2}{*}{R10} & SDV & \textbf{G} (v1.userCtrl==true $\wedge$  v1.const==true $\wedge$ v1.gearReq$>$v1.gear $\implies$ \textbf{F}$_{[0,200]}$ v1.acc==true)
 & valid & 6.98 & 1882.97& 2.95\\\cline{2-7}
    & \smc& Pr[$\leqslant$3000] ([ ] ($\neg$R10TR.fail)) $\geqslant$ 0.95 & valid & 29.16 & 38.92 & 24.52 \\
    \hline

    \multirow{3}{*}{R11} & SDV & \textbf{G} (v1.userCtrl==true $\wedge$  v1.const==true $\wedge$ v1.gearReq$<$v1.gear $\implies$ \textbf{F}$_{[0,200]}$ v1.dec==true)
 & valid & 259.98 & 1430 &1.85\\\cline{2-7}
    & \smc& Pr[$\leqslant$3000] ([ ] ($\neg$R11TR.fail)) $\geqslant$ 0.95 & valid & 29.03 & 39.24 & 24.57\\
    \hline

  \label{table_verification_result}%
  \end{longtable}
\end{sidewaystable}

\begin{sidewaystable}[htbp]
\begin{longtable}[htbp]{|c|c|p{320pt}|c|c|c|c|}
   \hline
    Req & Type & Expression & Result & Time (min) & Memory (Mb) & CPU (\%))\\
    \hline
     \multirow{3}{*}{R12} & SDV & \textbf{G} (v1.dx==1 $\wedge$ v2.dx==1 $\wedge$ v3.dx==1 $\implies$ v1.x$>$v2.x) & valid & 14.02 & 1320 &3.65\\\cline{2-7}
    & \smc& Pr[$\leqslant$3000]([ ] (dx[0]==1  $\wedge$  dy[0]==0  $\wedge$  dx[1]==1   $\wedge$  dy[1]==0$\newline$   $\wedge$  (in\_Str[0] $\vee$ in\_dec[0] $\vee$in\_str[0]$\vee$in\_acc[0]$\vee$in\_constSpeed[0]) $\wedge$ (in\_Str[1]$\vee$in\_str[1]) $\implies$ x[0] $>$ x[1])) $\geqslant$ 0.95 & valid & 13.27 & 57.06 & 24.31\\
    \hline

    {R13} & SDV &\textbf{G} (v1.dx==1 $\wedge$ v2.dx==1 $\wedge$ v3.dx==1 $\implies$ v2.x$>$v3.x) & valid & 35.37 & 2085 & 3.05\\\cline{2-7}
    & \smc& Pr[$\leqslant$3000]([ ] (dx[1]==1  $\wedge$  dy[1]==0  $\wedge$  dx[2]==1   $\wedge$  dy[2]==0$\newline$   $\wedge$  (in\_Str[1] $\vee$ in\_dec[1] $\vee$in\_str[1]$\vee$in\_acc[1]$\vee$in\_constSpeed[1]) $\wedge$  $\newline$(in\_Str[2]$\vee$in\_str[2]) $\implies$ x[1] $>$ x[2])) $\geqslant$ 0.95 & valid & 11.43 & 55.67 & 24.42\\
    \hline

\multirow{3}{*}{R14} & SDV & \textbf{G} ( v1auto   $\wedge$  signType==5 $\implies$ \textbf{F}$_{[0,5000]}$  (v1static   $\wedge$   v2static   $\wedge$  v3static) )  & valid & 4.67 & 1903 & 4.91\\\cline{2-7}
    & \smc& Pr[$\leqslant$3000] ([ ] ($\neg$allstop.fail)) $\geqslant$ 0.95
     & valid & 20.28 & 52.47 & 23.10\\
    \hline
	
    \multirow{3}{*}{R15} & SDV & \textbf{G} (v1.const==true $\wedge$ v2.const==true $\wedge$ v1.vel$>$v2.vel   $\implies$ \textbf{F}$_{[0,200]}$ v2.acc==true) & valid & 5.10& 2095& 4.20\\\cline{2-7}
 & \smc& Pr[$\leqslant$3000] ([ ] ($\neg$acc(1).fail)) $\geqslant$ 0.95 & valid & 21.52 & 43.68 & 22.43\\
    \hline

    \multirow{3}{*}{R16} & SDV & \textbf{G} (v2.const==true $\wedge$ v3.const==true $\wedge$ v2.vel$>$v3.vel   $\implies$ \textbf{F}$_{[0,200]}$ v3.acc==true) & valid & 0.15& 2038& 38.76\\\cline{2-7}
    & \smc& Pr[$\leqslant$3000] ([ ] ($\neg$acc(2).fail)) $\geqslant$ 0.95 & valid & 19.43 & 40.46 & 23.21\\
    \hline

    \multirow{1}{*}{R17} & SDV & \textbf{G} (v1.const==true $\wedge$ v2.const==true $\wedge$ v1.vel$<$v2.vel   $\implies$ \textbf{F}$_{[0,200]}$ v2.dec==true)  & valid & 0.15& 1963& 23.04\\ \cline{2-7}
    & \smc& Pr[$\leqslant$3000] ([ ] ($\neg$R17TR.fail)) $\geqslant$ 0.95 & valid & 27.3 & 39.10 & 24.60 \\
    \hline

    \multirow{1}{*}{R18} & SDV & \textbf{G} (v2.const==true $\wedge$ v3.const==true $\wedge$ v2.vel$<$v3.vel   $\implies$ \textbf{F}$_{[0,200]}$ v3.dec==true) & valid & 11.52& 2026& 3.99\\ \cline{2-7}
    & \smc& Pr[$\leqslant$3000] ([ ] ($\neg$R18TR.fail)) $\geqslant$ 0.95 & valid & 26.89 & 39.07 & 24.73\\
    \hline

   \multirow{3}{*}{R19} & SDV & \textbf{G} (v1.const==true $\wedge$ v2.const==true $\wedge$ dist12$>$500 $\implies$ v2.acc==true
 & valid & 0.04 & 1961& 4.38\\ \cline{2-7}
    & \smc& Pr[$\leqslant$3000] ([ ] ($\neg$R19TR.fail)) $\geqslant$ 0.95 & valid & 27.2 & 39.16 & 24.84\\
    \hline

    \multirow{3}{*}{R20} & SDV & \textbf{G} (v1.const==true $\wedge$ v2.const==true $\wedge$ dist12$>$500$\implies$ v3.acc==true
& valid & 0.01 & 1961& 4.38\\ \cline{2-7}
    & \smc& Pr[$\leqslant$3000] ([ ] ($\neg$R20TR.fail)) $\geqslant$ 0.95 & valid & 26.53 & 38.92 & 24.65\\
    \hline

    \multirow{3}{*}{R21} & SDV & \textbf{G} (v1.const==true $\wedge$ v2.const==true $\wedge$ dist12$<$safeDis $\implies$ v2.dec==true &valid & 11.13 & 2054& 4.09\\
    & \smc& Pr[$\leqslant$3000] ([ ] ($\neg$dec(1).fail)) $\geqslant$ 0.95 & valid & 22.1 & 43.23 & 22.60\\
    \hline

    \multirow{3}{*}{R22} & SDV & \textbf{G} (v2.const==true $\wedge$ v3.const==true $\wedge$ dist23$<$safeDis$\implies$ v3.dec==true
 & valid &0.03 & 2234& 3.92\\\cline{2-7}
    & \smc& Pr[$\leqslant$3000] ([ ] ($\neg$dec(1).fail)) $\geqslant$ 0.95 & valid & 20.59 & 43.27 & 22.81\\
    \hline

  \label{table_logic2}%
  \end{longtable}
\end{sidewaystable}

\begin{sidewaystable}[htbp]
\begin{longtable}[htbp]{|c|c|p{320pt}|c|c|c|c|}
   \hline
    Req & Type & Expression & Result & Time (min) & Memory (Mb) & CPU (\%))\\
    \hline

    \multirow{3}{*}{R23} & SDV & \textbf{G} (v1TurnLeft  $\implies$ \textbf{F}$_{[0,5000]}$ RunInSameLine)  & valid & 15.49 & 1492.96 & 1.82\\\cline{2-7}

    & \smc& Pr[$\leqslant$3000] ([ ] $\neg$turnleftTime12.fail)$\geqslant$ 0.95 & valid & 40.93 & 42.46 & 24.79
     \\ \hline
    \multirow{3}{*}{R24} & SDV & \textbf{G}(v2TurnLeft  $\implies$ \textbf{F}$_{[0,5000]}$ RunInSameLine)  & valid & 1.10 & 1918 &  3.48\\\cline{2-7}
    & \smc& Pr[$\leqslant$3000] ([ ] $\neg$turnleftTime23.fail)$\geqslant$ 0.95 & valid & 42.71 & 47.23 & 23.80\\
    \hline
 \multirow{3}{*}{R25} & SDV & \textbf{G} (v1.turnRight==true $\implies$ \textbf{F}$_{[0,5000]}$ v1.turnRight==false $\wedge$  v2.turnRight==false $\wedge$ v1.x==v2.x $\vee$ v1.y==v2.y)  & valid & 0.31 & 1930&  2.02\\\cline{2-7}
    & \smc& Pr[$\leqslant$3000] ([ ] $\neg$turnrightTime23.fail)$\geqslant$ 0.95 & valid & 39.13 & 55.11 & 24.63\\
    \hline

    \multirow{3}{*}{R26} & SDV & \textbf{G} (v2.turnRight==true $\implies$ F$_{[0,5000]}$ v2.turnRight==false $\wedge$  v3.turnRight==false $\wedge$ v2.x==v3.x $\vee$ v2.y==v3.y) & valid &31.26 & 2253&  3.86\\\cline{2-7}
    & \smc& Pr[$\leqslant$3000] ([ ] $\neg$turnrightTime23.fail)$\geqslant$ 0.95 & valid & 40.24 & 56.20 & 24.89\\
    \hline
    \multirow{3}{*}{R27.1} & SDV &  \textbf{G} ({{velIn}}==1 $\implies$ \textbf{F}$_{{{[40,\ 60]}}}$ {{velIn==1}}) &  valid & 28.4 & 1642 & 6.51\\\cline{2-7}
    & \smc& Pr[$\leqslant$3000] ([ ] $\neg$VehicleDynamic(0)$_{cumulative}$.fail)$\geqslant$ 0.95 & valid & 7.18 & 51.57 & 24.64\\
    \hline
    \multirow{3}{*}{R27.2} & SDV & \textbf{G} (period$_{pre}$==0 $\wedge$ period==1  $\implies$ \textbf{F}$_{[0,20]}$ velIn) &  valid & 8.42 & 1819.24 & 1.22\\\cline{2-7}
    & \smc& Pr[$\leqslant$3000] ([ ] $\neg$VehicleDynamic(0)$_{noncumulative}$.fail)$\geqslant$ 0.95 & valid &  9.17 & 51.19 & 23.25\\
    \hline
    \multirow{3}{*}{R28.1} & SDV &  \textbf{G} ({{vel2In}}==1 $\implies$ \textbf{F}$_{{{[40,\ 60]}}}$ {{vel2In==1}}) &  valid & 0.71 & 1285 & 0.5\\\cline{2-7}
    & \smc& Pr[$\leqslant$3000] ([ ] $\neg$VehicleDynamic(1)$_{cumulative}$.fail)$\geqslant$ 0.95 & valid & 7.28 & 42.70 & 24.65\\
    \hline
    \multirow{3}{*}{R28.2} & SDV & \textbf{G} (p2$_{pre}$==0 $\wedge$ p2==1  $\implies$ \textbf{F}$_{[0,\ 20]}$ vel2In) &  valid & 10.98 & 2002.17 & 8.10\\\cline{2-7}
    & \smc& Pr[$\leqslant$3000] ([ ] $\neg$VehicleDynamic(1)$_{noncumulative}$.fail)$\geqslant$ 0.95 & valid & 5.6 & 51.79 & 24.66\\
    \hline

    \multirow{3}{*}{R29.1} & SDV &  \textbf{G} ({{vel3In}}==1 $\implies$ \textbf{F}$_{{{[40,\ 60]}}}$ {{vel2In==1}}) &  valid & 0.65 & 54.69 &
   8.3\\\cline{2-7}
    & \smc& Pr[$\leqslant$3000] ([ ] $\neg$VehicleDynamic(2)$_{cumulative}$.fail)$\geqslant$ 0.95 & valid & 7.23 & 42.70 & 24.69\\
    \hline
    \multirow{3}{*}{R29.2} & SDV & \textbf{G} (p3$_{pre}$==0 $\wedge$ p3==1  $\implies$ \textbf{F}$_{[0,\ 20]}$ vel3In) &  valid & 6.97 & 2084 & 8.30\\\cline{2-7}
    & \smc& Pr[$\leqslant$3000] ([ ] $\neg$VehicleDynamic(2)$_{noncumulative}$.fail)$\geqslant$ 0.95 & valid & 6.13 & 51.79 & 24.49\\
    \hline

    \multirow{3}{*}{R30} & SDV & \textbf{G} ($\neg$msgNormal  $\implies$ $\neg$auto\textbf{U}$_{[2000,\infty]}$ auto==true) & valid & 4h & 1588.14 & 7.27\\\cline{2-7}
    & \smc& Pr[$\leqslant$3000] ([ ] ($\neg$sporadic(0).fail)) $\geqslant$ 0.95 & valid & 0.001 & 30.96 & 0.9\\
    \hline

    \multirow{3}{*}{R31} & SDV & \textbf{G} ($\neg$msgNormal2  $\implies$ $\neg$auto2\textbf{U}$_{[2000,\infty]}$ auto2==true)  & valid &49.15  & 2253&  38.85\\\cline{2-7}
    & \smc& Pr[$\leqslant$3000] ([ ] ($\neg$sporadic(1).fail)) $\geqslant$ 0.95 & valid & 0.002 & 30.96 & 1.2\\
    \hline
    \multirow{3}{*}{R32} & SDV &\textbf{G} ($\neg$msgNormal  $\implies$ $\neg$auto3\textbf{U}$_{[2000,\infty]}$ auto3==true)  & valid &59.15  & 2125&  36.40\\\cline{2-7}
    & \smc& Pr[$\leqslant$3000] ([ ] ($\neg$sporadic(2).fail)) $\geqslant$ 0.95 & valid & 23:03 & 1167.54 & 22.24\\
    \hline
    \multirow{3}{*}{R33} & SDV &  \textbf{G} (r$_{pre}$==0 $\wedge$ r$>$0 $\implies$ \textbf{F}$_{[100,300]}$ s==r)  &  valid & 19.7 & 1661 &7.31\\\cline{2-7}
    & \smc& Pr[$\leqslant$3000] ([ ] ($\neg$v1Controller$_{execution}$.fail)) $\geqslant$ 0.95  & valid & 7.7 & 51.67 & 24.71\\
    \hline

    \multirow{3}{*}{R34} & SDV &  \textbf{G} (r2$_{pre}$==0 $\wedge$ r2$>$0 $\implies$ \textbf{F}$_{[100,300]}$ s2==r2)  &  valid & 1.38 & 1493 &3.63\\\cline{2-7}
    & \smc & Pr[$\leqslant$3000] ([ ] ($\neg$v2Controller$_{execution}$.fail)) $\geqslant$ 0.95 & valid & 6.5 & 42.61 & 24.72\\
    \hline

    \multirow{3}{*}{R35} & SDV & \textbf{G} (r3$_{pre}$==0 $\wedge$ r3$>$0 $\implies$ \textbf{F}$_{[100,300]}$ s3==r3)  &  valid & 1.08 & 1551 &20.3\\\cline{2-7}
    & \smc & Pr[$\leqslant$3000] ([ ] ($\neg$v3Controller$_{execution}$.fail)) $\geqslant$ 0.95 & valid & 7.12 & 43.24 & 24.79\\
    \hline

    \multirow{3}{*}{R36} & SDV & \textbf{G} (rcd$_{pre}$==0 $\wedge$ rcd$>$0 $\implies$ \textbf{F}$_{[50,100]}$ scd==rcd)  &  valid & 1.83 & 1533.81 &1.49\\\cline{2-7}
    & \smc & Pr[$\leqslant$3000] ([ ] ($\neg$sensor1$_{execution}$.fail)) $\geqslant$ 0.95 &valid & 7.47 & 43.24 & 24.97\\
    \hline

  \label{table_logic3}%
  \end{longtable}
\end{sidewaystable}

\begin{sidewaystable}[htbp]
\begin{longtable}[htbp]{|c|c|p{320pt}|c|c|c|c|}
   \hline
    Req & Type & Expression & Result & Time (min) & Memory (Mb) & CPU (\%))\\
    \hline
     \multirow{3}{*}{R37} & SDV & \textbf{G} (rcd2$_{pre}$==0 $\wedge$ rcd2$>$0 $\implies$ \textbf{F}$_{[50,100]}$ scd2==rcd2)  &  valid & 1.23 & 12.69 &8.07\\\cline{2-7}
    & \smc& Pr[$\leqslant$3000] ([ ] ($\neg$sensor2$_{execution}$.fail)) $\geqslant$ 0.95 & valid & 7.03 & 42.72 & 24.63\\
    \hline

    \multirow{1}{*}{R38} & SDV & \textbf{G} (rcd3$_{pre}$==0 $\wedge$ rcd3$>$0 $\implies$ \textbf{F}$_{[50,\ 100]}$ scd3==rcd3)  &  valid & 2.26 & 1515.98 &3.41\\\cline{2-7}
    & \smc& Pr[$\leqslant$3000] ([ ] ($\neg$sensor3$_{execution}$.fail)) $\geqslant$ 0.95 & valid & 9.91 & 35.12 & 24.56\\
    \hline

   \multirow{1}{*}{R39} & SDV & \textbf{G} (sou1$_{pre}$==0 $\wedge$ sou2$>$0 $\implies$ (\textbf{F}$_{[300, 700]}$ tar1==cur1 $\wedge$ tar1$\neq$0))  & valid & 30.01 & 1718.32 & 3.78\\\cline{2-7}
    & \smc& Pr[$\leqslant$3000] ([ ] forall (e:endtoend) ($\neg$e.fail)) $\geqslant$ 0.95 & valid & 5.15 & 39.38 & 24.54\\
    \hline

    \multirow{1}{*}{R40} & SDV & \textbf{G} (sou2$_{pre}$==0 $\wedge$ sou2$>$0 $\implies$ (\textbf{F}$_{[300, 700]}$ tar2==cur2 $\wedge$ tar2$\neq$0))  & valid & 10.08 & 1204.86 & 5.96\\\cline{2-7}
    & \smc& Pr[$\leqslant$3000] ([ ] forall (e:endtoend2) ($\neg$e.fail)) $\geqslant$ 0.95 & valid & 6.27 & 42.56 & 24.68\\
    \hline
    \multirow{3}{*}{R41} & SDV & G(v1.const==true $\wedge$ v2.const==true $\wedge$ dist12$>$500$\implies$ v2.acc==true
 & valid & 0.04 & 1961& 4.38\\\cline{2-7}
    & \smc& Pr[$\leqslant$3000] ([ ] ($\neg$execv1.fail)) $\geqslant$ 0.95 & valid & 7.28 & 42.70 & 24.76\\
    \hline
    \multirow{3}{*}{R42} & SDV & G(v1.const==true $\wedge$ v2.const==true $\wedge$ dist12$>$500$\implies$ v3.acc==true
& valid & 0.01 & 1961& 4.38\\\cline{2-7}
    & \smc& Pr[$\leqslant$3000] ([ ] ($\neg$execv2.fail)) $\geqslant$ 0.95 &valid & 7.47 & 42.70 & 24.67\\
    \hline

    \multirow{3}{*}{R43} & SDV & G(v2.const==true $\wedge$ v3.const==true $\wedge$ dist23$<$safeDis$\implies$ v3.dec==true
 & valid &0.03 & 2234& 3.92\\\cline{2-7}
    & \smc& Pr[$\leqslant$3000] ([ ] ($\neg$execv3.fail)) $\geqslant$ 0.95 & valid & 7.33 & 42.70 & 24.77\\
    \hline

    \multirow{3}{*}{R44} & SDV &  \textbf{G} ($\neg$ r$_{pre}$==0 $\wedge$ r$>$0 $\implies$ \textbf{F}$_{[0,200]}$ posIn $\wedge$  \textbf{F}$_{[0,200]}$ AposIn  $\wedge$  \textbf{F}$_{[0,200]}$ velIn $\wedge$ \textbf{F}$_{[0,200]}$ AvelIn)  &  valid & 32.53 & 36.51 & 15.31\\\cline{2-7}
    & \smc& Pr[$\leqslant$3000] ([ ] $\neg$v1Controller$_{synchronization}$.fail)) $\geqslant$ 0.95 & valid & 5.98 & 50.84 & 24.51\\
    \hline

      \multirow{3}{*}{R45} & SDV & \textbf{G} ($\neg$ r2$_{pre}$==0 $\wedge$ r2$>$0 $\implies$ \textbf{F}$_{[0,200]}$ posIn2 $\wedge$  \textbf{F}$_{[0,200]}$ AposIn2  $\wedge$  \textbf{F}$_{[0,200]}$ velIn2 $\wedge$ \textbf{F}$_{[0,200]}$ AvelIn2)  &  valid &105.75 & 1197.67 & 6.56\\ \cline{2-7}
    & \smc& Pr[$\leqslant$3000] ([ ] $\neg$v2Controller$_{synchronization}$.fail)) $\geqslant$ 0.95  & valid & 7.17 & 42.37 & 24.74\\
    \hline

    \multirow{3}{*}{R46} & SDV & \textbf{G} ($\neg$ r3$_{pre}$==0 $\wedge$ r3$>$0 $\implies$ \textbf{F}$_{[0,200]}$ posIn3 $\wedge$  \textbf{F}$_{[0,200]}$ AposIn3  $\wedge$  \textbf{F}$_{[0,200]}$ velIn3 $\wedge$ \textbf{F}$_{[0,200]}$ AvelIn3)  &  valid & 1.04 & 1249.30 & 4.3\\\cline{2-7}
    & \smc& Pr[$\leqslant$3000] ([ ] $\neg$v3Controller$_{synchronization}$.fail)) $\geqslant$ 0.95 & valid & 7.33 & 42.37 & 24.52\\
    \hline

    \multirow{3}{*}{R47} & SDV &\textbf{G} (et$_{con1}$==w$_{con1}$ $\wedge$ et$_{con2}$==w$_{con2}$ $\wedge$ et$_{sen1}$==w$_{sen1}$ $\wedge$ et$_{sen2}$==w$_{sen2}$ $\implies$ w$_{con1}$ + w$_{con2}$ + w$_{sen1}$ + w$_{sen2}$ $\geqslant$ EndtoEnd) & valid &31.26 & 2253&  38.86\\\cline{2-7}
    & \smc& Pr[$\leqslant$3000]($<$\ $>$ $\neg$(et$_{con1}$==w$_{con1}$ $\wedge$ et$_{con2}$==w$_{con2}$ $\wedge$ et$_{sen1}$==w$_{sen1}$ $\wedge$ et$_{sen2}$==w$_{sen2}$ $\implies$ w$_{con1}$ + w$_{con2}$ + w$_{sen1}$ + w$_{sen2}$ $\geqslant$ EndtoEnd) $\leqslant$ 0.05 & valid & 165.65 & 175.24 & 23.53\\
    \hline

    \multirow{3}{*}{R48} & SDV & \textbf{G} (BrakeEnergy $<$ 30000)  &  valid & 32.53 & 36.51 & 1.31 \\ \cline{2-7}
    & SMC & E[$\leqslant$3000;100] (max: energy.braking\_energy) & 7944$\pm$1172 & 8.7 & 51.44 & 24.58\\\cline{2-7}

    & SMC & simulate 100 [$\leqslant$3000] (total\_energy) & valid & 7.67 & 45.50 & 24.63\\
    \hline

    \multirow{3}{*}{R49} & SDV & \textbf{G} (ConEnergy $\leqslant$ 30)  & valid &18.05 & 1000.8&  2.05\\\cline{2-7}
    & \smc& Pr[$\leqslant$3000] ([ ] energy\_con $\leqslant$ 30)) $\geqslant$ 0.95 & valid & 18.05  & 1000.8 & 0.9\\
    \hline
    \multirow{3}{*}{R50} & SDV & \textbf{G} (CDEnergy $\leqslant$ 5) & valid &31.26 & 2253&  38.86\\\cline{2-7}
    & \smc& Pr[$\leqslant$3000] ([ ] energy\_sen $\leqslant$ 5)) $\geqslant$ 0.95 & valid & 20.2 & 80.9 & 24.02\\
    \hline
    \label{table_logic4}%
  \end{longtable}
\end{sidewaystable}

\begin{figure}[htbp]
\centering
\subfigure[{\scriptsize \gt{signType}}]{
\label{CE1}
\includegraphics[width=3.8in]{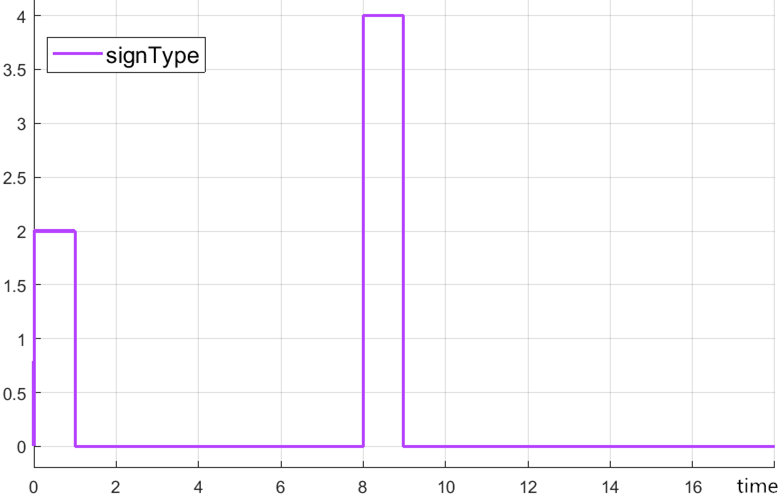}}
\subfigure[Vehicles' position]{
\label{CE2}
\includegraphics[width=3.8in]{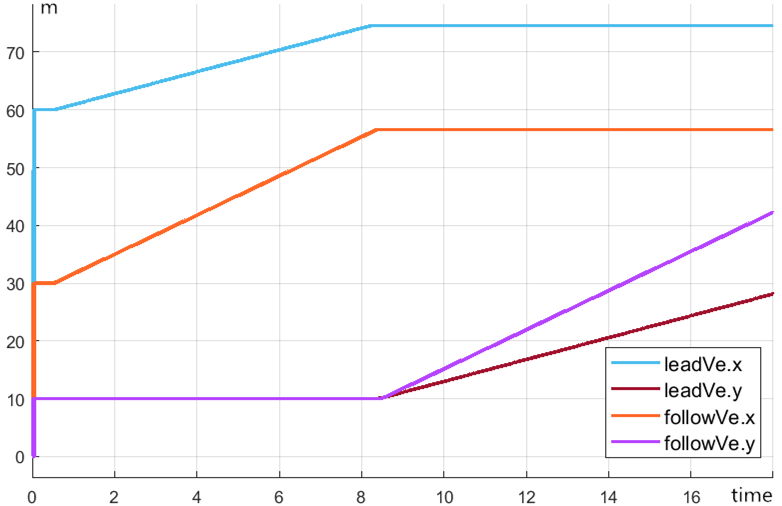}}
\caption{CE in SDV: In (b), \emph{leadVe.y} equals to \emph{followVe.y}) from 0 to 8s, which indicates that the two vehicles run in the same lane. In (a), {\gt{signType}} becomes ``turn left (4)'' at 8s and the lead vehicle begins to turn left. However, the following vehicle turns left at a different location (\emph{leadVe.x} is not equal to \emph{followVe.x} at 9s). Hence, the ``x'' is unequal after turning, which means that the two vehicles run on different lanes. Thus R23 and R24 is violated.}
\label{SDVCE}
\end{figure}

\begin{figure}[htbp]
  \centering
  \includegraphics[width=4in]{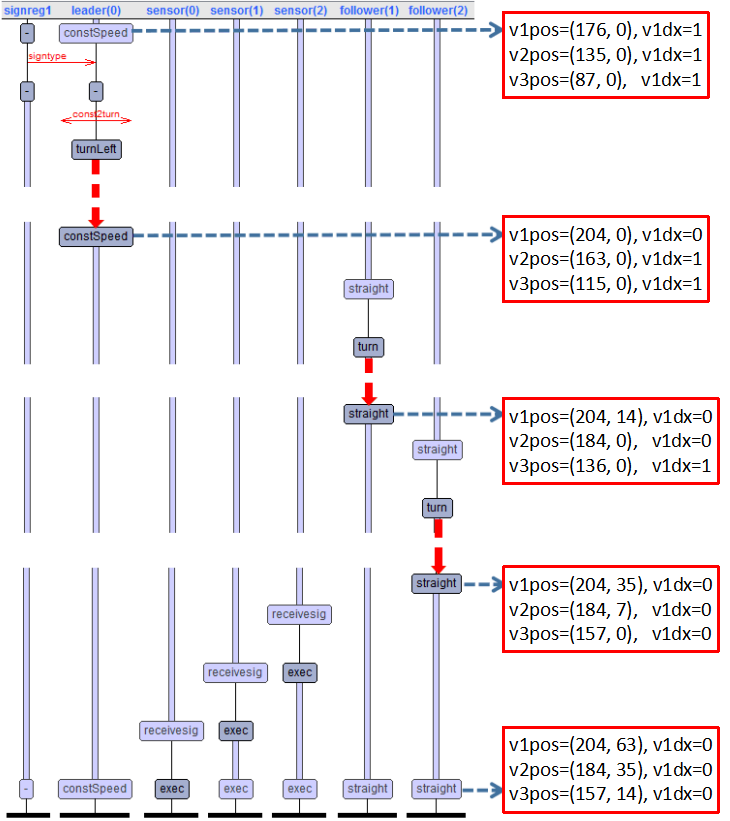}
  \caption{CE in SMC: At the beginning, three vehicles run towards direction (1,0). The three values of \emph{posy} are the same, which indicates that the three vehicles are in the same lane. After a left turn sign is detected,  the lead vehicle starts to turn left. Since the first follower turns left at (184, 0) and the second turns left at (157, 0), the three values of \emph{posx} are not equal after turning, which indicates the three vehicle are not running in the same lane. Thus R5 is violated.}
  \label{cesmc}
\end{figure}

The invalid property R23 is identified using SDV and \smc. The counter-examples (CE) generated by SDV is shown in Fig.\ref{SDVCE}.  We perform the simulation in \smc\ to obtain the error trace of CE, which is illustrated in Fig.\ref{cesmc}. \emph{DirectX} and \emph{DirectY} represent the directions of the three vehicles and \emph{posx} and \emph{posy} indicate the position of the three vehicles.
After analyzing CE, the cause of errors was found:
After the lead vehicle turns left, the following vehicle begins to turn left after it detects that the running direction of the lead vehicle is changed. The two vehicles turn left at different locations. We modify SDV and \smc\ models according to the CE: a turning location where the lead vehicle turns is recorded and sent to the following vehicle. Hence all the following vehicles should turn at the same turning location to ensure that the three vehicles are still in a lane after turning. After the refinement, R23 and R24 are satisfied in both SDV and \smc.

\begin{figure}[htbp]
  \centering
  \includegraphics[width=5in]{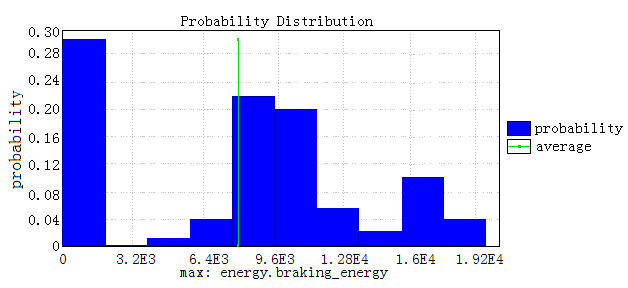}
  \caption{Probabilistic distribution of the expected energy consumption}
  \label{energybraking}
\end{figure}

\begin{figure}[htbp]
  \centering
  \includegraphics[width=5in]{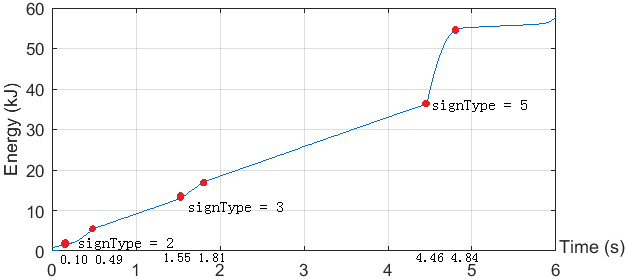}
  \caption{Total energy consumption estimation of the lead vehicle}
  \label{energyes}
\end{figure}


\emph{Probability\ Estimation} query is applied on R1-R47, R49-R50, indicating that the property can be verified as valid within [0.902, 1].
For estimate the maximum energy consumption for braking the vehicles (R48), we adapt \emph{Expected\ value} query and generate probability distribution diagram (Fig.\ref{energybraking}) of the expected energy consumption. It illustrates that the average of energy consumption is approx. 7944J and the expected energy consumption is always less than 30kJ. The total energy of the lead vehicle is estimated with \emph{Simulation} query for 100 times within 300 time units. Fig.\ref{energyes} shows the result of one time simulation. At the beginning, the lead vehicle runs at constant speed, hence the total energy consumption increases constantly. When it recognizes the minimum speed limit sign ({\gt{signType}} = 2), the vehicle starts to accelerate thus energy is consumed faster. The total energy then increases constantly with slight fluctuation due to energy consumption of the sensor and the camera. When the vehicle recognizes a turn right sign ({\gt{signType}} = 3), it turns right from 1.55s to 1.88s, during which the energy is consumed faster than when the vehicle is running constantly. The vehicle finally recognizes a stop sign ({\gt{signType}} = 5) at 4.46s and stops at 4.84s. The total energy consumed during braking is 18.33kJ, which is less than 30kJ, which validates R48. After the vehicle stops, the sensors and the camera still work, thus the consumed energy increases slowly.

\chapter{Verification of Comparison Constraint in \smc}
For comparison constraint (R47), which states that the execution time interval measured from {\gt{v1Controller}} to {\gt{v2Controller}} needs to be greater than or equal to the sum of the worst-case execution time of \fp s, can be specified as:
 \begin{center}
 $\varphi$ = [ ](et$_{con1}$==w$_{con1}$ $\wedge$ et$_{con2}$==w$_{con2}$ $\wedge$ et$_{sen1}$==w$_{sen1}$ $\wedge$ et$_{sen2}$==w$_{sen2}$ $\implies$ $\newline$ w$_{con1}$ + w$_{con2}$ + w$_{sen1}$ + w$_{sen2}$ $\geqslant$ \emph{EndtoEnd}.
 \end{center}
We verify this requirement in \smc\ with the query:
\begin{eqnarray}
Pr[\leqslant 3000] \varphi \geqslant 0.95
\end{eqnarray}
  which means that the probability that $\varphi$ is satisfied should be greater than or equal to 0.95. However, \smc\ cannot give the decidable result because of the large state space.
Hence we transform the ``always'' operator with its dual operator ``eventually'' and then verify that the probability that the negation of $\varphi$ is satisfied is less than or equal to 0.05, i.e.,
\begin{eqnarray}
Pr[\leqslant 3000] \neg \varphi \leqslant 0.05
\end{eqnarray}

In CTL, A[ ] $\psi$ $\equiv$ E$<$ $>$ $\neg\ \psi$. Here we want to validates the equivalence between (9.1) and (9.2) by specifying same requirements into the two different types of formulas and compare the verification results.
We applied the existed case studies of Fischer protocol and train gate system provided in \smc\.
In Fischer protocol example, we check the mutual exclusion property that two processes can get into the critical section at the same time.  We verify

\begin{eqnarray}
 Pr[\leqslant 3000] ([\ ] \forall (i:id) \forall (j:id) P(i).cs \wedge P(j).cs \implies \newline i=j) \geqslant p
\end{eqnarray}

\begin{eqnarray}
 Pr[\leqslant3000] (<\ >\neg(\forall (i:id) \forall (j:id) P(i).cs \wedge P(j).cs \implies i=j)) \leqslant 1-p
\end{eqnarray}

The verification results of (9.3) and (9.4) are shown in Table.\ref{table_fischer}. The verification results of (9.3) and (9.4) are the same. We then conduct the same verification for the mutual exclusion property in train gate system, i.e., only one train can cross the gate at one time.

\begin{eqnarray}
Pr[\leqslant100] ([ ] \forall (i:id) \forall (j:id) Train(i).cross \wedge Train(j).cross \implies i=j) \geqslant p
\end{eqnarray}

\begin{equation}
Pr[\leqslant100] (< >\neg(\forall (i:id) \forall (j:id) Train(i).cross \wedge Train(j).cross \\ \implies i=j)) \leqslant 1-p
\end{equation}

Table.\ref{table_train} shows that verification result of (9.5) and (9.6) are the same.
From the above experimental results, we can conclude that (9.1) and (9.2) are equivalent. By verifying the negation of R47 is satisfied with the probability at most 0.05, we can get the decidable result of R47.

\begin{table*}[t]
  \small
  \centering
  \caption{Verification Results of Mutual Exclusion in Fischer Protocol Example}
    \begin{tabular}{|p{250pt}|c|c|c|c|}
    \hline
    Expression & Result & Time (s) & No. of Runs & Mem (Kb) \\
    \hline
    Pr [$\leqslant$3000] ([ ] forall (i:id\_t) forall (j:id\_t) P(i).cs $\wedge$ P(j).cs $\implies$ i == j) $\geqslant$ 0.95 & valid & 0.003 & 140 & 26188\\
    \hline
    Pr [$\leqslant$3000] ($<\ >$ $\neg$ (forall (i:id\_t) forall (j:id\_t) P(i).cs $\wedge$ P(j).cs $\implies$ i == j)) $\leqslant$ 0.05 & valid & 8.672 & 140 & 26188\\\hline
    Pr [$\leqslant$3000] ([ ] forall (i:id\_t) forall (j:id\_t) P(i).cs $\wedge$ P(j).cs $\implies$ i == j) $\geqslant$ 0.9 & valid & 0.009 & 67 & 26752\\
    \hline
    Pr [$\leqslant$3000] ($<\ >$ $\neg$ (forall (i:id\_t) forall (j:id\_t) P(i).cs $\wedge$ P(j).cs $\implies$ i == j)) $\leqslant$ 0.1 & valid & 4.175 & 67 & 26752\\\hline
    Pr [$\leqslant$3000] ([ ] forall (i:id\_t) forall (j:id\_t) P(i).cs $\wedge$ P(j).cs $\implies$ i == j) $\geqslant$ 0.8 & valid & 0.007 & 60 & 27688\\
    \hline
    Pr [$\leqslant$3000] ($<\ >$ $\neg$ (forall (i:id\_t) forall (j:id\_t) P(i).cs $\wedge$ P(j).cs $\implies$ i == j)) $\leqslant$ 0.2 & valid & 3.081 & 60 & 27688\\\hline
    Pr [$\leqslant$3000] ([ ] forall (i:id\_t) forall (j:id\_t) P(i).cs $\wedge$ P(j).cs $\implies$ i == j) $\geqslant$ 0.7 & valid & 0.002 & 53 & 26996\\
    \hline
    Pr [$\leqslant$3000] ($<\ >$ $\neg$ (forall (i:id\_t) forall (j:id\_t) P(i).cs $\wedge$ P(j).cs $\implies$ i == j)) $\leqslant$ 0.3 & valid & 3.263 & 53 & 26996\\\hline
    Pr [$\leqslant$3000] ([ ] forall (i:id\_t) forall (j:id\_t) P(i).cs $\wedge$ P(j).cs $\implies$ i == j) $\geqslant$ 0.6 & valid & 0.007 & 31 & 27584\\
    \hline
    Pr [$\leqslant$3000] ($<\ >$ $\neg$ (forall (i:id\_t) forall (j:id\_t) P(i).cs $\wedge$ P(j).cs $\implies$ i == j)) $\leqslant$ 0.4 & valid & 1.755 & 31 & 27584\\\hline
    Pr [$\leqslant$3000] ([ ] forall (i:id\_t) forall (j:id\_t) P(i).cs $\wedge$ P(j).cs $\implies$ i == j) $\geqslant$ 0.5 & valid & 0.005 & 26 & 27724\\
    \hline
    Pr [$\leqslant$3000] ($<\ >$ $\neg$ (forall (i:id\_t) forall (j:id\_t) P(i).cs $\wedge$ P(j).cs $\implies$ i == j)) $\leqslant$ 0.5 & valid & 1.487 & 26 & 27740\\\hline

    \end{tabular}%
  \label{table_fischer}%
\end{table*}

\begin{table*}[t]
  \small
  \centering
  \caption{Verification Results of Mutual Exclusion in Train Gate Example}
    \begin{tabular}{|p{300pt}|c|c|c|c|}
    \hline
    Expression & Result & Time (s) & No. of Runs & Mem (Kb) \\
    \hline
    Pr [$\leqslant$100] ([ ] forall (i:id\_t) forall (j:id\_t) Train(i).Cross $\wedge$ Train(j).Cross $\implies$ i == j) $\geqslant$ 0.95 & valid & 0.004 & 140 & 26988\\
    \hline
    Pr [$\leqslant$100] ($<\ >$ $\neg$ (forall (i:id\_t) forall (j:id\_t) Train(i).Cross $\wedge$ Train(j).Cross $\implies$ i == j)) $\leqslant$ 0.05 & valid & 0.078 & 140 & 26988\\\hline
    Pr [$\leqslant$100] ([ ] forall (i:id\_t) forall (j:id\_t) Train(i).Cross $\wedge$ Train(j).Cross $\implies$ i == j) $\geqslant$ 0.85 & valid & 0.003 & 126 & 26980\\
    \hline
    Pr [$\leqslant$100] ($<\ >$ $\neg$ (forall (i:id\_t) forall (j:id\_t) Train(i).Cross $\wedge$ Train(j).Cross $\implies$ i == j)) $\leqslant$ 0.15 & valid & 0.003 & 126 & 26996\\\hline
Pr [$\leqslant$100] ([ ] forall (i:id\_t) forall (j:id\_t) Train(i).Cross $\wedge$ Train(j).Cross $\implies$ i == j) $\geqslant$ 0.75 & valid & 0.004 & 111 & 27080\\
    \hline
    Pr [$\leqslant$100] ($<\ >$ $\neg$ (forall (i:id\_t) forall (j:id\_t) Train(i).Cross $\wedge$ Train(j).Cross $\implies$ i == j)) $\leqslant$ 0.25 & valid & 0.003 & 111 & 27436\\\hline
Pr [$\leqslant$100] ([ ] forall (i:id\_t) forall (j:id\_t) Train(i).Cross $\wedge$ Train(j).Cross $\implies$ i == j) $\geqslant$ 0.65 & valid & 0.003 & 96 & 27436\\
    \hline
    Pr [$\leqslant$100] ($<\ >$ $\neg$ (forall (i:id\_t) forall (j:id\_t) Train(i).Cross $\wedge$ Train(j).Cross $\implies$ i == j)) $\leqslant$ 0.35 & valid & 0.002 & 96 & 27260\\\hline
Pr [$\leqslant$100] ([ ] forall (i:id\_t) forall (j:id\_t) Train(i).Cross $\wedge$ Train(j).Cross $\implies$ i == j) $\geqslant$ 0.55 & valid & 0.003 & 81 & 27272\\
    \hline
    Pr [$\leqslant$100] ($<\ >$ $\neg$ (forall (i:id\_t) forall (j:id\_t) Train(i).Cross $\wedge$ Train(j).Cross $\implies$ i == j)) $\leqslant$ 0.45 & valid & 0.057 & 81 & 27272\\\hline
Pr [$\leqslant$100] ([ ] forall (i:id\_t) forall (j:id\_t) Train(i).Cross $\wedge$ Train(j).Cross $\implies$ i == j) $\geqslant$ 0.45 & valid & 0.003 & 67 & 27308\\
    \hline
    Pr [$\leqslant$100] ($<\ >$ $\neg$ (forall (i:id\_t) forall (j:id\_t) Train(i).Cross $\wedge$ Train(j).Cross $\implies$ i == j)) $\leqslant$ 0.55 & valid & 0.032 & 67 & 27328\\\hline
Pr [$\leqslant$100] ([ ] forall (i:id\_t) forall (j:id\_t) Train(i).Cross $\wedge$ Train(j).Cross $\implies$ i == j) $\geqslant$ 0.35 & valid & 0.005 & 52 & 26752\\
    \hline
    Pr [$\leqslant$100] ($<\ >$ $\neg$ (forall (i:id\_t) forall (j:id\_t) Train(i).Cross $\wedge$ Train(j).Cross $\implies$ i == j)) $\leqslant$ 0.65 & valid & 0.041 & 52 & 26752\\\hline
Pr [$\leqslant$100] ([ ] forall (i:id\_t) forall (j:id\_t) Train(i).Cross $\wedge$ Train(j).Cross $\implies$ i == j) $\geqslant$ 0.25 & valid & 0.003 & 37 & 27896\\
    \hline
    Pr [$\leqslant$100] ($<\ >$ $\neg$ (forall (i:id\_t) forall (j:id\_t) Train(i).Cross $\wedge$ Train(j).Cross $\implies$ i == j)) $\leqslant$ 0.75 & valid & 0.007 & 37 & 27840\\\hline
Pr [$\leqslant$100] ([ ] forall (i:id\_t) forall (j:id\_t) Train(i).Cross $\wedge$ Train(j).Cross $\implies$ i == j) $\geqslant$ 0.15 & valid & 0.007 & 23 & 27840\\
    \hline
    Pr [$\leqslant$100] ($<\ >$ $\neg$ (forall (i:id\_t) forall (j:id\_t) Train(i).Cross $\wedge$ Train(j).Cross $\implies$ i == j)) $\leqslant$ 0.85 & valid & 0.007 & 23 & 27840\\\hline
Pr [$\leqslant$100] ([ ] forall (i:id\_t) forall (j:id\_t) Train(i).Cross $\wedge$ Train(j).Cross $\implies$ i == j) $\geqslant$ 0.05 & valid & 0.0.007 & 8 & 27840\\
    \hline
    Pr [$\leqslant$100] ($<\ >$ $\neg$ (forall (i:id\_t) forall (j:id\_t) Train(i).Cross $\wedge$ Train(j).Cross $\implies$ i == j)) $\leqslant$ 0.95 & valid & 0.007 & 8 & 27840\\\hline
    \end{tabular}%
  \label{table_train}%
\end{table*}

\chapter{Related Work}
\label{sec:r-work}
In the context of \ed, efforts on the integration of \ed\ and formal techniques based on timing constraints were investigated in several works \cite{Qureshi2012,kress13,ksafecomp11}, which are however, limited to the executional aspects of system functions without addressing energy-aware behaviors. Kang \cite{ksac14}, Goknil et al. \cite{Goknil2013} and Seceleanu et al. \cite{seceleanu2017analyzing} defined the execution semantics of both the controller and the environment of industrial systems in CCSL \cite{abs-ccsl} which are also given as mapping to \uppaal\ models amenable to model checking. In contrast to our current work, those approaches lack precise stochastic annotations specifying continuous dynamics in particular regarding different energy consumption rates during execution. Though, Kang et al. \cite{kiciea16,kapsec15} and Marinescu et al. \cite{Marinescu3762} present both simulation and model checking approaches of \simu\ and \smc\ on \ed\ models, neither formal specification nor verification of extended \ed\ timing constraints with probability were conducted. Our approach is a first application on the integration of \ed\ and formal V\&V techniques based on a composition of energy- and probabilistic extension of \ed/\tdl\ constraints. Gholami  \cite{gholami2016verifying} extended the block library of \simu\ in order to specify basic LTL properties and to transform them into assertions that are verifiable by \sdv. However, this work is limited to functional properties (i.e., Neither energy nor timing constraints are addressed) and restricted to specify a subset of LTL.

\chapter{Conclusion}
\label{sec:conclusion}

We present an approach to perform V\&V on nonfunctional properties of
(cooperative)-automotive systems using SDV and \smc\ at the early design
phase:  \begin{inparaenum} \item ET requirements are translated into
POMs to perform formal verification using \sdv; \item Probabilistic
extension of \ed\ constraints is defined and the semantics of the \xtc\
is translated into analyzable \smc\ models for formal verification;
\item A set of mapping rules is proposed to facilitate the guarantee of
translation and V\&V are performed on the \xtc. \end{inparaenum} As
ongoing work, a dedicated plugin that directly provides translation of
LTL2\simu\ and of S/S to \smc\ in fully automatic will supplement our
tool chain, A-BeTA (A$\beta$: EAST-\textbf{A}DL \textbf{B}ehavioral
Modeling and \textbf{T}ranslation into \textbf{A}naly-zable Model).

\chapter*{Acknowledgment}
This work is supported by the National Natural Science Foundation of China and International Cooperation \& Exchange Program (46000-41030005) within the project EASY.

\end{document}